%% file: pathos4.tex
\documentclass[a4paper,fleqn,usenatbib]{mnras}

\usepackage[normalem]{ulem}

\usepackage{newtxtext,newtxmath}
\usepackage{siunitx}
\usepackage[T1]{fontenc}
\usepackage{ae,aecompl}

\usepackage{graphicx}	% Including figure files
\usepackage{amsmath}	% Advanced maths commands
\usepackage{pdflscape}
\usepackage{hyperref}
\defcitealias{2019MNRAS.490.3806N}{I}
\defcitealias{2020MNRAS.495.4924N}{II}
\defcitealias{2020MNRAS.498.5972N}{III}
\title[Frequencies and Age vs. $R_{\rm P}$ for exoplanets in OCs]{A PSF-based Approach to TESS High quality data Of Stellar
  clusters (PATHOS) - IV.  Candidate exoplanets around stars in open
  clusters: frequency and age--planetary radius distribution.}

\author[D.\ Nardiello]{D.\ Nardiello$^{1,2}$\thanks{E-mail: domenico.nardiello@lam.fr}, 
M.\ Deleuil$^{1}$,
G.\ Mantovan$^{3,2}$,
L.\ Malavolta$^{3,2}$,
G.\ Lacedelli$^{3,2}$,
\newauthor
M.\ Libralato$^{4}$,
L.\ R.\ Bedin$^{2}$,
L.\ Borsato$^{2}$,
V.\ Granata$^{3,2}$,
G.\ Piotto$^{3,2}$\\
$^{1}$Aix Marseille Univ, CNRS, CNES, LAM, Marseille, France \\
$^{2}$Istituto Nazionale di Astrofisica - Osservatorio Astronomico di Padova, Vicolo dell'Osservatorio 5, IT-35122, Padova, Italy \\
$^{3}$Dipartimento di Fisica e Astronomia ``Galileo Galilei'', Universit\`a di Padova, Vicolo dell'Osservatorio 3, IT-35122, Padova, Italy  \\
$^{4}$AURA for the European Space Agency (ESA), ESA Office, Space Telescope Science Institute, 3700 San Martin Drive, Baltimore MD 21218, USA \\
}

\date{Accepted 2021 May 19. Received 2021 May 18; in original form 2021 February 15.}

\pubyear{2020}
\begin{document}
\label{firstpage}
\pagerange{\pageref{firstpage}--\pageref{lastpage}}
\maketitle

\begin{abstract}

The knowledge of the ages of stars hosting exoplanets allows us to
obtain an overview on the evolution of exoplanets and understand the
mechanisms affecting their life. The measurement of the ages of stars
in the Galaxy is usually affected by large uncertainties. An
exception are the stellar clusters: for their coeval members, born from
the same molecular cloud, ages can be measured with extreme accuracy.
In this context, the project PATHOS is providing candidate exoplanets
orbiting members of stellar clusters and associations through the
analysis of high-precision light curves obtained with cutting-edge
tools.
In this work, we exploited the data collected during the second year
of the {\it TESS} mission. We extracted, analysed, and modelled the
light curves of $\sim 90\,000$ stars in open clusters located in the
northern ecliptic hemisphere in order to find candidate exoplanets. 
We measured the frequencies of candidate exoplanets in open clusters
for different orbital periods and planetary radii, taking into
  account the detection efficiency of our pipeline and the false
  positive probabilities of our candidates. We analysed the Age--$R_{\rm P}$
distribution of candidate and confirmed exoplanets with periods
$<100$~days and well constrained ages. While no peculiar trends are
observed for Jupiter-size and (super-)Earth-size planets, we found
that objects with $4\,R_{\rm Earth} \lesssim R_{\rm P} \lesssim
13\,R_{\rm Earth}$ are concentrated at ages $\lesssim 200$~Myr;
different scenarios (atmospheric losses, migration, etc.) are
  considered to explain the observed age-$R_{\rm P}$ distribution.
 \end{abstract}

\begin{keywords}
  techniques: image processing -- techniques: photometric -- Galaxy: open clusters and associations: general -- stars: variables: general -- planets and satellites: general
\end{keywords}

%%%%%%%%%%%%%%%%%%%%%%%%%%%%%%%%%%%%%%%%%%%%%%%%%%

%%%%%%%%%%%%%%%%% BODY OF PAPER %%%%%%%%%%%%%%%%%%

\section{Introduction}

In summer 2020, the {\it Transiting Exoplanet Survey Satellite} ({\it
  TESS}, \citealt{2015JATIS...1a4003R}) concluded its main mission
after about two years of observations. In this period, the spacecraft
has observed millions of stars in about $\gtrsim 70$~\% of the sky with
an unprecedented photometric precision and temporal coverage, and new
data from the extended mission, characterised by (in part) a new
observing strategy, are coming.

Stellar clusters and associations offer the unique opportunity to
derive precise stellar parameters (like radius, mass, chemical
content, and especially age) for their members simply using
theoretical models. During the main mission {\it TESS} observed many
hundreds stellar (open and globular) clusters and associations in
sectors of $\sim 27$ days. However, the low resolution of the four
cameras ($\sim 21$~arcsec/pixel) makes difficult the extraction of
high-precision light curves for the stars located in these dense
regions.

The project ``a PSF-based Approach to TESS High quality data Of
Stellar clusters'' (PATHOS, \citealt[hereafter
  Paper~I]{2019MNRAS.490.3806N}) was born to exploit the {\it TESS}
data in order to extract high-precision photometry for members of
stellar clusters and associations adopting an innovative approach,
based on the use of empirical Point Spread Functions (PSFs) and
neighbour subtraction.  Scope of the project is the discovery and
characterisation of new candidate exoplanets in stellar clusters and
the analysis of possible correlations between well measured star properties
and candidate exoplanet characteristics.  Field stars' age is usually affected by large uncertainties,
but it is also an essential information to constrain the formation and understand the evolution of exoplanets, like for example how and on which
temporal scales the mechanisms that bring to the atmosphere
evaporation of low-mass close-in exoplanets happen
(\citealt{2003ApJ...598L.121L,2005A&A...436L..47B,2009ApJ...693...23M,2012MNRAS.425.2931O,2017ApJ...847...29O,2019ApJ...874...91W,2020MNRAS.498.5030O}).
In this context, the PATHOS project is providing interesting
candidate exoplanets orbiting stars with well constrained
ages. Moreover, the high-precision light curves generated in our
project, and publicly available to the astronomical community, allow
us to obtain results not only in the research field of exoplanets, but
also in other fields like asteroseismology
(\citealt{2021MNRAS.tmp..131M}), or the analysis of the spin axis
orientations of cluster members (\citealt{2020ApJ...903...99H}).

\begin{figure*}
\centering
\includegraphics[bb=1 30 501 256, width=0.9\textwidth]{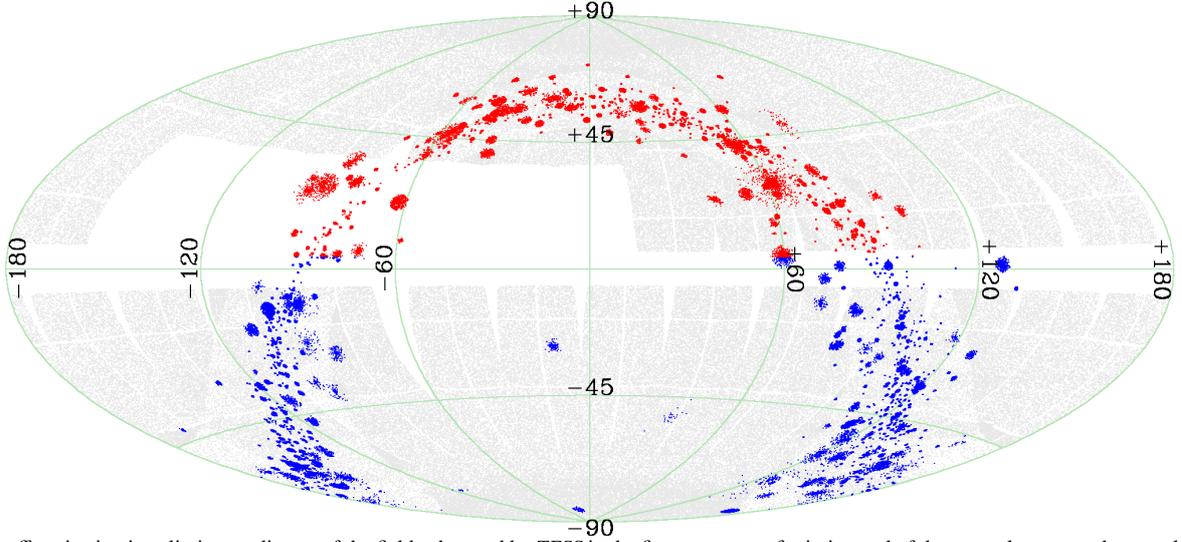}
\caption{Aitoff projection in ecliptic coordinates of the fields
  observed by {\it TESS} in the first two years of mission and of the
  open cluster members analysed in the PATHOS project: grey points
  represent the sources observed in 2-min cadence mode in Sectors
  1-26, blue and red points are the stars in the input list used in
  Paper~\citetalias{2020MNRAS.495.4924N} and in this work,
  respectively. \label{fig:1}}
\end{figure*}

We already successfully applied the PATHOS pipeline in
Paper~\citetalias{2019MNRAS.490.3806N}, when we studied the stars in
an extremely crowded region containing the globular cluster 47~Tuc and
the Small Magellanic Cloud.  In \citet[hereafter
  Paper~II]{2020MNRAS.495.4924N} we extracted and analysed the light curves of open
cluster members located in the southern ecliptic hemisphere, finding 33 objects of interest and
deriving a first estimate of exoplanet frequency in open clusters.
\citet[hereafter Paper~III]{2020MNRAS.498.5972N} studied the
light curves of the members of five young associations, having ages
$\lesssim 10$~Myr; in particular, the author performed a gyrochronological
analysis of association members to constraint the age of the stars,
analysed the dust in the circumstellar discs of the young members and
identified and characterised six strong candidate exoplanets.

In the present work we exploited the {\it TESS} data collected
during Cycle~2 (Sectors 14-26) to obtain high-precision light curves
of cluster members in the northern ecliptic hemisphere by using our
cutting-edge tools (Section~\ref{sec:obs}), find and characterise
candidate exoplanets in stellar clusters (Section~\ref{sec:cand}), and
analyse their frequency and properties as a function of host stars'
characteristics (Sections~\ref{sec:res}). We summarised and discussed
the joined results obtained in this work and in
Paper~\citetalias{2020MNRAS.495.4924N} in Section~\ref{sec:sum}.

\section{Observation and data reduction}
\label{sec:obs}
In this work we extracted and analysed the light curves of the stars
likely members of northern ecliptic hemisphere open clusters observed
by {\it TESS} during the second year of the mission. The observations used
in this work were carried out between 2019 July 18 and 2020 July 4
($\sim$352 days), and are divided into thirteen sectors (Sectors
14--26); in Sectors 21, 22, and 23 no open clusters fell in the {\it TESS} field of view and therefore the final
number of analysed sectors is 10.

For the light curve extraction and correction, we used the PATHOS
pipeline described in detail in Papers~\citetalias{2019MNRAS.490.3806N} and
\citetalias{2020MNRAS.495.4924N}. We extracted the light curves of
stars in a given catalogue from {\it TESS} Full Frame Images (FFIs) by
using the light curve extractor \texttt{IMG2LC}. This software was
developed by \citet{2015MNRAS.447.3536N,2016MNRAS.455.2337N} for
ground-based observations, and it is a versatile tool that can be used
with photometric time-series collected also with space-based
observatories (see, e.g.,
\citealt{2016MNRAS.456.1137L,2016MNRAS.463.1780L,2016MNRAS.463.1831N}).

The three main inputs of our PSF-based approach are: (i) FFIs, (ii)
PSFs, and (iii) input catalogue. For each star in the input catalogue,
the light curve extractor searches for the neighbours within a radius of
20 {\it TESS} pixels in the Gaia~DR2 catalogue
(\citealt{2018A&A...616A...1G}), transforms their positions and
luminosities in the reference system of the FFI, models them by using
a local PSF and then subtracts them from the FFI. Finally, it extracts
PSF-fitting and aperture (1-, 2-, 3-, and 4-pixel radius) photometries
of the target star from the neighbour-subtracted FFI. This approach
has two advantages: (i) it minimises the dilution effects due to the
neighbour contaminants, and (ii) it allows us the extraction of
high-precision photometry for stars in the {\it TESS} faint regime of
magnitudes ($T\gtrsim 15$, see, e.g.,
\citealt{2021ApJ...906...64A}). We corrected the extracted raw light
curves for systematic effects by fitting and applying the
Cotrending Basis Vectors, as widely discussed in
Papers~\citetalias{2019MNRAS.490.3806N} and
\citetalias{2020MNRAS.495.4924N}.

As in Paper~\citetalias{2020MNRAS.495.4924N}, we used as input list the catalogue
of cluster members published by \citet{2018A&A...618A..93C}; this
catalogue contains the positions, colours, magnitudes, proper motions,
parallaxes, and membership probabilities of likely members in 1229
stellar clusters. From this catalogue, we selected all the stars that
satisfy these two conditions: (i) magnitude $G<17.5$, because stars
with larger magnitude are too faint to be detected by {\it TESS}; (ii) ecliptic
latitude $\beta>4\degr$, which corresponds to the part of the northern
ecliptic hemisphere covered by {\it TESS}\footnote{As also shown in
  Fig.~\ref{fig:1}, during Sectors 14-16 and 24-26, the {\it TESS}
  pointings were modified in order to avoid
  excessive contamination by stray Earth- and Moon-light in cameras 1
  and 2.}. 

\begin{figure*}
\includegraphics[bb=19 317 572 695,  width=0.9\textwidth]{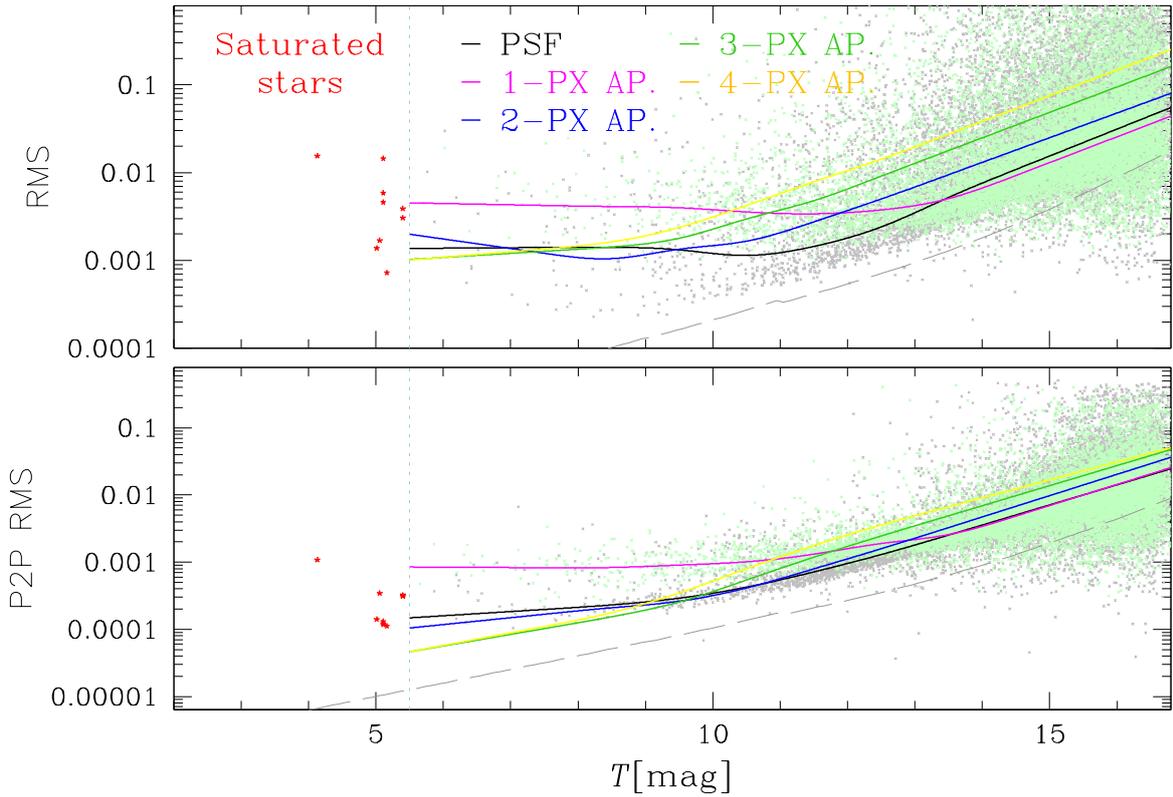}
\caption{Mean trends of the photometric \texttt{RMS} (top-panel) and
  \texttt{P2P RMS} (bottom panel) as a function of the {\it TESS}
  magnitude $T$ for different photometric methods: black lines are
  associated with PSF-fitting photometry, magenta, blue, green, and
  yellow lines are associated with 1-, 2-, 3-, 4-pixel aperture
  photometries, respectively. As an example, the rms distributions
  obtained with 3-pixel and PSF-fitting photometry are shown in light
  green and grey crosses, respectively (for clarity, only 10\% of the
  stars are plotted). Red starred symbols represent the saturated stars. The
  dashed line is the theoretical limit calculated as in
  Paper~\citetalias{2020MNRAS.495.4924N}. \label{fig:2}}
\end{figure*}

Figure~\ref{fig:1} shows the 126\,372 stars (red points) in the input
catalogue overlapped with the {\it TESS} fields of view (grey points):
about 1/3 of them fall outside the {\it TESS} observations. We
extracted 150\,216 light curves of 89\,858 stars in 411 clusters;
about $50.3$\,\%, i.e. 45\,182 stars, were observed in only one
sector, 30\,957 stars ($\sim 34.4$\,\%) were observed in two sectors,
11\,844 ($\sim 13.2$\,\%) in three sectors, 1\,875 ($\sim 2.1$\,\%) in
four or more sectors. 

Light curves are released on the Mikulski Archive for Space Telescopes
(MAST) as a High Level Science Product (HLSP) under the project
PATHOS\footnote{\url{https://archive.stsci.edu/hlsp/pathos}} (DOI:
10.17909/t9-es7m-vw14). A detailed description of the light curves
(that are both in \texttt{ascii} and \texttt{fits} format) is reported
in Papers \citetalias{2019MNRAS.490.3806N} and
\citetalias{2020MNRAS.495.4924N} and in the MAST web-page of the PATHOS
project.

\begin{figure*}
\includegraphics[bb=1 138 586 718, width=0.8\textwidth]{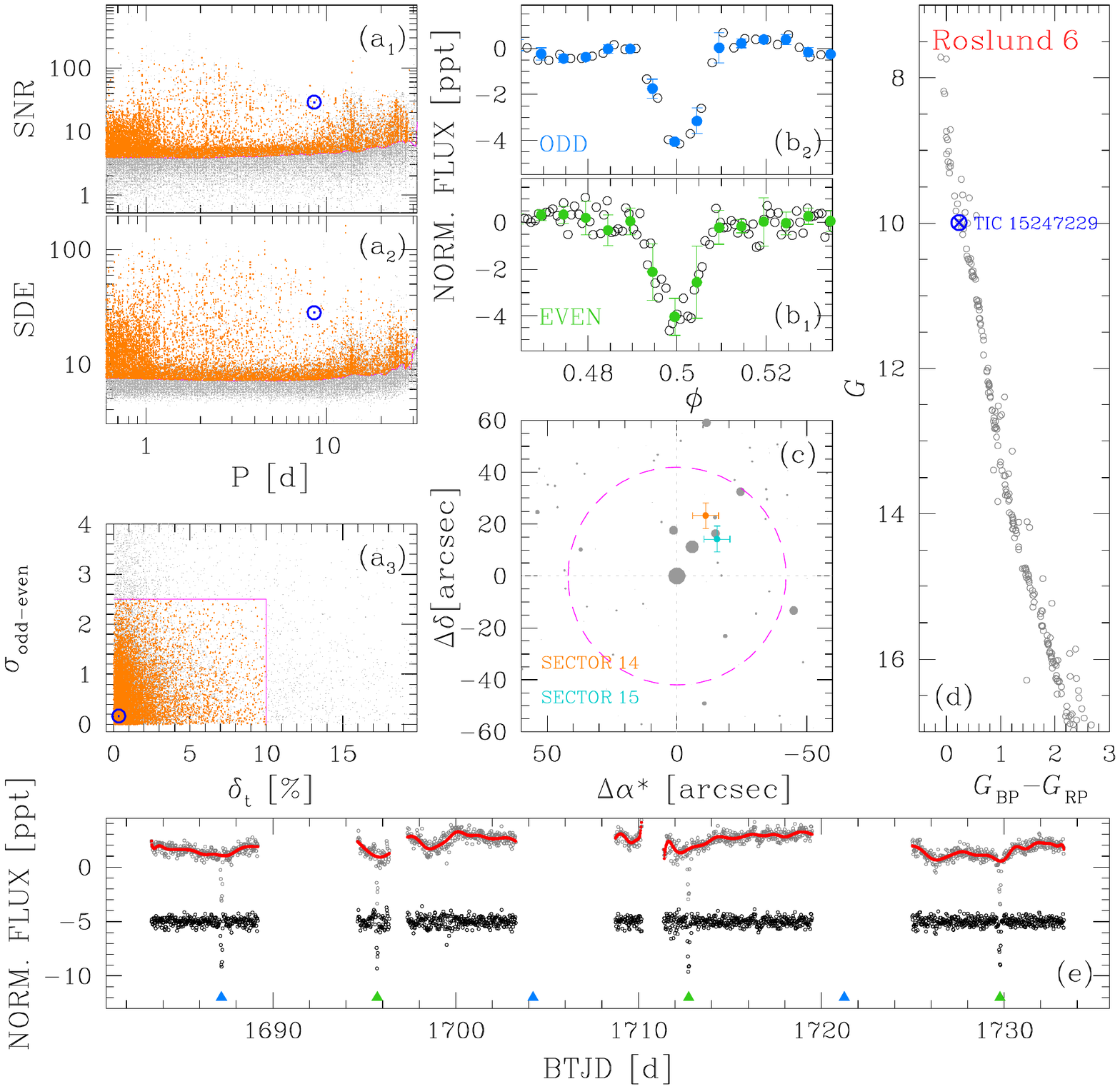}
\caption{Candidate selection and vetting in the case of the star
  TIC~15247229 (TOI-1188). Panels (a$_1$) and (a$_2$) show the TLS SNR
  and SDE versus detected Period, respectively, while panel (a$_3$)
  reports the $\sigma_{\rm odd-even}$ as a function of $\delta_{\rm
    t}$; grey points are all the analysed stars, orange points the
  light curves that passed the selection described in the text, and
  the blue circle is TOI-1188. In panels (b$_1$) and (b$_2$) we
  compare the binned odd (azure) and even (green) transits of the
  candidate. Panel (c) represents the analysis of the
  in-/out-of-transit difference centroid: the centroid is shifted on a
  neighbour star that contaminates the target. Panel (d) is the $G$
  versus $(G_{\rm BP}-G_{\rm RP})$ colour-magnitude diagram of
  Roslund~6, the open cluster that hosts TOI-1188 (blue circle). Panel
  (e) explains the procedure of flattening of the light curve: grey
  points form the original, cleaned light curve; red line is the model
  defined by a spline on knots spaced by 13.0~h; black points are the
  flattened light curve. Azure and green triangles indicate the odd
  and even transits, respectively. \label{fig:3}}
\end{figure*}

\subsection{Photometric precision}
We explored two different quality parameters, already defined in
Papers~\citetalias{2019MNRAS.490.3806N},
\citetalias{2020MNRAS.495.4924N}, and
\citetalias{2020MNRAS.498.5972N}, to identify for each star the
photometric method that gives the best light curve. 

The first quality parameter is the simple \texttt{RMS}, defined as the
68.27th percentile of the 3.5$\sigma$-clipped sorted residual from the
median value. This parameter is sensitive to the (high) variability of
some stars, and for this reason is not recommended to estimate the
photometric precision of the light curve. The mean trends of the
\texttt{RMS} as a function of the {\it TESS} magnitude $T$ for the
five photometric methods are reported in Fig.~\ref{fig:2} (top panel).

The second quality parameter is the \texttt{P2P RMS}, defined as the
 68.27th percentile of the 3.5$\sigma$-clipped sorted residual from
the median value of the vector $\delta F_j = F_j -F_{j+1}$, with $F$
the flux at a given epoch $j$. This parameter is not sensitive to the
intrinsic luminosity variations of the stars, and we used it to define
the interval in which each photometric method works, on average,
better than the others. From the mean trends shown in bottom panel of
Fig.~\ref{fig:2}, we found (confirming the results obtained in the
previous works) that for stars with $5.5 \lesssim T \lesssim 7.0$ the
aperture photometry with radius 4-pixel gives the best results; the
best photometric methods for the intervals $7.0 \lesssim T \lesssim
9.0$, $9.0 \lesssim T \lesssim 10.0$, and $10.0 \lesssim T \lesssim
13.0$ are 3-pixel, 2-pixel aperture and PSF-fitting photometries,
respectively. For faint stars with $T\gtrsim 13.0$ the 1-pixel
aperture photometry gives the lower \texttt{P2P RMS}.

In the following analysis, we used, for each star of magnitude
$T_\star$, the light curve associated with the photometric method that
has the lower mean \texttt{P2P RMS} in $T_\star$. We excluded from the
analysis all the stars whose mean light curve instrumental magnitude
($T_{\rm instr}$) is too different from the expected magnitude $T_{\rm
  calib}$, following this procedure: we extracted the $\delta T=T_{\rm
  instr}-T_{\rm calib}$ distribution, we calculated its mean
($\bar{\delta T}$) and the standard deviation ($\sigma_{\delta T}$),
and we excluded the $i$-th light curve if $|\delta T_i-\bar{\delta
  T}|>4\sigma_{\delta T}$. We also excluded all the light curves that
have $<75\,\%$ of well-measured points (i.e. \texttt{DQUALITY$=0$} and
\texttt{FLUX$\ne 0$}). The final number of analysed light curves is
138\,924 associated with 84\,967 stars.

\begin{table}
  \caption{Cluster parameters}
  \resizebox{0.5\textwidth}{!}{
      \input{figure/table1.tex}

}
      \label{tab:1}
\end{table}

\begin{table*}
  \renewcommand{\arraystretch}{1.5}
  \caption{Results of transit modelling}
  \label{tab:3}
    \resizebox{.99\textwidth}{!}{
      \input{figure/table3.tex}

    }
\end{table*}

\section{Candidate exoplanets: searching, vetting, and characterisation}
\label{sec:cand}
We searched for signals of transiting objects among the selected
light curves following the procedure described in Papers
\citetalias{2020MNRAS.495.4924N} and \citetalias{2020MNRAS.498.5972N}.
Briefly, we removed the intrinsic stellar variability interpolating 
the light curve with a 5-th order spline defined on $N_{\rm knots}$
knots. In order to model short- and long-period variability, we
considered two different grids of knots, with knots spaced by 6.5~h and
13.0~h, respectively. The grids of knots are defined on continuous parts of light
curves that does not present ``breaks'' $>0.5$ days, in order to avoid
the introduction of artifacts in the flattened light curve. We also
removed from the light curves the photometric points associated with
high values of the local sky ($>5\sigma_{\rm SKY}$ from the mean local
background), with \texttt{DQUALITY>0} and $4\sigma$ above the median
normalised flux. We extracted the transit-fitting least-squares (TLS)
periodograms of the flattened light curves adopting the
\texttt{PYTHON} package \texttt{TLS}\footnote{\texttt{TLS v. 1.0.24}
  \url{https://github.com/hippke/tls}}
(\citealt{2019A&A...623A..39H}), and searched for transit signals
with period $0.6~{\rm d} \le P \le T_{\rm LC}$, with $T_{\rm LC}$ the
temporal length of the light curve. We used the output parameters for
the first selection of candidate transiting objects, as follows:
(i) we selected the stars associated with a depth of the transit
$\delta_{\rm t}< 10\,\%$ and to a significance between odd and even
transits $\sigma_{\rm odd-even}<2.5$ ; (ii) we divided the signal
detection efficiency (SDE) and the signal-to-noise ratio (SNR)
distributions in bins of periods $\delta P = 0.5$~d, we calculated the
3.5$\sigma$-clipped mean and standard deviation of SDE ($\bar{\rm
  SDE}$ and $\sigma_{\rm SDE} $) and SNR ($\bar{\rm SNR}$ and
$\sigma_{\rm SNR} $) within each bin. Then we interpolated the binned
points $\bar{\rm SDE} + 3.5\sigma_{\rm SDE} $ and $\bar{\rm SNR} +
3.5\sigma_{\rm SNR} $ with a spline, and we selected as good
candidates all the stars above these splines. Panels (a) of
Fig.~\ref{fig:3} show an example of selections based on the output
parameters of TLS. We visually inspected the light curves that passed
the above selections to exclude false positives due to the presence of
artifacts. We applied the above procedure both to the light curves
obtained in each single sector, and then to the stacked light curves
of stars observed in more than one sector, in order to increase the
efficiency of transit detection. The number of stars that passed the
first-step selection is 279 ($\sim 0.3$~\%).

These candidates were subjected to a series of vetting tests widely
described in the previous papers of the PATHOS series. They are: (i)
check for the transit depths in the light curves obtained with
different photometric methods; (ii) check for the presence of secondary
eclipses in the light curves phased with a period $0.5\times P$,
$1.0\times P$, and $2.0\times P$, with $P$ the period found by the TLS
routine; (iii) check for the transit depths by comparing binned odd/even
transits (panels (b) of Fig.~\ref{fig:3}); (iv) check for 
contamination through the analysis of the in/out-of-transit difference
centroid (panel (c) of Fig.~\ref{fig:3}). After this second-step
selection, 39 transiting objects of interest survived ($\sim
0.05$\,\%); one of these objects showed only one transit in its light curve.

\begin{figure*}
\includegraphics[bb=21 365 520 705, width=0.8\textwidth]{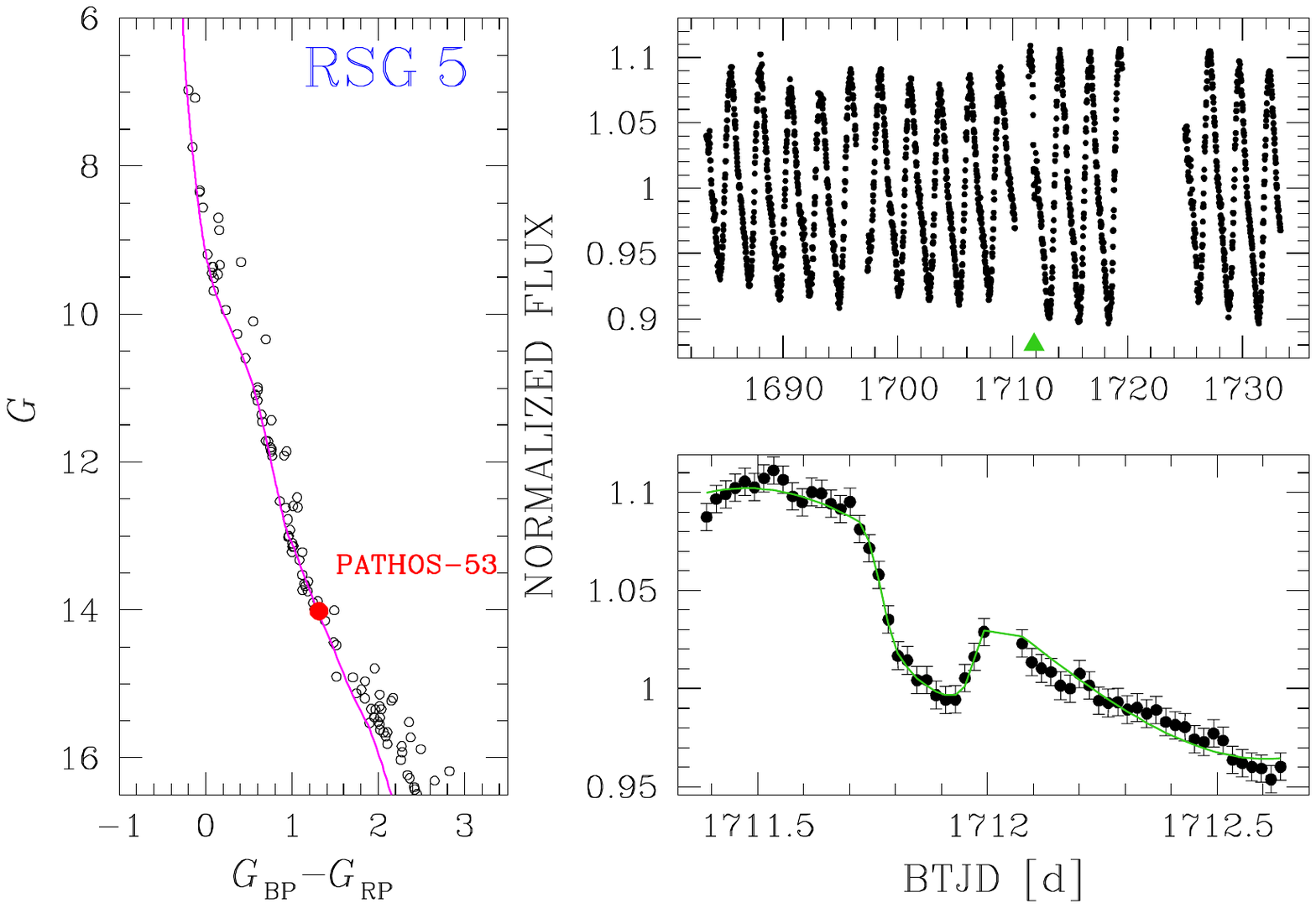}
\caption{Overview on the transit modelling of PATHOS-53: left-hand
  panel shows the $G$ versus $(G_{\rm BP}-G_{\rm RP})$ CMD of the
  members of the open cluster RSG~5 and the isochrone fit (in magenta,
  $t_{\rm AGE}=50$~Myr) used to derive the stellar parameters of
  PATHOS-53 (red point). Top right-hand panel is the light curve of
  PATHOS-53 collected in Sectors 14 and 15; only one transit is
  detected (pointed by the green triangle). Bottom right-hand panel
  shows the model fit (green line) performed by \texttt{PyORBIT} on
  the single transit. \label{fig:4}}
\end{figure*}

\subsection{TESS Objects of Interest}
We cross-matched the {\it TESS} Objects of Interest (TOI)
list\footnote{\url{https://tess.mit.edu/toi-releases/go-to-alerts/}}
with our input catalogue of cluster members. Four candidates, found by
the Quick-Look Pipeline (QLP, \citealt{2020RNAAS...4..204H}) are also
in our input catalogue, but only one of them (TOI-1535,
TIC\,420288086) is in our final list of transiting objects of
interest.  Two of them (TOI-1497, TIC\,371673488 and TOI-1321,
TIC\,195199644) were not detected by our pipeline because no transit
signals are present in the light curves we analysed, even if the mean
scatter of our light curves is lower than the scatter of the light
curves shown in the QLP data validation report. We checked the notes
about these two candidates on
ExoFOP\footnote{\url{https://exofop.ipac.caltech.edu/tess}}: (i) the
depth-aperture correlation for TOI-1497.01, reported in a note on
ExoFOP, is confirmed by our analysis; (ii) for TOI-1321 a
depth-aperture correlation is also reported; moreover, a note associated with
a photometric follow-up with MuSCAT2 reports a deep transit signal
from a nearby star at $\sim 1$~arcmin from TOI-1321. Therefore, both
these candidates are likely contaminated by neighbour sources. The
fourth QLP candidate, TOI-1188 (TIC~15247229) was excluded from our
final list after the centroid analysis. Its vetting tests are reported
in Fig.~\ref{fig:3}. Our conclusion is also supported by the notes
reported in the photometric follow-up section of the ExoFOP website.

\subsection{Stellar parameters}

We fitted theoretical models from the last release of BaSTI (‘a Bag of
Stellar Tracks and Isochrones’) models (\citealt{2018ApJ...856..125H})
to the colour-magnitude diagrams (CMDs) of the 32 open clusters that
host the stars associated with the 39 transiting objects of interest. In
this way, we were able to extract primary information (stellar radius,
mass, density, effective temperature) of the stars that host candidate
transiting objects. Because metallicity measurements are not available
for the large part of the clusters and because open clusters have, on
average, metallicities similar to that of the Sun, in our fit we used
isochrones with [Fe/H]=$0.0 \pm 0.3$ as already done in
Paper~\citetalias{2020MNRAS.495.4924N}, and we added the contribution
of the uncertainties on the metallicity to the final errors on the
stellar parameter estimates.

We transformed the isochrones from the theoretical to the
observational plane using the distance modulus of the clusters
obtained by \citet{2018A&A...618A..93C}, and the reddening and ages
measured by \citet{2016A&A...585A.101K}, \citet{2016A&A...595A..22R},
and \citet{2019A&A...623A.108B}; since some of the catalogues do not
provides error estimates, for homogeneity we used a conservative error
of 10~\% on the age and reddening values. Gulliver~49 is an open
cluster discovered by \citet{2018A&A...618A..93C}, and no age estimate
is provided in literature: we followed the technique by
\citet{2015MNRAS.451..312N} based on the use of the
$\chi^2$-minimisation between isochrones and fiducial lines to derive
an estimate of the cluster age. We found an age of $200 \pm 20$~Myr.

Clusters' parameters adopted for the isochrone fitting are reported in
Table~\ref{tab:1}. Stellar parameters of the transiting candidates'
hosts obtained from isochrone fitting were used as priors in 
transit modelling described in the next section, and are reported in Table~\ref{tab:2}.

\subsection{Transit modelling}

We modelled the transits of the objects of interest using the
\texttt{PYTHON} package
\texttt{PyORBIT}\footnote{\url{https://github.com/LucaMalavolta/PyORBIT}}(\citealt{2016A&A...588A.118M,2018AJ....155..107M},
see also
\citealt{2019AA...630A..81B,2020arXiv201113795C,2020arXiv200902332L}),
based on the combined use of the package \texttt{BATMAN}
(\citealt{2015PASP..127.1161K}), the global optimisation algorithm
\texttt{PyDE}\footnote{\url{https://github.com/hpparvi/PyDE}}
(\citealt{1997..............S}), and the affine invariant Markov chain
Monte Carlo sampler \texttt{emcee} (\citealt{2013PASP..125..306F}).

For the transit modelling, we included the central time of the first
transit ($T_0$), the period ($P$), the impact parameter ($b$), the
planetary-to-stellar-radius ratio ($R_{\rm P}/R_\star$), the stellar
density ($\rho_\star$), and the dilution factor ($df$). The latter
quantity is included as a free parameter, with a Gaussian prior
obtained considering all the stars in the Gaia~DR2 catalogue that fall
in the same pixel of the target\footnote{In such crowded environments,
  the completeness of the Gaia~DR2 catalogue is $\gtrsim 80$-$90$~\%
  for stars with $16 \lesssim G \lesssim 18$ and $\gtrsim 95\,\%$ for
  brighter stars (see Gaia DR2 Documentation release 1.2,
  Sect. 10.7.4), and therefore the values of dilution factor obtained
  in this work represent, within the errors, a good estimate of the
  real dilution factor. In the case of close binaries with
    separation $<1-2$~arcsec, \citet{2018AJ....156..259Z} demonstrated
  that Gaia satellite is not always able to resolve the components. If
  two (or more stars) are not resolved by Gaia, the result will be a
  single source whose flux will approximately be equal to the sum of
  the fluxes coming from the different not-resolved components; this
  does not affect the subtraction process in the extraction of the
  light curve, because we have considered the multiple components as a
  single source whose flux is the sum of components' fluxes. It might
  affect the estimate of the dilution factor used for the modelling of
  the transits, and for this reason further AO follow-up of the
  candidates are mandatory}, and transforming their Gaia magnitudes in
  {\it TESS} magnitudes adopting the equations by
  \citet{2019AJ....158..138S}. Host star parameters, like the stellar
  radius ($R_\star$), mass radius ($M_\star$), gravity ($\log{g}$) and
  effective temperature ($T_{\rm eff}$) come from the isochrone fits
  described in the previous section. On the basis of $\log{g}$ and
  $T_{\rm eff}$, we obtained information on the limb-darkening (LD) by
  using the grid of values published by \citet{2018A&A...618A..20C};
  we adopted the LD parametrization described by
  \citet{2013MNRAS.435.2152K}. In the modelling process the routine
  takes into account the local variability of the star by fitting a
  second-degree polynomial to the out-of-transit part of the light
  curve. The routine modelled the transits with a fixed circular
  orbital eccentricity ($e=0$), and taking into account the 30-minutes
  cadence of the {\it TESS} FFIs (\citealt{2010MNRAS.408.1758K}).

The adopted priors on stellar parameters are reported in
Table~\ref{tab:2}. The package \texttt{PyORBIT} explored all the
parameters in linear space. In the \texttt{emcee} run, the number of walkers used is 10 times the
number of free parameters. We ran, for each model, the sampler for
80\,000 steps, excluding the first 15\,000 steps as burn-in and using
a thinning factor of 100. Figure~\ref{fig:4} shows an example of the
modelling process in the case of PATHOS-53, a mono-transit object of
interest.

In Table~\ref{tab:3} we report the results of the transit
fitting; in Appendix \ref{app:cand}, Figs.~\ref{fig:A1a}, \ref{fig:A1b},
\ref{fig:A1c} give an overview on the main properties of each
transiting object of interest (position on the CMD, proper motions,
in-/out-of-transit centroid analysis, transit modelling).

\begin{figure}
\includegraphics[bb=50 222 306 688,
  width=0.45\textwidth]{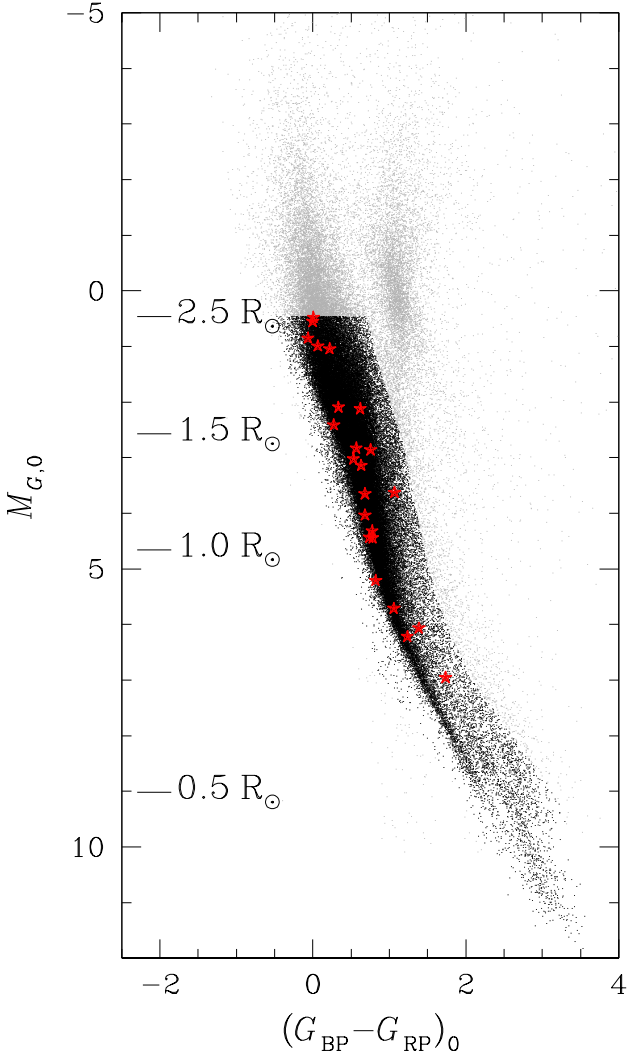}
\caption{The $M_{G,0}$ versus $(G_{\rm BP}-G_{\rm RP})_0$ CMD of all
  the stars analysed in Paper~\citetalias{2020MNRAS.495.4924N} and in
  this work. Black points are the main sequence stars with
    $R_{\star}\leq 2.5 R_{\odot}$, grey points are the stars excluded
  from the analysis of candidate exoplanets' frequency (see text for
  details). For clarity, only 50\% of the stars are plotted. Red
  starred symbols indicate the positions on the CMD of the 23
  candidate transiting exoplanets identified in stellar clusters from
  {\it TESS} data. \label{fig:6}}
\end{figure}

\begin{table}
  \renewcommand{\arraystretch}{0.99}
  \caption{Detection efficiencies of the PATHOS pipeline}
  \resizebox{0.49\textwidth}{!}{
    \input{figure/table7.tex} 
    }
   \label{tab:7}
 \end{table}

\begin{figure*}
\includegraphics[bb=18 169 580 405,  width=0.875\textwidth]{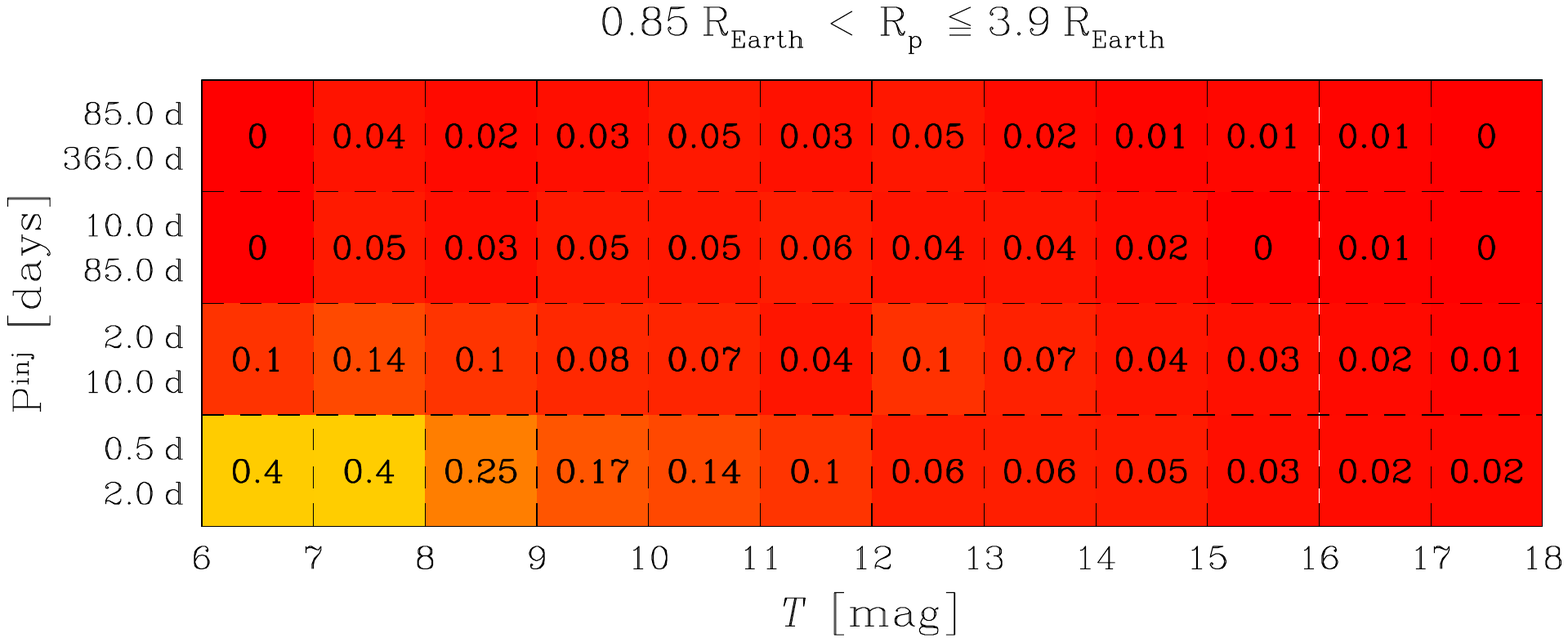} \\
\includegraphics[bb=18 169 580 405,  width=0.875\textwidth]{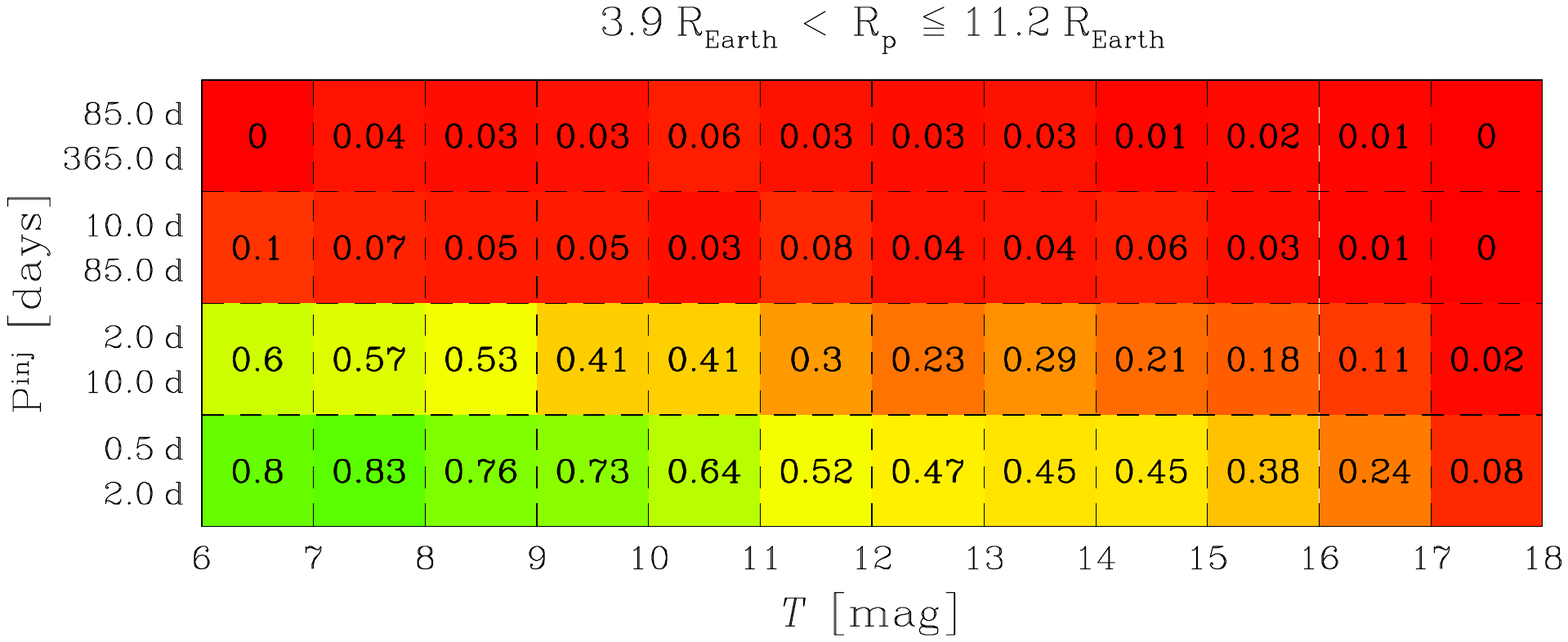} \\
\includegraphics[bb=18 169 580 405,  width=0.875\textwidth]{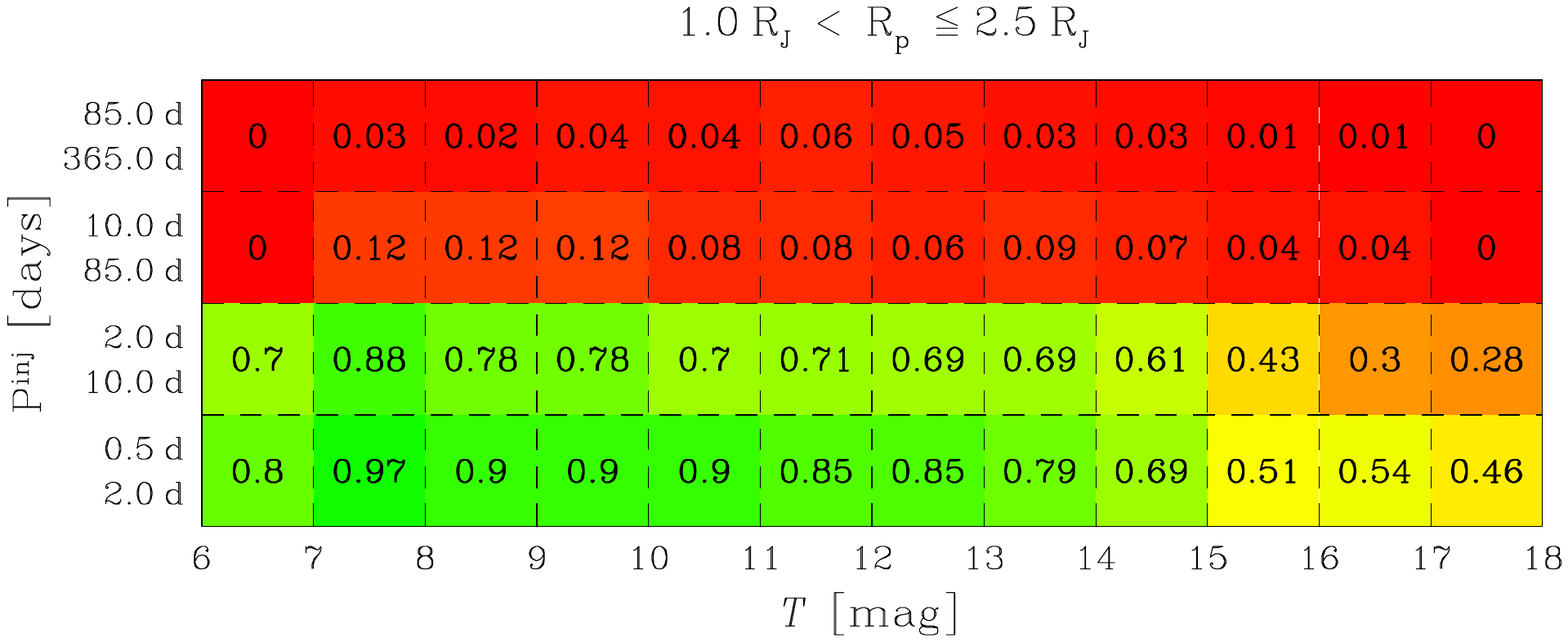}
\caption{Transit detection efficiency (normalised to 1) of our
    pipeline in the injected period $P^{\rm inj}$ versus $T$ magnitude
    plane for different size simulated planets: Earth and super-Earth
    size planets (top panel), Neptune size planets (middle panel), and
    Jupiter size planets (bottom panel). \label{fig:7}}
\end{figure*}

\begin{figure*}
\includegraphics[bb=43 166 576 717,
  width=0.8\textwidth]{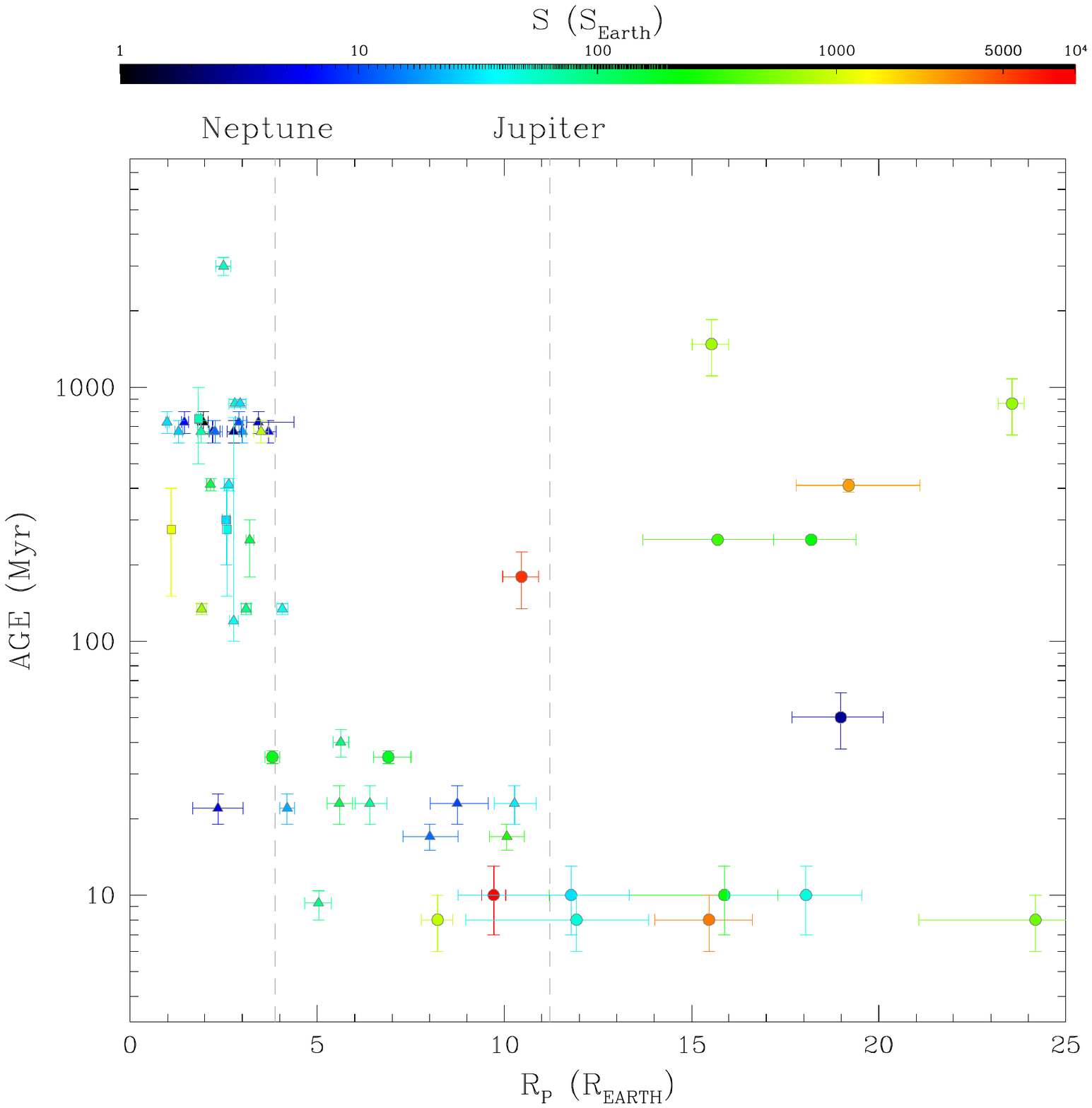}
\caption{Stellar age versus planetary radius $R_{\rm P}$ for candidate
  exoplanets identified in Papers~\citetalias{2020MNRAS.495.4924N},
  \citetalias{2020MNRAS.498.5972N}, and in this work (circles), for
  candidates and confirmed exoplanets from literature (triangles), and
  for the objects under investigation in the GAPS-YO programme
  (squares). Coloured points represent the exoplanets orbiting stars
  with well constrained ages (from isochrone fitting and/or
  gyrochronological analysis of cluster/association members that host
  the stars).  Different colours are associated with different
  insolation fluxes on the basis of the bar shown on the top.
  \label{fig:5}}
\end{figure*}

%\begin{table*}
%  \caption{Star parameters and priors for the modelling.}
%  \resizebox{0.95\textwidth}{!}{
%  \input{figure/table2.tex}
%  }
%  \label{tab:2}
%\end{table*}

\section{Results}
\label{sec:res}
We improved the calculation of the frequencies of exoplanets
  reported in Paper~\citetalias{2020MNRAS.495.4924N}, taking into
  consideration the detection efficiency of our method and the False
  Positive Probabilities (FPPs) of our candidates. In this analysis we
  considered candidates identified in this work and in
  Paper~\citetalias{2020MNRAS.495.4924N}, excluding all the PATHOS
  objects with $R_{\rm P}>2.5 R_{\rm J}$ (because of their doubtful
  planet nature): 23 candidates survived after this selection (14 and
  9 in the Southern and Northern ecliptic hemispheres, respectively).

\subsection{Detection Efficiency}
\label{sec:simul}
We calculated the detection efficiency of our finding pipeline
  injecting transit signals of planets having radii $R_{\rm P}$ in a
  sample of light curves extracted randomly from the collection of
  light curves analysed in Paper~\citetalias{2020MNRAS.495.4924N} and
  in this work.

To inject the transit signals in each light curve, we roughly
  estimated the stellar radius and mass of each the 232\,669 stars
  analysed in Paper~\citetalias{2020MNRAS.495.4924N} and in this work
  by using their absolute $M_{G,0}$ versus $(G_{\rm BP}-G_{\rm RP})_0$
  CMD.  To obtain the absolute CMD, we transformed the apparent
  magnitudes into absolute magnitudes by using the Gaia DR2 distances
  obtained by \citet{2018AJ....156...58B}. For each star, we corrected
  the effects of the extinction on the colour and the magnitude of the
  stars calculating the reddening value [$E(B-V)$] by using the
  \texttt{PYTHON} routine
  \texttt{mwdust}\footnote{\url{https://github.com/jobovy/mwdust}}
  (\citealt{2016ApJ...818..130B}) and the \texttt{Combined19} dustmap
  (\citealt{2003A&A...409..205D,2006A&A...453..635M,2019ApJ...887...93G}),
  and the colour-dependent equation and the coefficients reported by
  \citet{2019A&A...623A.108B}. Figure~\ref{fig:6} shows the $M_{G,0}$
  versus $(G_{\rm BP}-G_{\rm RP})_0$ CMD for all the stars in our
  sample; red starred points indicate the 23 candidates.  We selected
  the likely main sequence stars as follows: we performed a first
  guess selection of the main sequence stars by hand, excluding all
  the stars that clearly are evolved stars (sub-giant and red-giant
  stars); in a second step, we calculated the fiducial line of the
  likely main sequence stars by using the naive estimator
  (\citealt{1986desd.book.....S}; see also
  \citealt{2015MNRAS.451..312N} for the description of the method) and
  we selected the stars whose colours are within $2\sigma$ from the
  mean colour of the fiducial line. We calculated the radius and the
  mass of each star in our catalogue by using the main sequence points
  of the \texttt{PARSEC}
  (\citealt{2002A&A...391..195G,2012MNRAS.427..127B,
    2017ApJ...835...77M}) isochrones with ages between 10 and
  1000~Myr. In Fig.~\ref{fig:6} we reported as reference the mean
  $M_{G,0}$ absolute magnitude for stars having radius $R_{\star}=0.5,
  1.0, 1.5, 2.5\,R_{\odot}$. 

Because our candidate exoplanets orbit stars with $R_{\star}
  \lesssim 2.5~R_{\odot}$, we first selected the light curves
  associated to these stars; we divided our sample of light curves on
  the basis of the {\it TESS} magnitude of the associated stars,
  considering bins of size $\Delta T=1.0$ in the magnitude range $6.0
  \leq T \leq 18.0$. In each bin we randomly caught 350 light
  curves\footnote{If the number of light curves in the considered
    magnitude bin was $<350$, we considered all the light curves
    belonging to that bin.} associated to stars having the previously
  estimated radii and masses, and, by using the routine
  \texttt{INJECTTRANSIT} implemented in \texttt{VARTOOLS} v.1.39
  (\citealt{2016A&C....17....1H}), we injected in each light curve a
  periodic transit signal of a planet having radius $R^{\rm inj}_{\rm
    P}$, orbital period $P^{\rm inj}$, random inclination $i^{\rm
    inj}$ (with the constraint that there must be a transit), and
  eccentricity $e=0$. We considered 12 different cases in which
$R^{\rm inj}_{\rm P}$ and $P^{\rm inj}$ randomly vary between a given
minimum and a maximum; (i) we first injected (super-)Earth-size planets
having radii $ 0.85\,R_{\rm Earth} \leq R^{\rm inj}_{\rm P} \leq
3.9\,R_{\rm Earth}$; (ii) the second case we considered was for
(super-)Neptune planets having radii $ 3.9\,R_{\rm Earth} < R^{\rm
  inj}_{\rm P} \leq 11.2\,R_{\rm Earth} \sim 1 R_{\rm J}$; finally we
considered giant planets with radii $ 1.0\,R_{\rm J} < R^{\rm
  inj}_{\rm P} \leq 2.5\,R_{\rm J}$. For each $R^{\rm inj}_{\rm P}$
case we considered random orbital periods in the following intervals:
(a) short periods $0.5\,{\rm d} \leq P^{\rm inj} \leq 2.0\,{\rm d}$;
(b) short-medium periods $2.0\,{\rm d} < P^{\rm inj} \leq 10.0\,{\rm
  d}$; (c) long-medium periods $10.0\,{\rm d} < P^{\rm inj} \leq
85.0\,{\rm d}$; and (d) long periods $85.0\,{\rm d} < P^{\rm inj} \leq
365.0\,{\rm d}$. We have not made any selection on the time length of
the light curves, so, on the basis of the associated star
observability, the light curve can span randomly between $\sim 27$ and
$\sim 365$~days.

We injected the transit signals in the raw light curves and we
  followed the same pipeline for the correction of the light curves
  and the detection of transit signals described in
  Sections~\ref{sec:obs} and \ref{sec:cand}. We considered the
  injected planet as recovered if it passes the selections illustrated
  in Sect.~\ref{sec:cand} and in Fig.~\ref{fig:3} and if $|P^{\rm
    inj}-P^{\rm inj}_{\rm TLS}|<4.0\sigma(P^{\rm inj}_{\rm TLS})$,
  where $P^{\rm inj}_{\rm TLS}$ is the period obtained by the
  extraction of the TLS periodogram, and $\sigma(P^{\rm inj}_{\rm
    TLS})$ its error. We also considered the planet as recovered if
  $P^{\rm inj}_{\rm TLS}$ is equal to $0.5 \times P^{\rm inj}$ or $2.0
  \times P^{\rm inj}$, within $4.0\sigma(P^{\rm inj}_{\rm TLS})$.  On
  a sample of 100 light curves selected randomly from those that
  passed the selections, we performed the vetting tests described in
  Section~\ref{sec:cand}\footnote{The analysis of the centroid was
    excluded from these vetting tests because it is obtained analysing
    the images, where there are no signals for the simulated
    transits.}: all the selected objects passed the tests.  We finally
  calculated the detection efficiency in each magnitude interval
  $\Delta T$ as the ratio $N_{\rm rec}/N_{\rm inj}$, where $N_{\rm
    rec}$ is the number of simulated planets recovered, and $N_{\rm
    inj}$ the number of light curves in which we injected transit
  signals. In Fig.~\ref{fig:7} and in Table~\ref{tab:7}, we reported
  the detection efficiency in the $P^{\rm inj}$ versus $T$ grid for
  the three different planetary radius intervals: for Earth-size
  planets with $R_{\rm P} \lesssim 3.9~R_{\rm Earth}$, the detection
  efficiency is low ( $\lesssim 40~\%$), mainly because the large part
  of low mass stars, for which the detection of small size planets is
  easier, is concentrated at magnitudes $T \gtrsim 11$ (see Fig.~8 of
  Paper~\citetalias{2020MNRAS.495.4924N}); for short period
  Neptune-size planets ($P<10$~d) orbiting stars with magnitude $T
  \lesssim 12$ the detection efficiency is $\gtrsim 30$--$50\%$, and
  decreases at fainter magnitude to 20--30\%; we obtained a detection
  efficiency always $\gtrsim 30~\%$ for giant planets with periods
  $\lesssim 10$~d. For periods $P\gtrsim 10$~d the detection
  efficiencies are always $\lesssim 20~\%$: this is an effect caused
  by the fact that only $\sim 15\%$ of the stars in our sample are
  observed in $>2$ Sectors, making difficult the detection of long
  period planets. Finally, even if we used different grids to flat the
  light curves to take into account the different kinds of
  variability, the detection efficiency can be lower for very active
  stars in young stellar clusters.

\subsection{FPP estimation}
We used the tool
\texttt{VESPA}~v.~0.6\footnote{https://github.com/timothydmorton/VESPA}
(\citealt{2012ApJ...761....6M,2015ascl.soft03011M}) to estimate the
FPP in a Bayesian framework of each of the 23 candidate exoplanets
previously selected. This software estimates the probability that the
transit signal detected in a light curve is due to a real planet or to
a different source. Different scenarios are taken into consideration
to explain the signal: (i) simple eclipsing binary (EB, single/double
period); (ii) a hierarchical triple system where two components
eclipse (HEB, single/double period); (iii) a blended eclipsing binary
inside the photometric aperture of the target (BEB, single/double
period); (iv) a transiting planet on the target star (pl). We
  want to highlight that \texttt{VESPA} does not consider the scenario in which
  transit signals are due to non-astrophysical sources, even if in our
  case the probability it happens is low, because the analysed
  candidates passed a series of vetting tests that allow to exclude
  signals due to systematic effects.
 For each configuration, \texttt{VESPA} simulates a
  representative stellar population, constrained by the information we
  have from the isochrone fitting and the light curve modelling; in
  particular, we put constraints on the $G$, $G_{\rm BP}$, $G_{\rm
    RP}$, $J_{\rm 2MASS}$, $H_{\rm 2MASS}$, $K_{\rm 2MASS}$ magnitudes
  of the target star, the $(\alpha, \delta)$ coordinates of the
  target, the effective temperature $T_{\rm eff}$, the density
  $\rho_{\star}$, and the surface gravity $\log(g)$ of the star, the
  orbital period $P$ of the candidate exoplanet, the planet-to-star
  radius ratio $R_{\rm P}/R_{\star}$. Moreover, we gave as an input
  the light curve of the stars hosting the candidates.  The software
  takes into account different false positive scenarios for those
  populations, and uses them to define a prior likelihood that a
  specific configuration actually exists and the likelihood of transit
  for those configurations. On the basis of the results obtained for
  different scenarios, at the end it calculates the FPP that the
  transit signals are due to a false positive: lower is the FPP, larger is
  the probability that the signal is due to a planet.  We reported in
  Table~\ref{tab:5} the FPPs and the most likely scenario (with the
  respective probability) derived with \texttt{VESPA} for the 23
  candidate exoplanets: $\sim 40$~\% of them have a probability
  $\gtrsim 50$~\% to be a real exoplanet. It means that $\sim 3/5$ of
  the signals we detected are likely false positives. 

To test the reliability of the FPP estimation performed by
  \texttt{VESPA}, we used this tool to validate 10 simulated
  transiting exoplanets extracted randomly from the sample of the
  recovered candidates described in Sect.~\ref{sec:simul} (considering
  all the scenarios). We ran \texttt{VESPA} on the simulated
  exoplanets following the same procedure we adopted for the real
  candidate exoplanets. In an ideal case, the sum of all the FPPs
  obtained for the simulated exoplanets is equal to 0, but we obtained
  that $\sum {\rm FPP}_i \sim 0.5$ ; it means that, on average, the
  measured FPP is overestimated of $\sim 5~\%$. We used this result to
  correct the FPPs measured for the real candidate exoplanets, as
  reported in Table~\ref{tab:5} (FPP$_{\rm c}$). 

We also reported in Table~\ref{tab:5} the Renormalised Unit
  Weight Error (RUWE) index calculated as described by
  \citet{2020MNRAS.496.1922B}; its value is $\sim 1$ for sources that
  can be fitted with single-star models, and its value is larger
  ($\gtrsim 1.4$) if the astrometric fit of the source by using
  single-star model is not the best solution. Among the candidates
  labelled as ``pl'', only PATHOS-33 shows a high RUWE value ($\sim
  2.6$), even if the FPP obtained with \texttt{VESPA} is $\sim 0$.

\begin{table}
  \renewcommand{\arraystretch}{0.99}
  \caption{False positive probabilities and most likely scenarios for
    PATHOS candidate exoplanets} \resizebox{0.45\textwidth}{!}{
    \input{figure/table5.tex} }
   \label{tab:5}
 \end{table}

\subsection{Candidate exoplanets’ frequency in open clusters}

We calculated the frequency of candidate exoplanets
  ($f_{\star}$) in open clusters taking into consideration the
  detection efficiency (de) of our pipeline and the FPPs of our
  candidates. We calculated the frequencies as a function of $R_{\rm
    P}$ and $P$, considering the 12 cases illustrated in
  Sect.~\ref{sec:simul}. For each period interval $P_{\rm min}< P_{\rm
    mid} \leq P_{\rm max}$, we calculated three frequencies associated
  to orbital periods $P_{\rm min}$, $P_{\rm max}$, and $P_{\rm mid} =
  0.5~(P_{\rm min}+P_{\rm max})$.  To calculate the frequency of
  candidate exoplanets in open clusters we used the following formula:
  \begin{equation}
   f_{\star}(R_{\rm P}, P) = \dfrac{\sum\limits_{i=1}^{N_{\rm cand}} [1-{\rm FPP^i_c(R_{\rm P}, P)}]} {\sum\limits_{j=1}^{N_{\rm stars}} [{\rm de^j(R_{\rm P}, P) \times {\rm Pr}^j_{\rm transit}(P) }  ]}
 \end{equation}
where $N_{\rm cand}$ and $N_{\rm stars}$ are associated to the
candidate exoplanets and the stars selected in the previous sections;
${\rm FPP^i_c(R_{\rm P}, P)}$ are the corrected false positive
probabilities of the candidate exoplanets having period $P$ and radius
$R_{\rm P}$ in the considered ($P$,$R_{\rm P}$) bin; ${\rm de}^j$ is
the detection efficiency associated to $j$-th star having {\it TESS}
magnitude $T_j$, obtained interpolating the values reported in
Fig.~\ref{fig:7} for each specific interval ($P$,$R_{\rm P}$); ${\rm
  Pr}^j_{\rm transit}(P) \simeq R_{\star}/a(P)$ is the transit
probability associated to the $j$-th star having radius $R_{\star}$,
with $a$ the orbital semi-major axis calculated on the basis of the
third law of Kepler and using the three different periods $P_{\rm
  min}$, $P_{\rm mid}$, and $P_{\rm max}$. When the numerator is $<1$,
we calculated an upper limit of the frequency considering the
numerator equal to 1. The frequencies $f^k_{\star}$, with $k={\rm min,
  mid, max}$ for the different scenarios ($P$,$R_{\rm P}$) are
reported in Table~\ref{tab:6}. This is a new approach to
  calculate the frequencies of exoplanets around a sample of stars, in
  which we directly used the FPP values obtained with \texttt{VESPA},
  including also those candidates with a not clear solution
  (FPPs$\sim$0.3--0.7). For this kind of candidates, independent
  statistical validation methods (like that proposed by \citealt{2020arXiv200810516A}) are mandatory.

\begin{table*}
  \renewcommand{\arraystretch}{0.99}
  \caption{Candidate exoplanets' frequencies} 
  \resizebox{0.99\textwidth}{!}{
    \input{figure/table6a.tex}

}
   \label{tab:6}
 \end{table*}

 \begin{table*}
   \renewcommand{\arraystretch}{1.25}
   \caption{Confirmed and candidate transiting exoplanets from literature}
   \resizebox{0.99\textwidth}{!}{
   \input{figure/table4a.tex}
   }
   \label{tab:4}
 \end{table*}
 
 \subsection{Age--Planetary radius distribution}
 
 Figure~\ref{fig:5} shows the stellar age versus planetary
  radius distribution for the candidate transiting exoplanets in open
  clusters identified in this work and in
  Papers~\citetalias{2020MNRAS.495.4924N}, and reported as ``pl'' in
  Table~\ref{tab:5}, and the candidates around stars in young
  associations identified in Paper~\citetalias{2020MNRAS.498.5972N};
the colours indicate the insolation flux $S$ calculated on the basis
of the stellar radius and effective temperature and of the planetary
semi-major axis. Moreover, we included confirmed and candidate
exoplanets from literature, orbiting stars in stellar clusters and
associations, i.e. with a well-constrained age (coloured
triangles).  The list of
literature objects plotted in Fig.~\ref{fig:5} and the
corresponding references are reported in Table~\ref{tab:4}. Finally,
we also added objects (coloured squares, Nardiello et al. in
preparation) that are under investigation in the context of the ``GAPS
Young Objects'' (GAPS-YO) project (\citealt{2020A&A...638A...5C}),
aimed at the monitoring of young and intermediate-age stars for the
discovery and characterisation of young planets. All the objects
showed in Fig.~\ref{fig:5} have orbital periods $<100$ days, i.e
semi-major axis $a \lesssim 0.5$ AU.
 
 We found that Jupiters with $R_{\rm P} \gtrsim
 1~R_{\rm J}$ are distributed randomly between $\sim 10$~Myr up to
 $\sim 10$~Gyr. Objects having Neptune-sizes or smaller
 (Earths/super-Earths) are concentrated at ages $>200$-$300$~Myr; anyway
 the lack of this kind of objects around young stars might be an
 observational bias due to the difficulty of detecting their transits in
 the highly variable light curves of active young stars.  Objects
 having a planetary radius $4\,R_{\rm Earth} \lesssim R_{\rm P}
 \lesssim 10\,R_{\rm Earth} $ are concentrated at ages
 $<100$-$200$~Myr. Is it an observational bias, or the lack of
 super-Neptune/sub-Jovian-size planets orbiting (on short periods)
 stars with (well-measured) ages $\gtrsim 200$~Myr is due to an evolutionary
 effect of the planets? 
 
 The majority of low-mass close-in exoplanets (with $R_{\rm P}$=
 1--4\,$R_{\rm Earth}$) detected until today orbit field stars with
 ages $\gtrsim 1-3$ Gyr. Some of these planets are almost totally
 rocky (e.g. Kepler-93, \citealt{2015ApJ...800..135D}), others have
 low densities that can be explained by the presence of an extended
 H/He atmosphere. Among them, the Kepler-36 system is particular
 because formed by two planets with very similar semimajor axis, but
 totally different densities (\citealt{2012Sci...337..556C}), with the
 inner planet less massive than the external
 planet. \citet{2013ApJ...776....2L} proved that both the exoplanets
 in the Kepler-36 system were born with H/He atmospheres and were more
 massive in the early stages of their life, and that, given the lower
 core mass of the inner planet, the latter has lost large part of its
 atmosphere, despite the outer planet that was able to retain about
 half of its initial atmosphere. As explained in detail in the review
 by \citet[see also references therein]{2019AREPS..47...67O}, there
 are strong evidence that atmospheric escape is the mechanism that
 prevails in the first stages of a low-mass close-in exoplanet's
 evolution, and it depends (in a first approximation) on the
 characteristics of the host star (and its high-energy emissions), the
 distance of the planet from the star, and its core mass. During their
 formation, these planets accreted large amount of H/He into their
 expanded atmospheres, inflating their radius at 5-13 $R_{\rm Earth}$;
 in few hundreds Myr, if irradiated strongly enough, these planets
 lose the large part of their atmosphere, resulting in
 (super-)Earth/sub-Neptune-size planets
 (\citealt{2020SSRv..216..129O}). Even if Fig.~\ref{fig:5} does not
 show any particular dependence on the insolation flux, the
 distribution of short-period (candidate) exoplanets with radii $1
  13\,R_{\rm Earth} \lesssim R_{\rm P}\lesssim 13\,R_{\rm Earth}$ seems to confirm the
 idea of atmospheric escape on timescales of $\lesssim 100-200$~Myr;
 anyway, as demonstrated by \citet{2020MNRAS.498.5030O}, even if
 challenging, for many of these candidates mass measurements are
 mandatory in order to understand the mechanisms and constraint the
 time scales of planetary atmospheric evolution.  

Other possible explanations for the lack of planets older than
  $\sim 200$Myr in the interval $4 R_{\rm Earth} \lesssim  R_{\rm P}
  \lesssim 10 R_{\rm Earth}$ may be linked to the dynamical evolution of
  this kind of planets after their formation, such as migration due
  to interaction with the protoplanetary disk or planet-planet
  scattering (\citealt{2007ApJ...669.1298F,
    2008ApJ...686..580C,2012ARA&A..50..211K}). These phenomena are
  expected to modify the orbital characteristics of the planets, such
  as inclination, eccentricity and semi-major axis, making the
  detection of a possible transit of these planets more
  difficult. These phenomena occur on scales of a few tens of Myr or
  less, and may not yet have come into play for the objects reported
  in this analysis.  

%\defcitealias{2019MNRAS.490.3806N}{I}
%\defcitealias{2020MNRAS.495.4924N}{II}
%\defcitealias{2020MNRAS.498.5972N}{III}

\section{Summary and conclusion}
\label{sec:sum}
The aim of the PATHOS project is the discovery and first
characterisation of transiting objects around stars in stellar
clusters and associations observed by {\it TESS}. Stellar clusters and
associations offer the rare opportunity to obtain precise measurements
of the ages of the stars (usually affected by large uncertainties), in
addition to stars' physical parameters such as radius, mass, effective
temperature, etc, and analyse planet characteristics as a function of
its host's properties. Because of the low resolution of {\it TESS}
cameras, stellar clusters appear as very crowded regions on FFIs, and
appropriate tools are necessary to obtain high-precision light curves
for cluster members. In order to obtain the best light curves, we
developed a cutting-edge technique for the extraction of high-precision
photometry of stars in crowded fields, based on the use of empirical
PSFs and neighbour subtraction, that allows us to minimise neighbour
contamination and extract light curves for very faint objects.

In this work, the fourth of the PATHOS series, we extracted and
corrected 150\,216 light curves of 89\,858 open cluster members listed
in the \citet{2018A&A...618A..93C}'s catalogue and observed during the
Cycle 2 (Sectors 14--26) of the {\it TESS} mission. By using the
pipeline already tested in the previous papers of the series, we
searched for transit signals among the light curves, finding 39
transiting objects of interest, which are added to the 33 objects
orbiting open cluster members identified in
Paper~\citetalias{2020MNRAS.495.4924N}. We modelled their light curves
to extract planet parameters. From the two lists of objects of
  interest, we isolated 23 candidates with planetary radius $R_{\rm P}
  \lesssim 2.5\,R_{\rm J}$, and we calculated their false
  positive probabilities to be a planet, finding that about 3/5 of
  them are likely false positives.

Taking into account of the transit detection efficiency of our
  pipeline and of the rate of false positives, we calculated the
  frequencies of candidate exoplanets in open clusters ($f_\star$) for
  different (Earth/Neptune/Jupiter) sizes planets in different orbital
  periods intervals (between 0.5~d and 365~d). Because we did not
  detect any strong candidate with orbital period $<2$~d, we
  calculated the frequency upper limit, finding  $f_{\star} \lesssim 0.06$~\%,
  $\lesssim 0.007$~\%, and $\lesssim 0.004$~\% for Earth-, Neptune-,
  and Jupiter-size candidate exoplanets, respectively.  For candidate
  exoplanets with periods $2.0~{\rm d} < P < 10.0$~d, we found for
  Earth-size objects an upper limit $f_{\star} \lesssim 0.2$~\%, while
  for candidate planets with $R_{\rm P}\gtrsim 3.9~R_{\rm Earth}$ the
  frequency is $\sim 0.06$~\%; in the range $10.0~{\rm d} < P <
  85.0$~d, we obtained that the frequency of Earth and Neptune-size
  candidates is $\lesssim 2.~\%$, while the measured frequency for
  giant planets is $\sim 1.6~\%$; finally, for long period planets
  ($>85$~d), we found that frequency of candidate exoplanets is
  $<5-7~\%$. The large part of the measured and upper limit
  frequencies are lower than the values reported by
  \citet{2013ApJ...766...81F} in the same period intervals for the
  corresponding size exoplanets around field stars (even if bias can
  be introduced by the poor statistic of our sample); an exception are
  the giant planets with 10--85~d period: we found a mean frequency of
  $1.6 \pm 0.9$~\%, in agreement with the value obtained by
  \citet[$1.5 \pm 0.2~\%$]{2013ApJ...766...81F}, but also by
  \citet[$f_\star=0.90 \pm 0.26$\%]{2016A&A...587A..64S}, and by
  \citet[$f_\star=1.86 \pm 0.68$\%]{2018A&A...619A..97D}.

We investigated the stellar age versus planetary radius distribution
by using the results obtained in the PATHOS project and the results
obtained in other works for exoplanets orbiting stars with well
constrained ages. We only used planets with orbital periods $<100$~d
and we divided the distribution on the basis of the planetary radius
$R_{\rm P}$: (1) for (candidate) exoplanets with $R_{\rm P}\gtrsim
1~R_{\rm J}$ we found no particular trends, and the planets are
randomly distributed on over the range of ages; (2) objects with
$R_{\rm P} \lesssim 4~R_{\rm Earth}$ are concentrated at ages
$>100-200$~Myr, but it might be an observational bias due to the
difficulty to detect (super-)Earth transits in the light curve of
young active stars; (3) in the range $4\,R_{\rm Earth} \lesssim R_{\rm
  P} \lesssim 13\,R_{\rm Earth}$ there is a concentration of objects
around young stars with ages $<100$~Myr. A possible explanation of
such concentration is that these objects are young planets with rocky
cores that, in the early stages of their formation, have accreted
large amount of hydrogen and helium in their atmospheres, inflating
their radius. On the timescales of $\sim 100-200$~Myr they lose large
part of their atmosphere mainly because of the strong irradiation of
the host star (see \citealt{2019AREPS..47...67O} for a review of all
the mechanisms that contribute to the atmosphere escape), and within
few hundreds Myr their radius decreases to that typical of
(super-)Earth/Neptune planets. Other explanations can be
  related to the dynamical evolution of these exoplanets
  (planet-planet scattering, migration, etc.).

For this reason the analysis of light curves of members of young
associations (ages $\lesssim 200-300$~Myr), the subject of the next
PATHOS works, will be essential to shed light on the mechanisms of
planet formation and evolution of close-in sub-Jovian planets.

%#######################################################
\section*{Acknowledgements}
DN acknowledges the support from the French Centre National d'Etudes
Spatiales (CNES). GL acknowledges support by CARIPARO Foundation,
according to the agreement CARIPARO-Università degli Studi di Padova
(Pratica n. 2018/0098). LB acknowledges the funding support from
Italian Space Agency (ASI) regulated by ``Accordo ASI-INAF
n. 2013-016-R.0 del 9 luglio 2013 e integrazione del 9 luglio 2015
CHEOPS Fasi A/B/C''. The authors thank the anonymous referee for
carefully reading the manuscript and for the useful suggestions that
improved this work. This paper includes data collected by the
{\it TESS} mission. Funding for the {\it TESS} mission is provided by
the NASA Explorer Program. This work has made use of data from the
European Space Agency (ESA) mission {\it Gaia}
(\url{https://www.cosmos.esa.int/gaia}), processed by the {\it Gaia}
Data Processing and Analysis Consortium (DPAC,
\url{https://www.cosmos.esa.int/web/gaia/dpac/consortium}). Funding
for the DPAC has been provided by national institutions, in particular
the institutions participating in the {\it Gaia} Multilateral
Agreement. Some tasks of the data analysis have been carried out using
\texttt{VARTOOLS} v~1.39 (\citealt{2016A&C....17....1H}) and
\texttt{TLS} \texttt{PYTHON} routine (\citealt{2019A&A...623A..39H}).

\section*{Data Availability}
The data underlying this article are available in MAST at
\url{doi:10.17909/t9-es7m-vw14} and at
\url{https://archive.stsci.edu/hlsp/pathos}. 


\bibliographystyle{mnras}
\bibliography{biblio}

\appendix

\section{Light curve modelling}
\label{app:cand}

\begin{figure*}
\includegraphics[bb=77 360 535 691, width=0.33\textwidth]{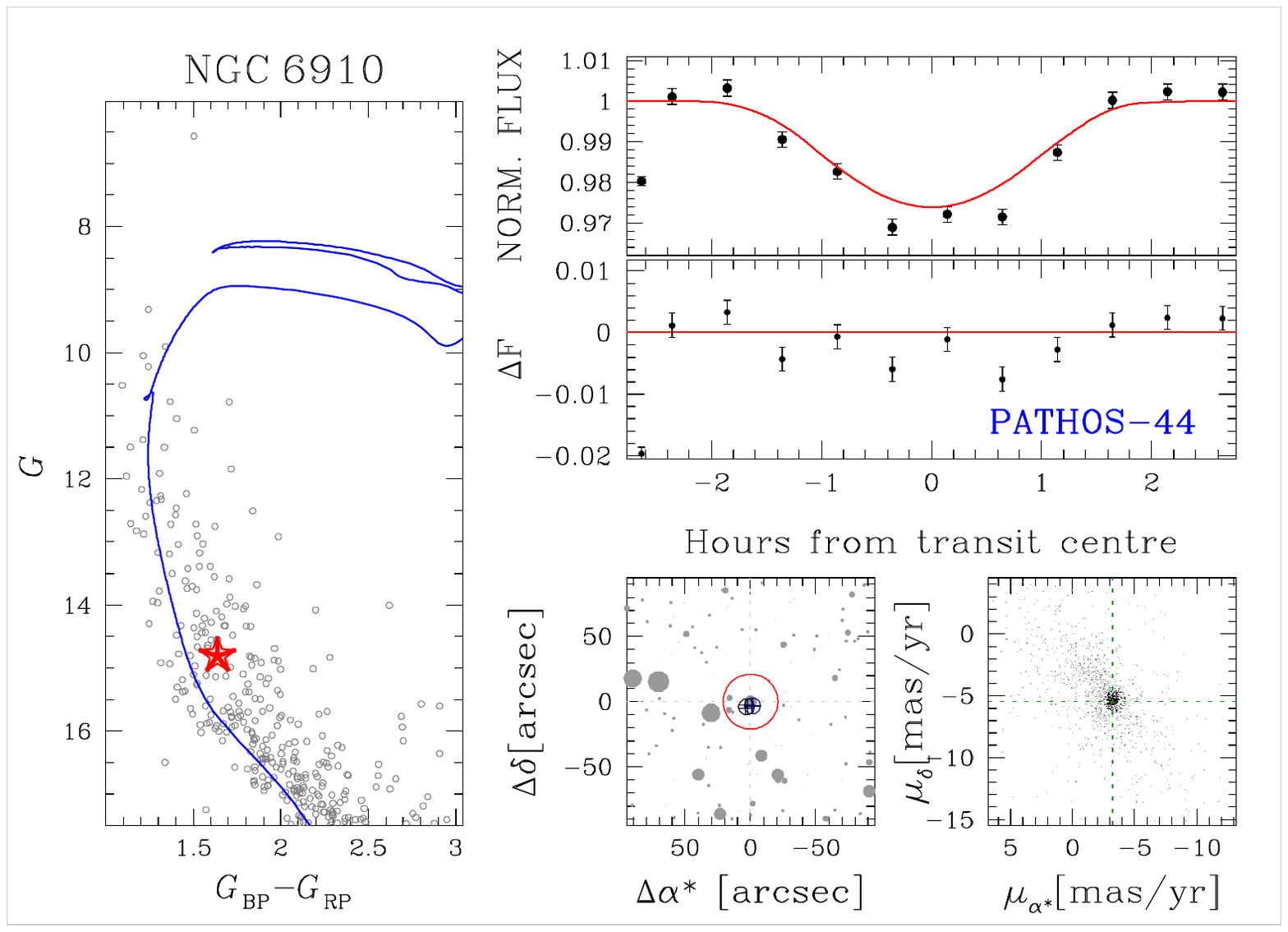}
\includegraphics[bb=77 360 535 691, width=0.33\textwidth]{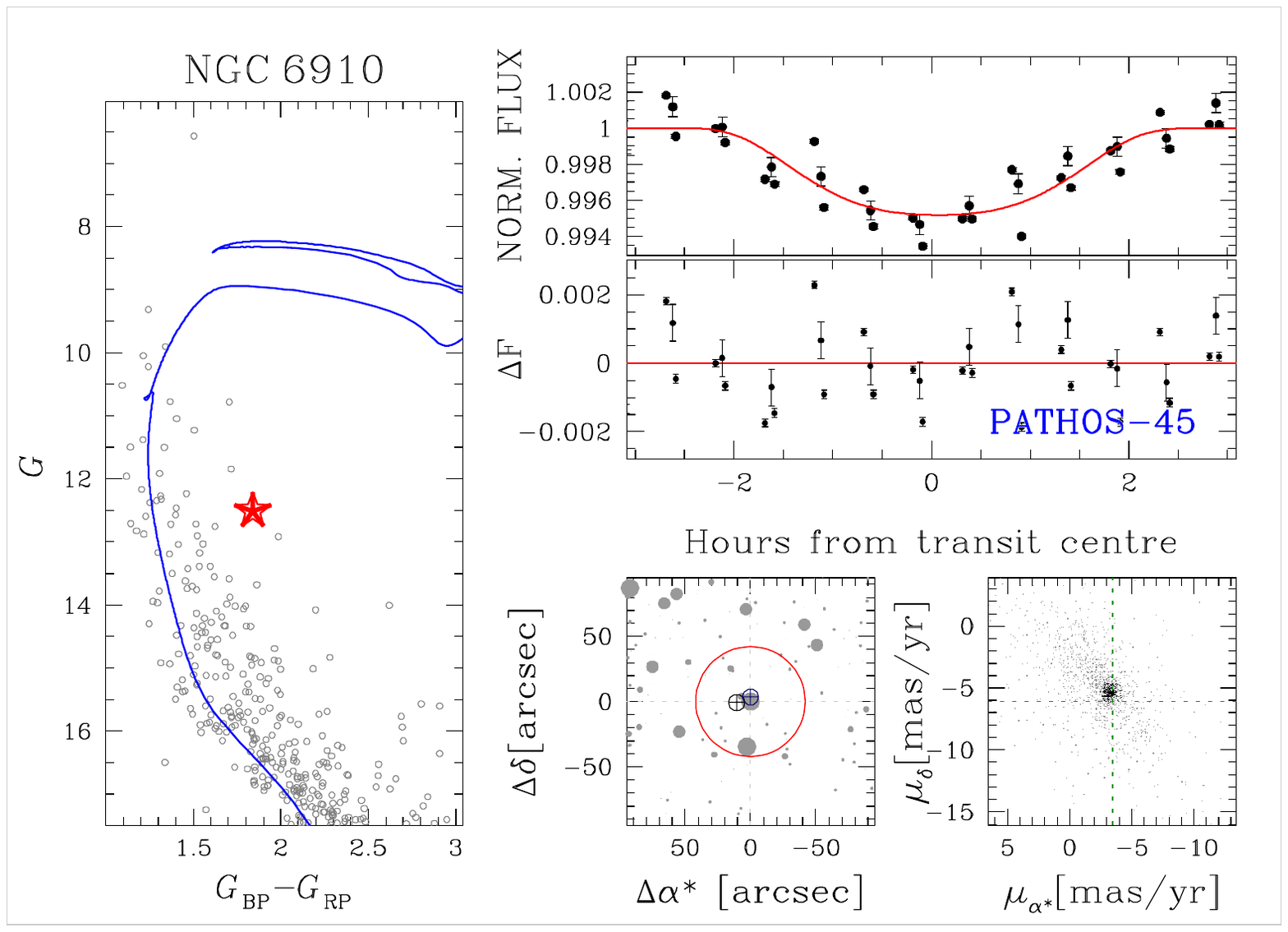}
\includegraphics[bb=77 360 535 691, width=0.33\textwidth]{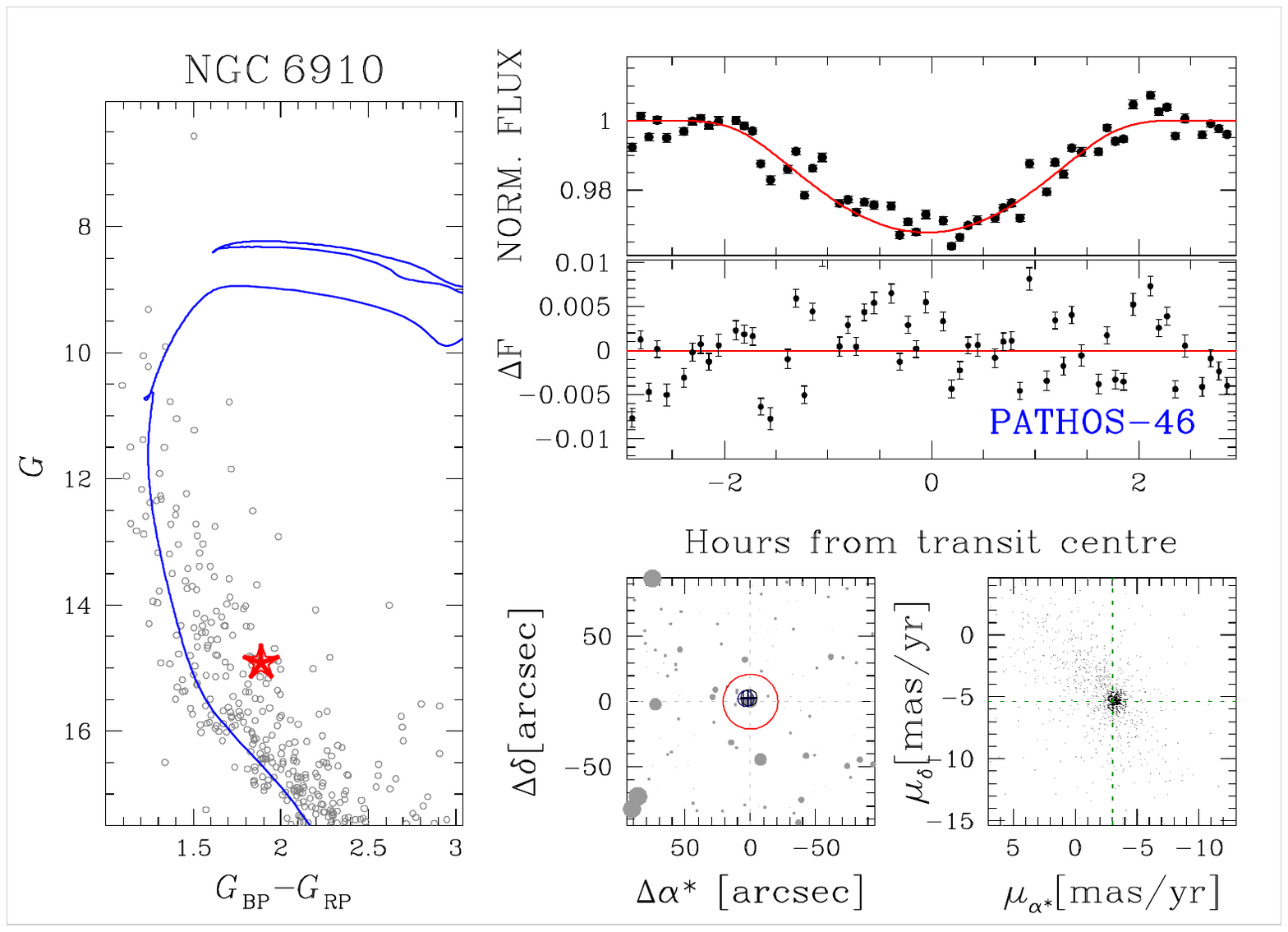} \\
\includegraphics[bb=77 360 535 691, width=0.33\textwidth]{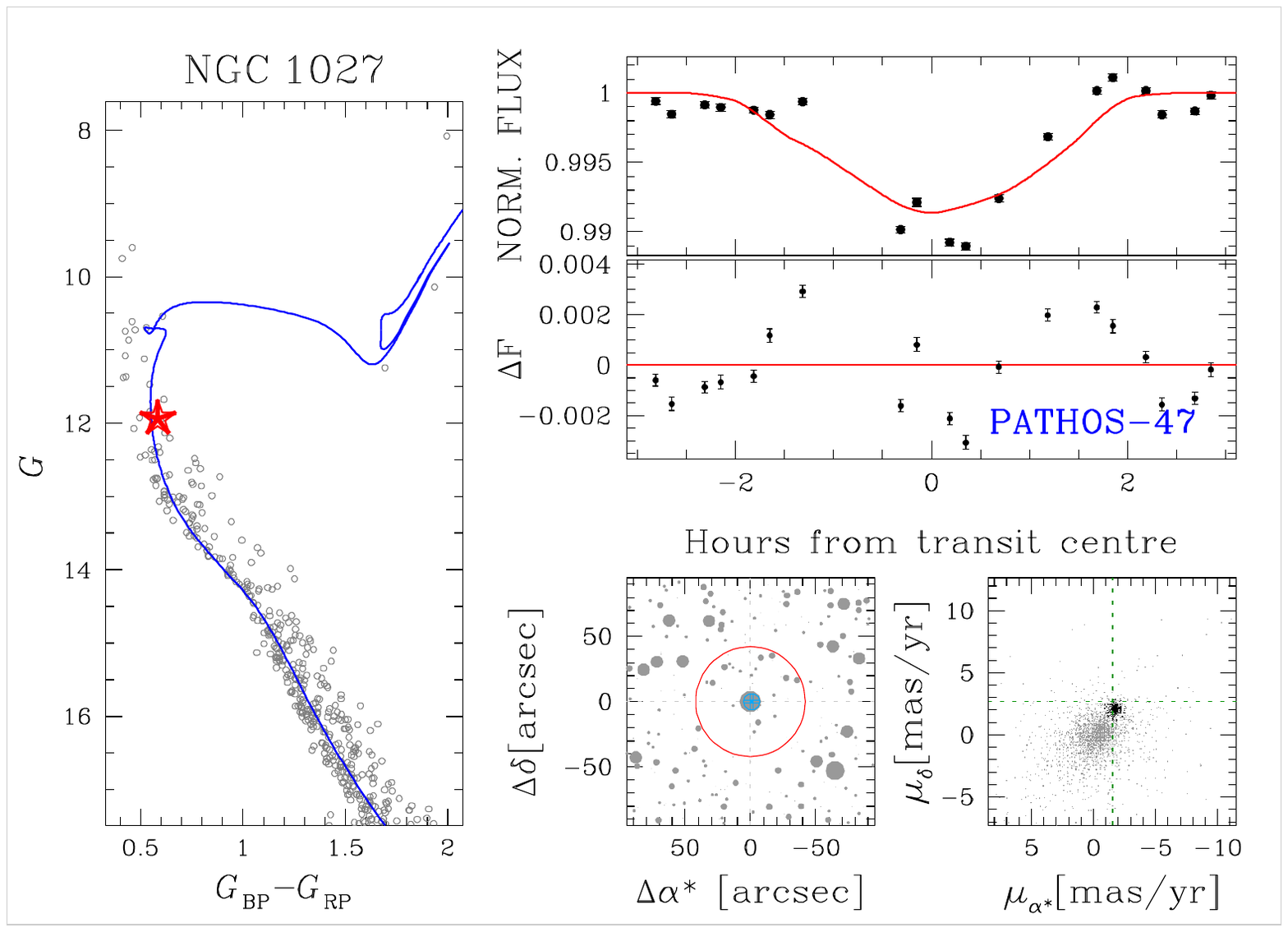}
\includegraphics[bb=77 360 535 691, width=0.33\textwidth]{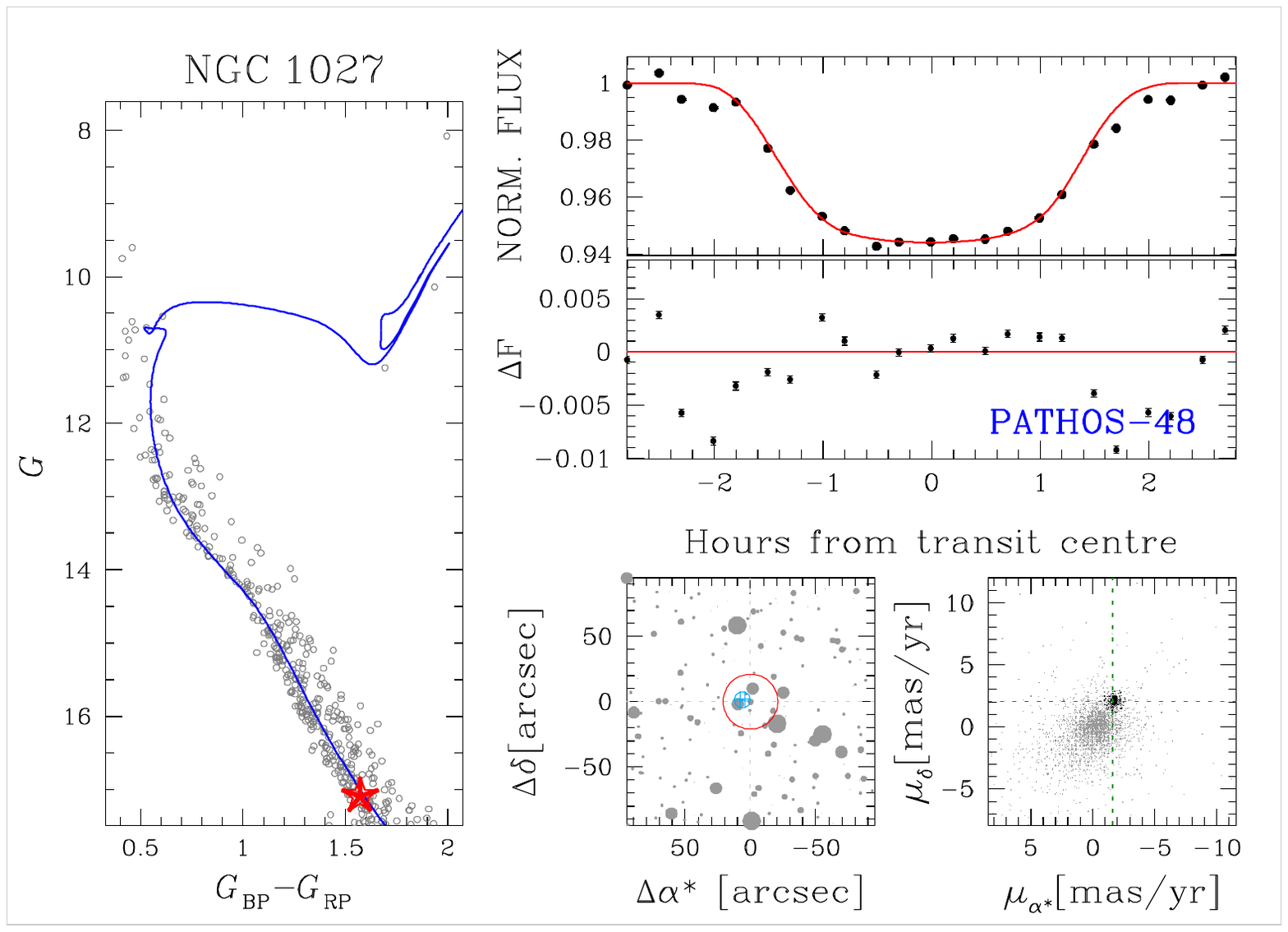}
\includegraphics[bb=77 360 535 691, width=0.33\textwidth]{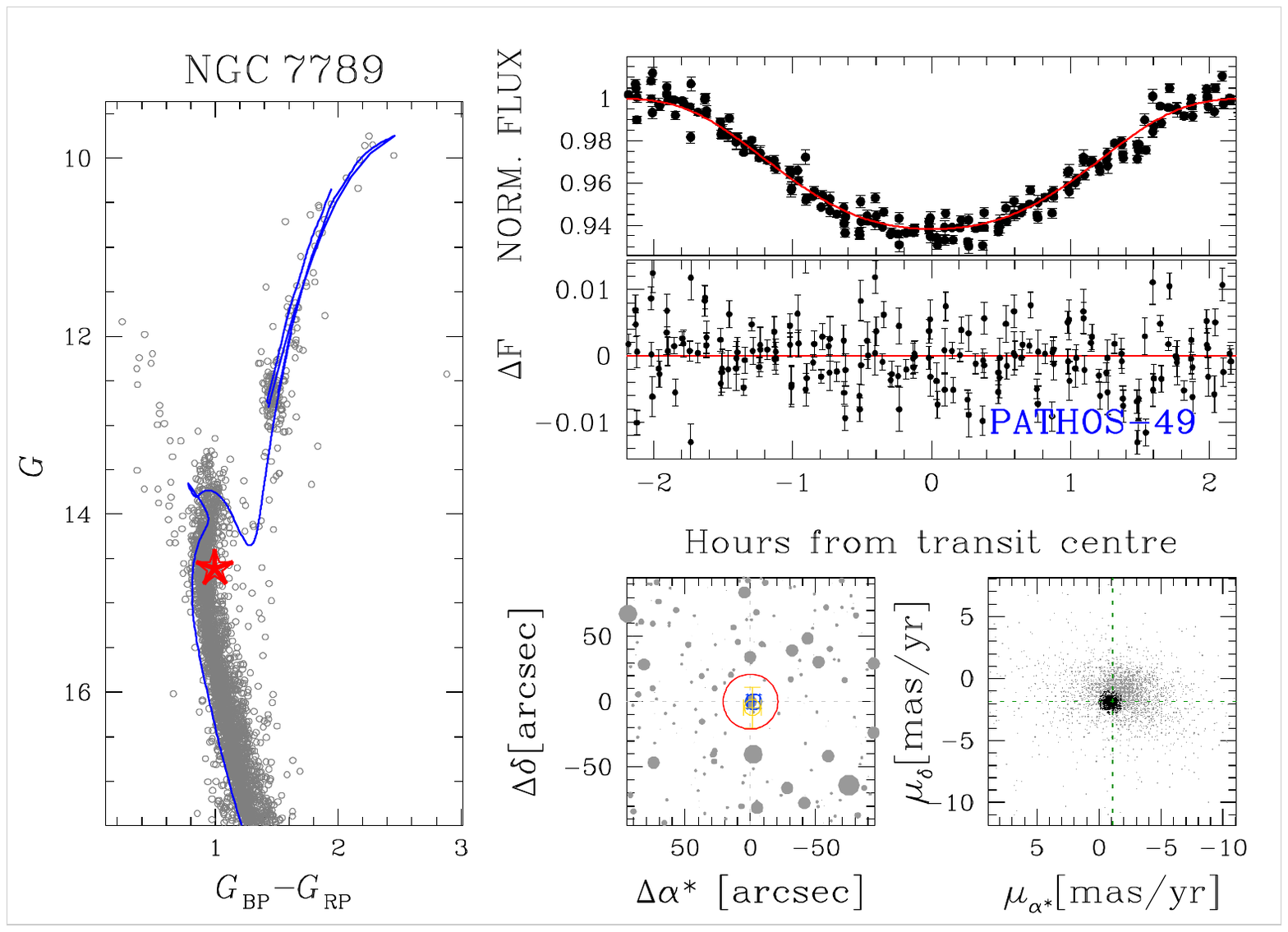} \\
\includegraphics[bb=77 360 535 691, width=0.33\textwidth]{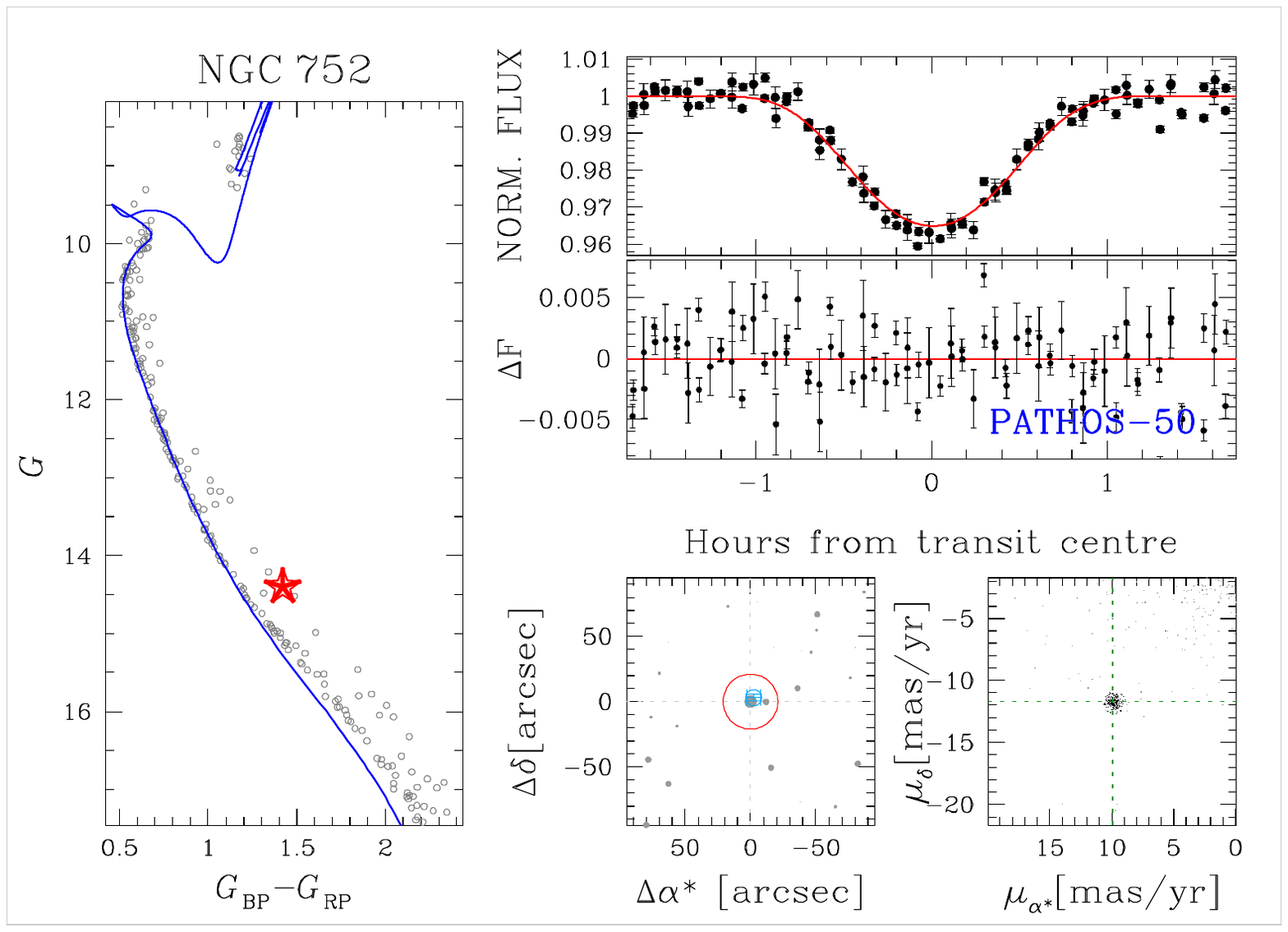}
\includegraphics[bb=77 360 535 691, width=0.33\textwidth]{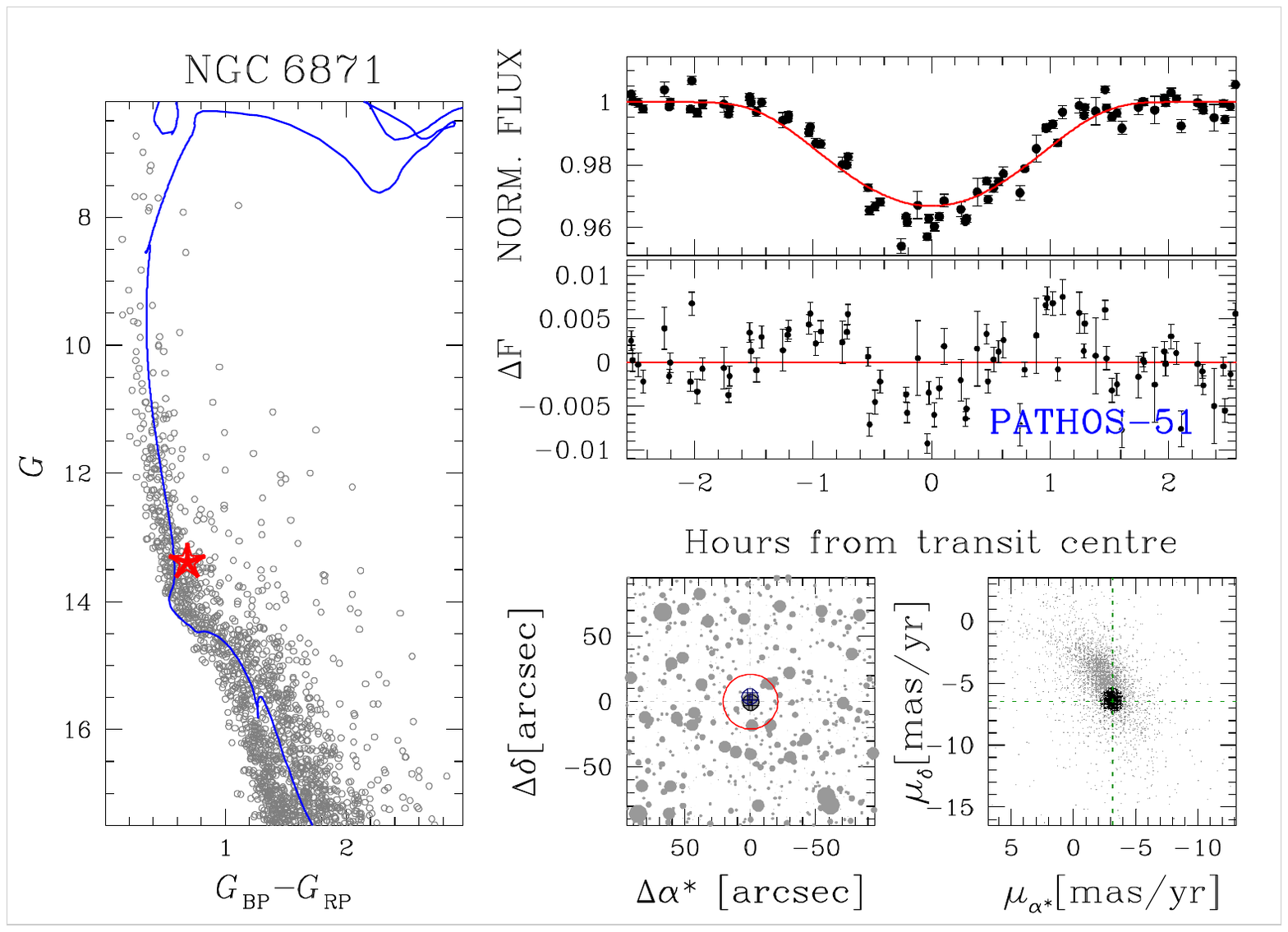}
\includegraphics[bb=77 360 535 691, width=0.33\textwidth]{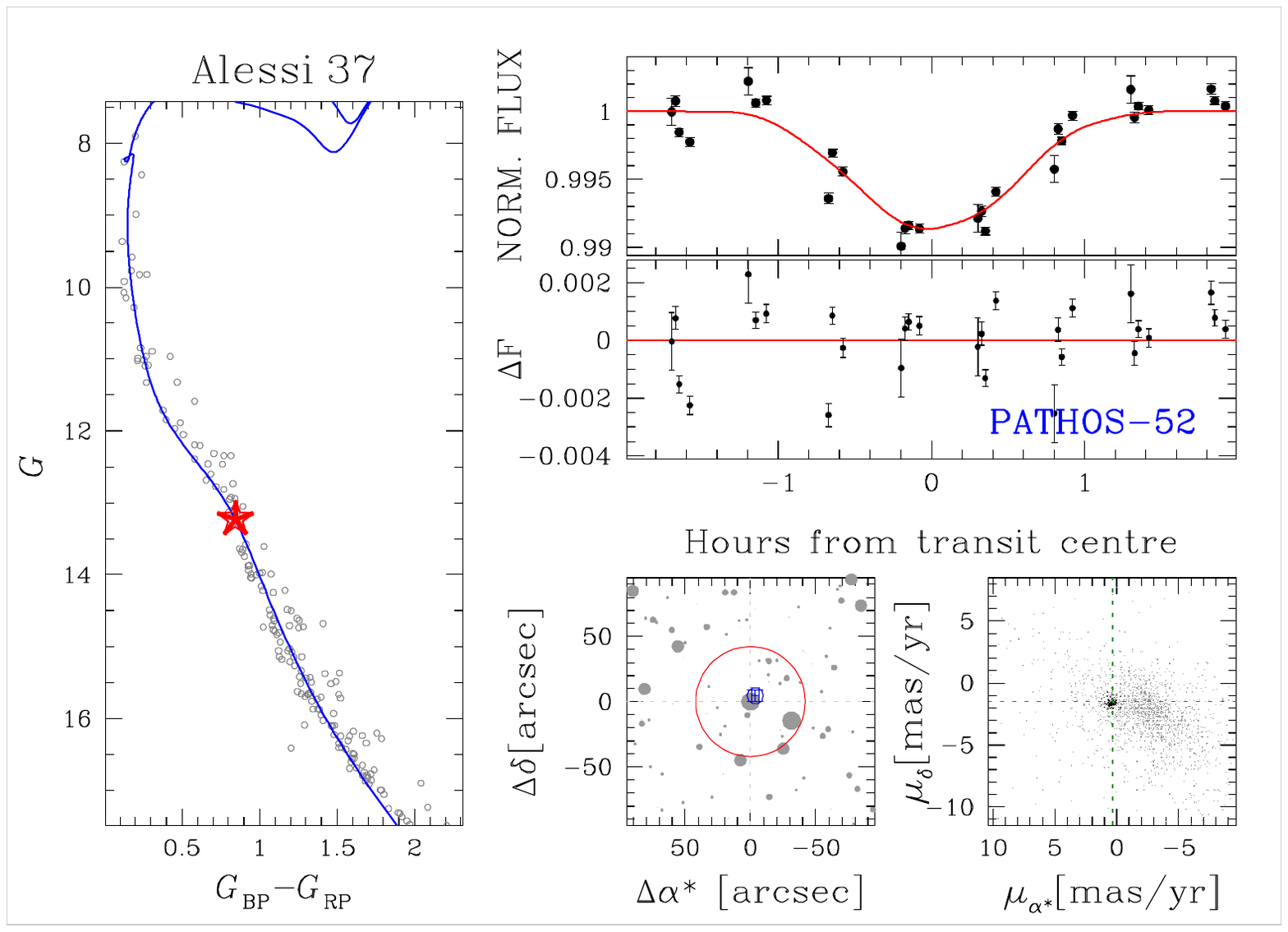} \\
\includegraphics[bb=77 360 535 691, width=0.33\textwidth]{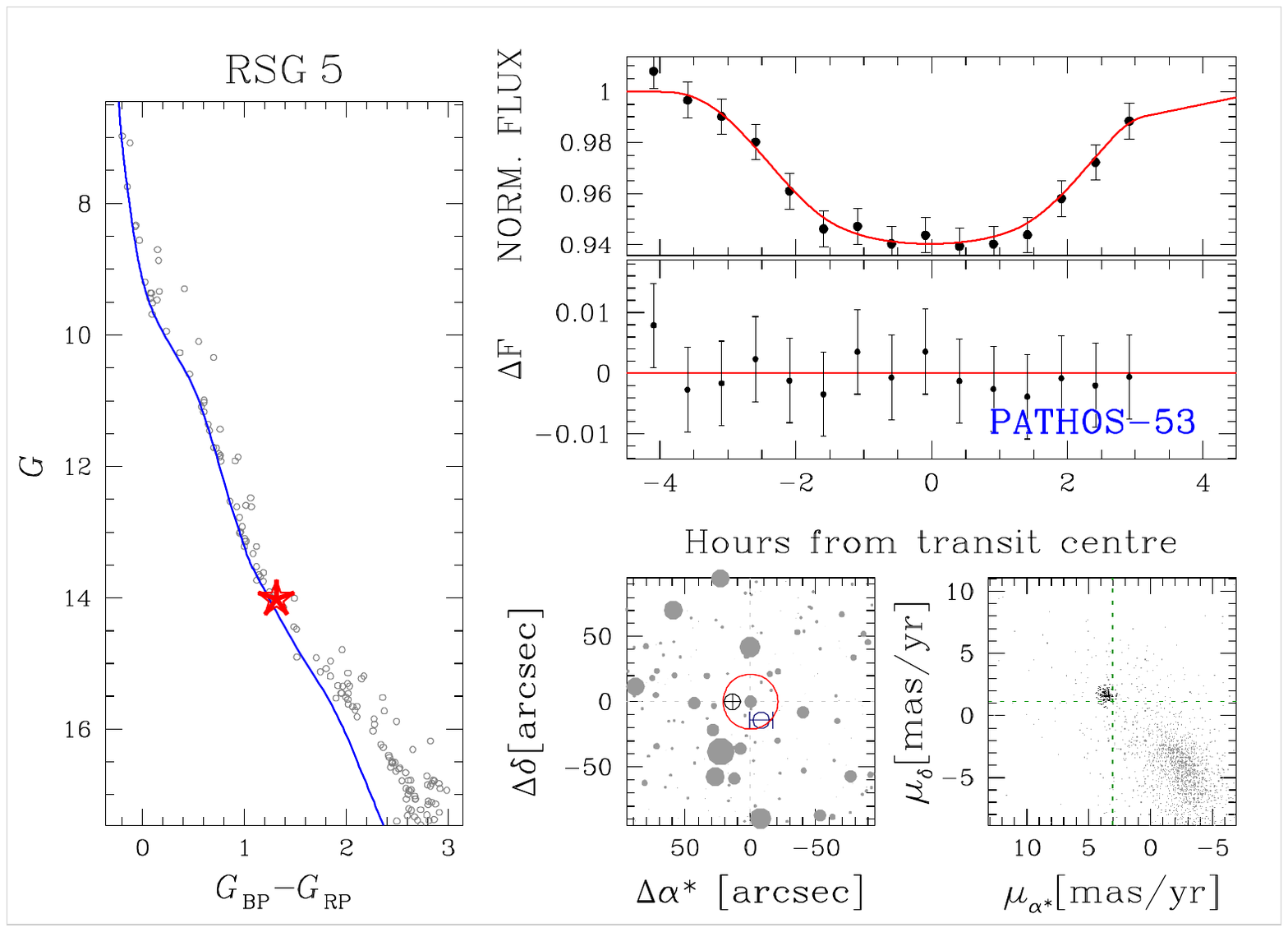}
\includegraphics[bb=77 360 535 691, width=0.33\textwidth]{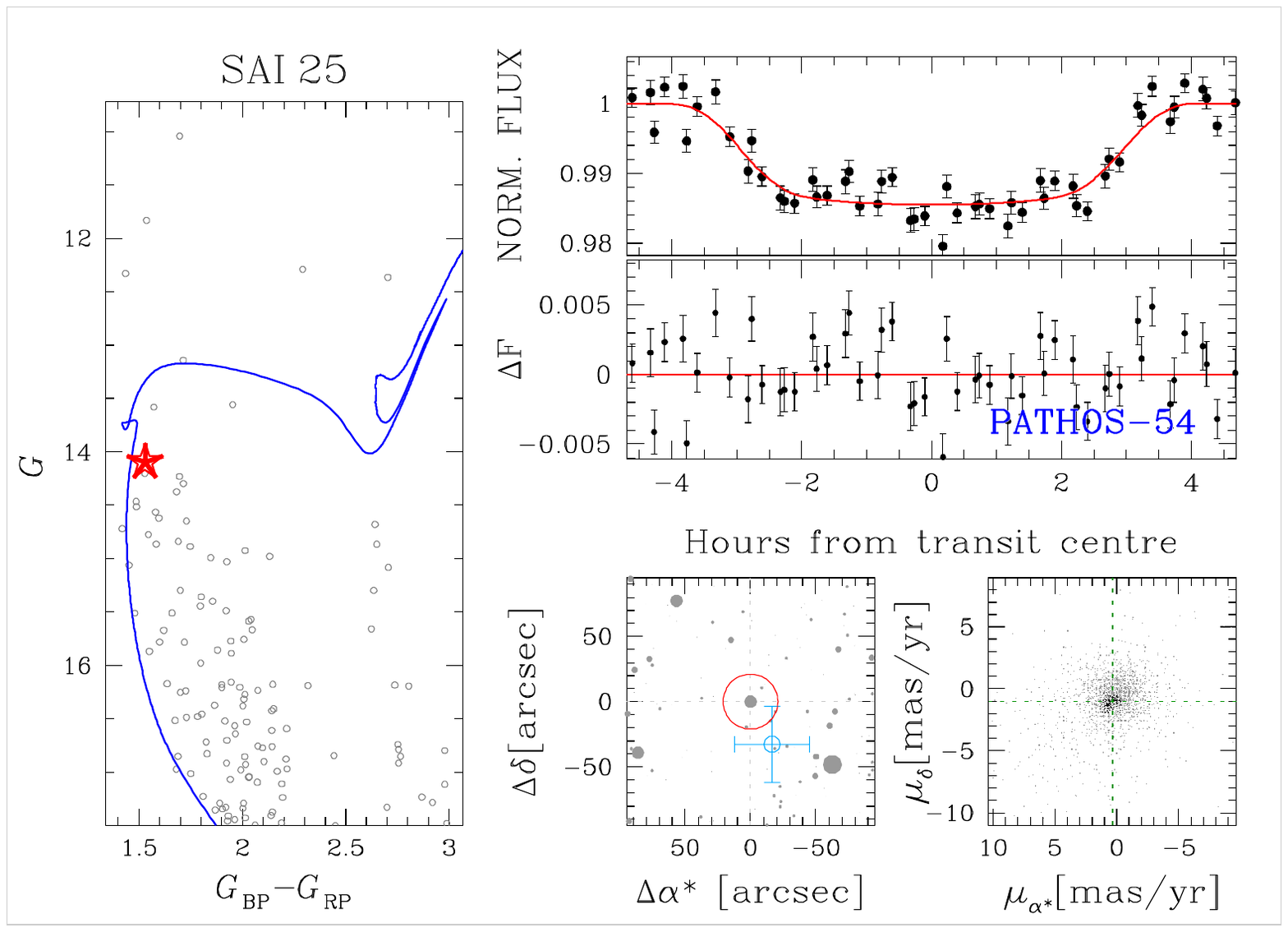}
\includegraphics[bb=77 360 535 691, width=0.33\textwidth]{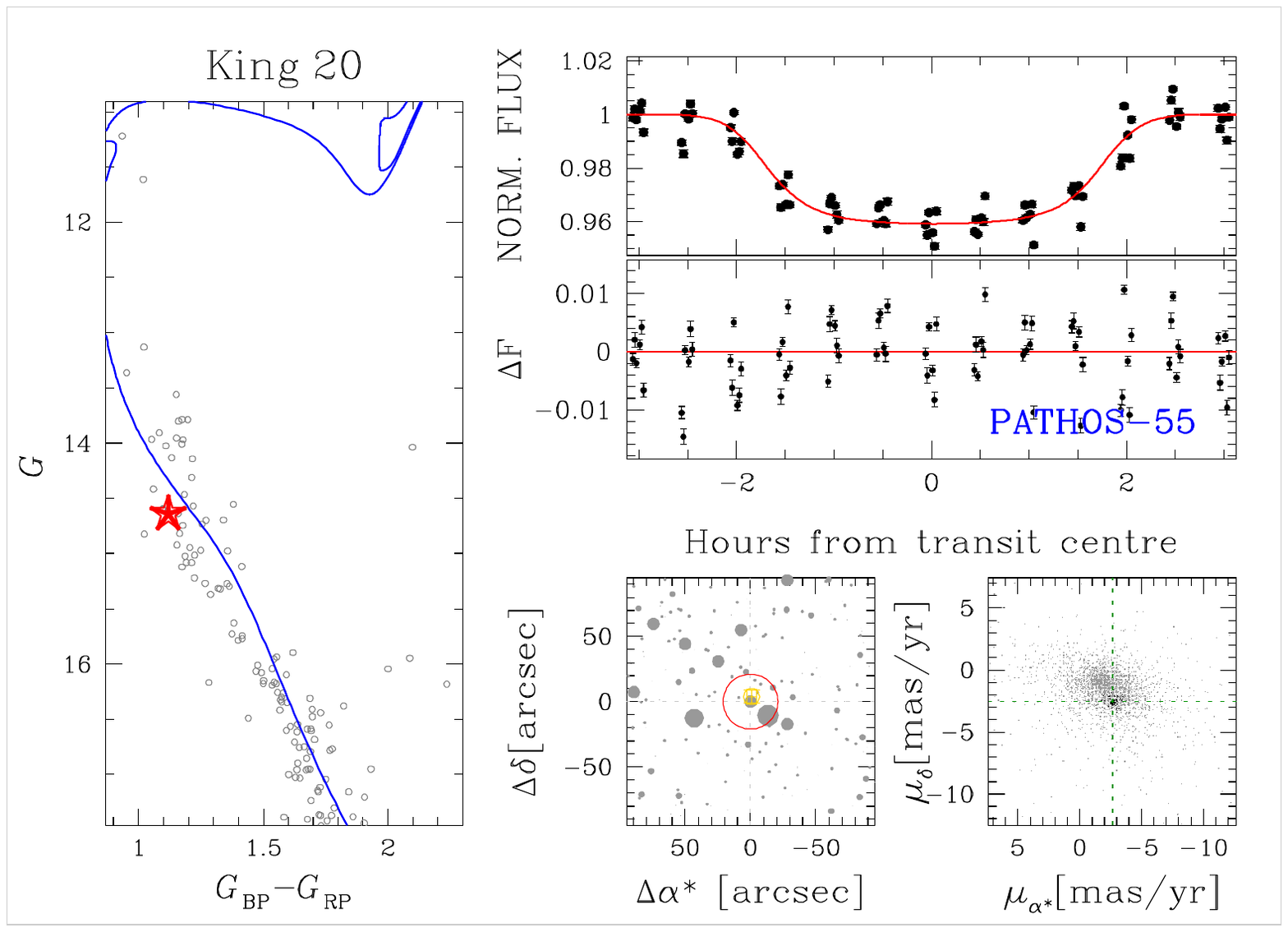} \\
\includegraphics[bb=77 360 535 691, width=0.33\textwidth]{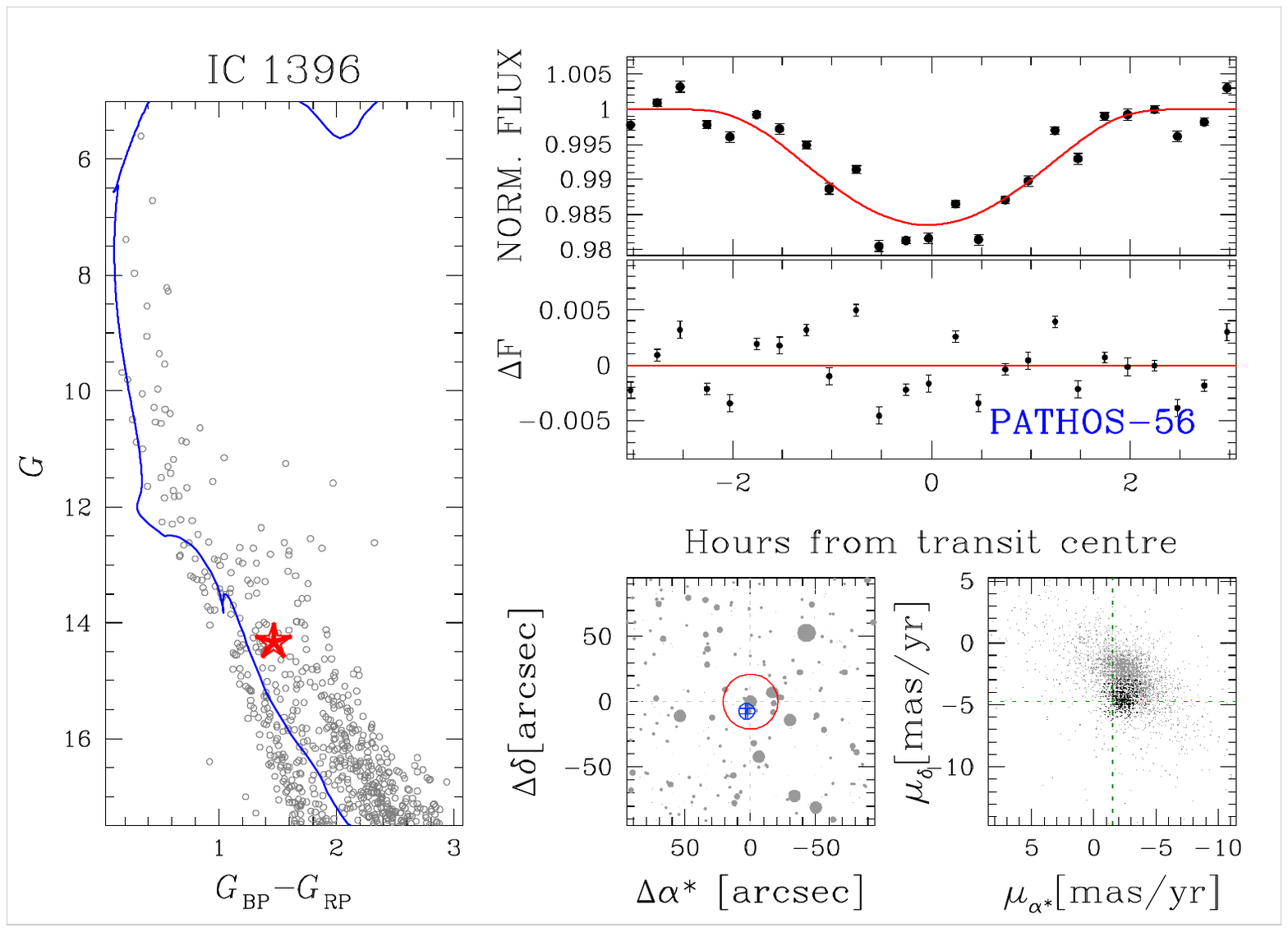}
\includegraphics[bb=77 360 535 691, width=0.33\textwidth]{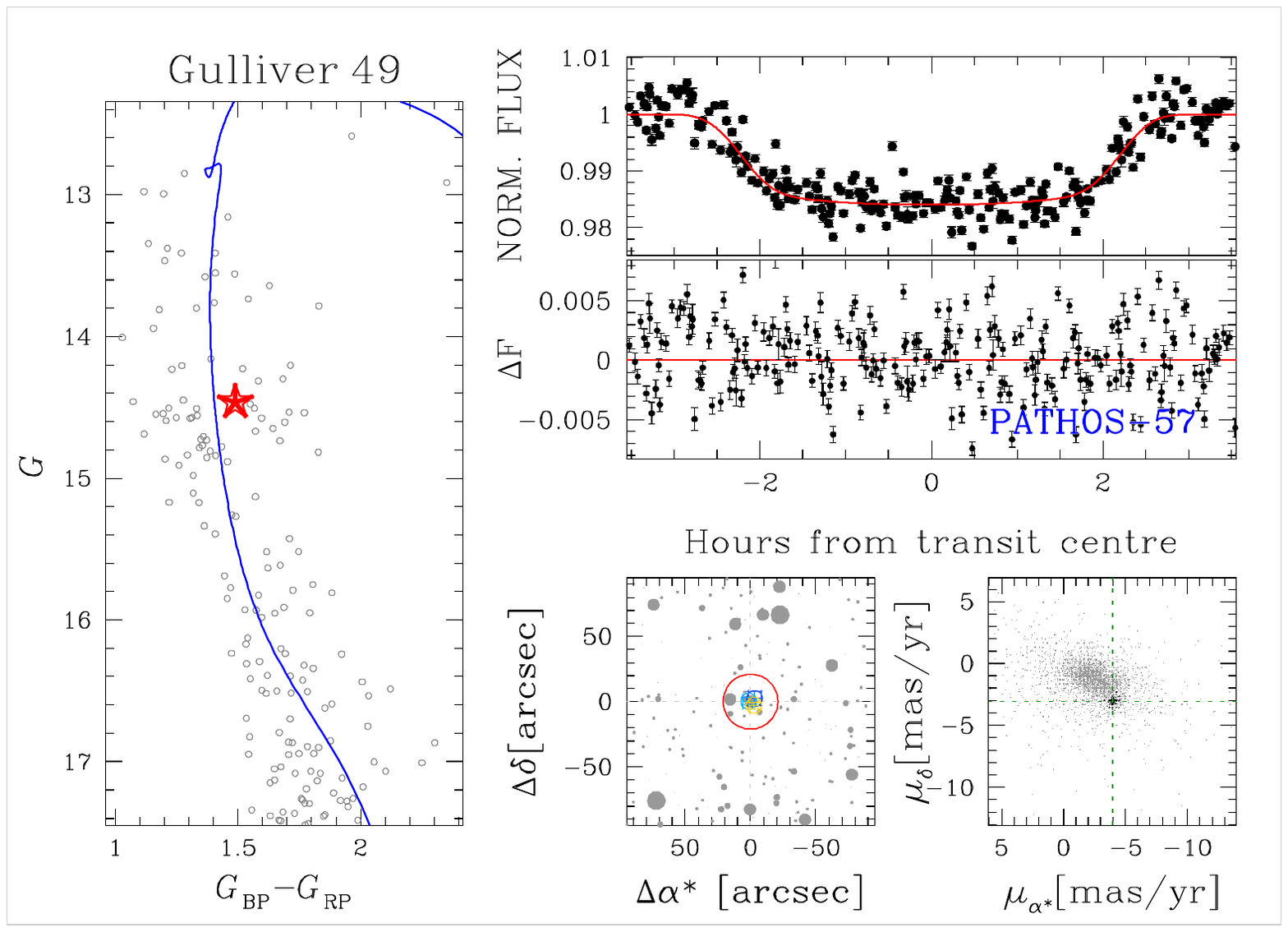}
\includegraphics[bb=77 360 535 691, width=0.33\textwidth]{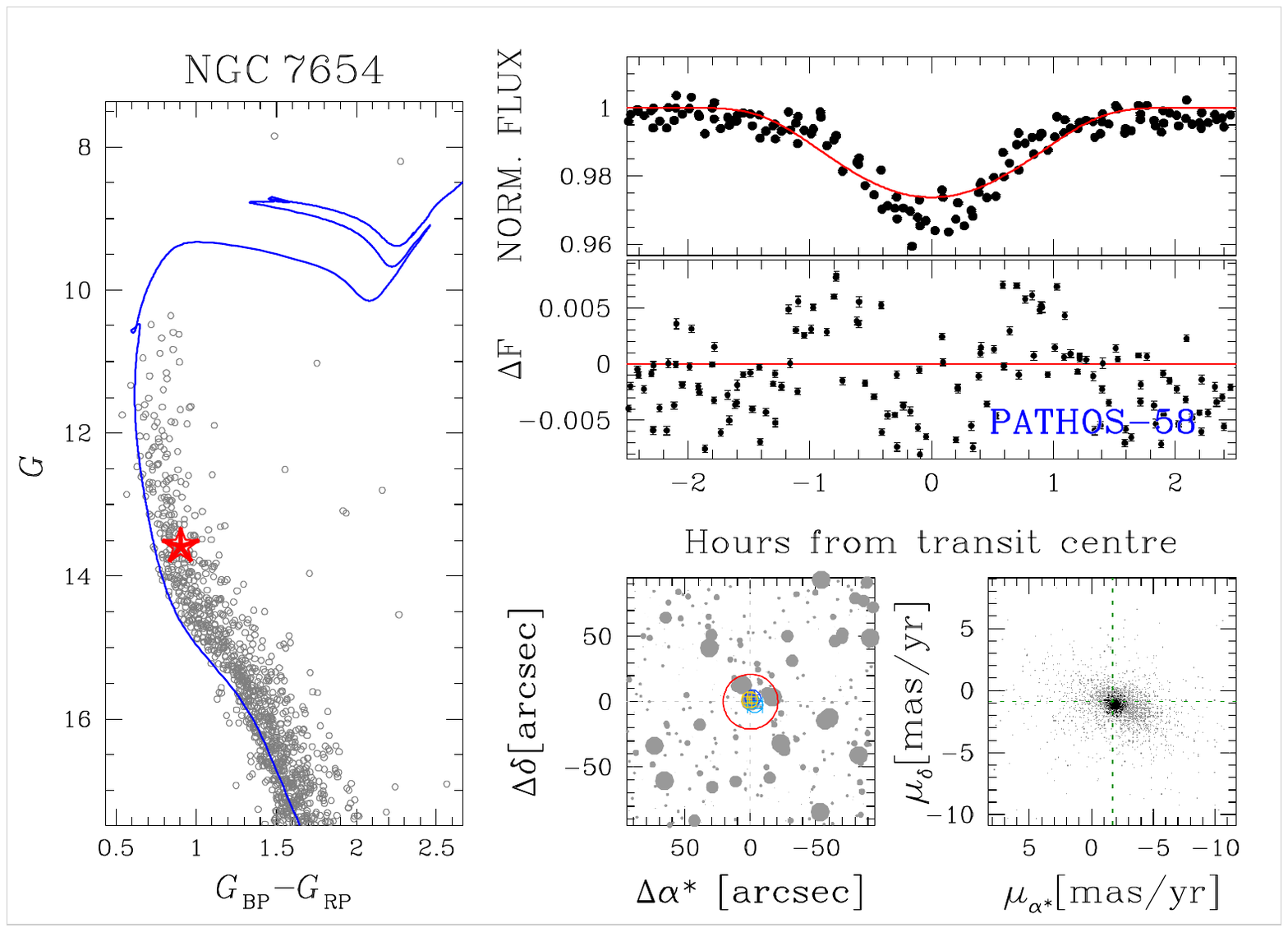} \\
\caption{Overview on the candidate exoplanets PATHOS-44--
  PATHOS-58. On the left-hand, the $G$ versus $G_{\rm BP}-G_{\rm RP}$
  CMD of the cluster that hosts the target star (red star or dashed
  line when $G_{\rm BP}-G_{\rm RP}$ colour is not available) and the
  isochrone (blue) fitted with the cluster parameters listed in
  Table~\ref{tab:1}. On the right-hand, top-panel shows the folded
  light curve (grey points) of the candidate and the model (in red)
  found with PyORBIT; middle panel shows the difference between the
  observed points and the model. Bottom left-hand panel shows the $95
  \times 95$\,arcsec$^2$ finding chart centred on the target star; red
  circle is the aperture adopted to extract photometry, crosses are
  the in-/out-of transit difference centroid. Bottom-right panel is
  the vector-point diagram, centred on the target star, for all the
  stars that are within 10\,arcmin from the target star; black points
  are the cluster members listed in the catalogue by
  \citet{2018A&A...618A..93C}.
  \label{fig:A1a}}
\end{figure*}

\begin{figure*}
\includegraphics[bb=77 360 535 691, width=0.33\textwidth]{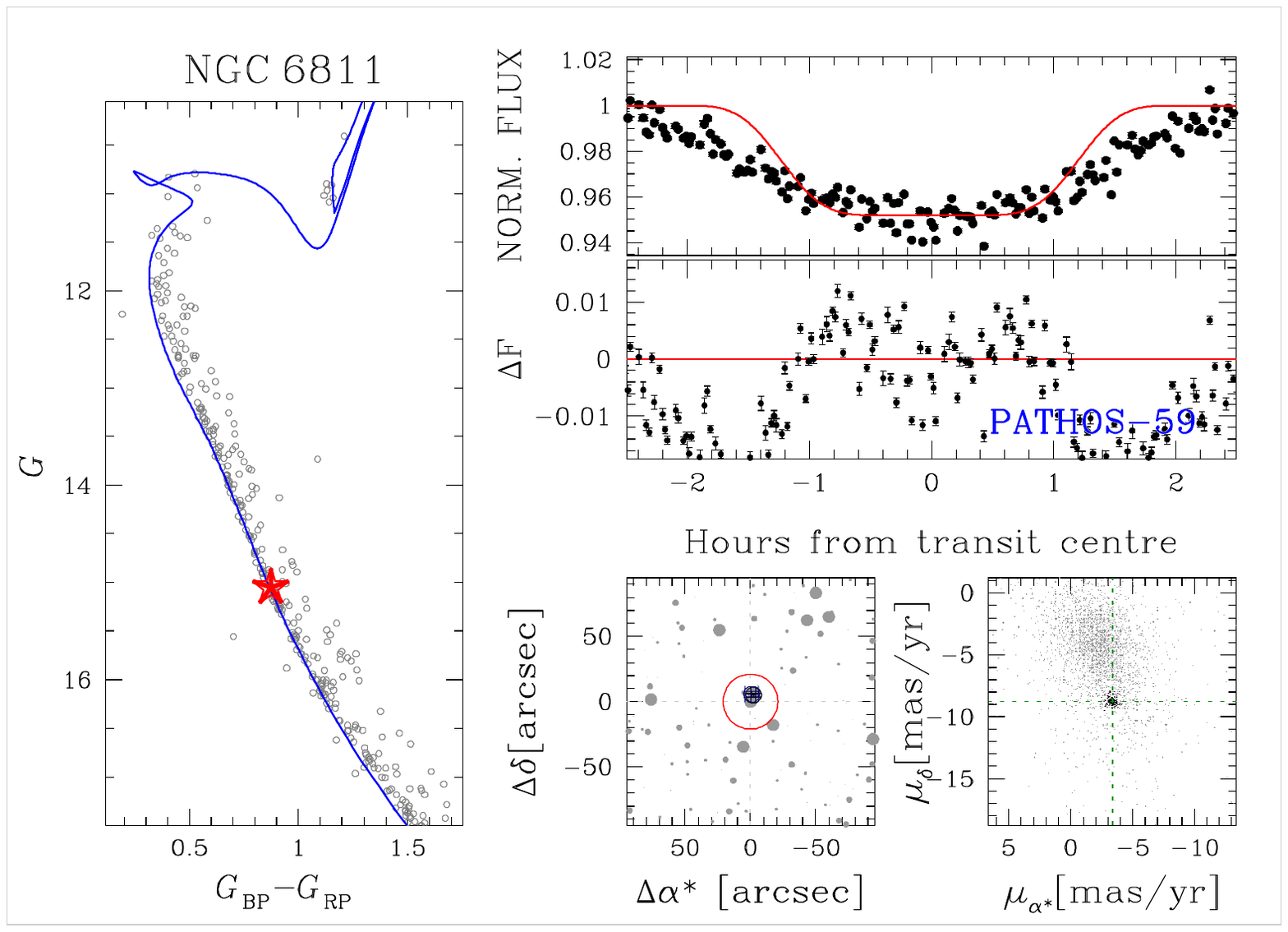}
\includegraphics[bb=77 360 535 691, width=0.33\textwidth]{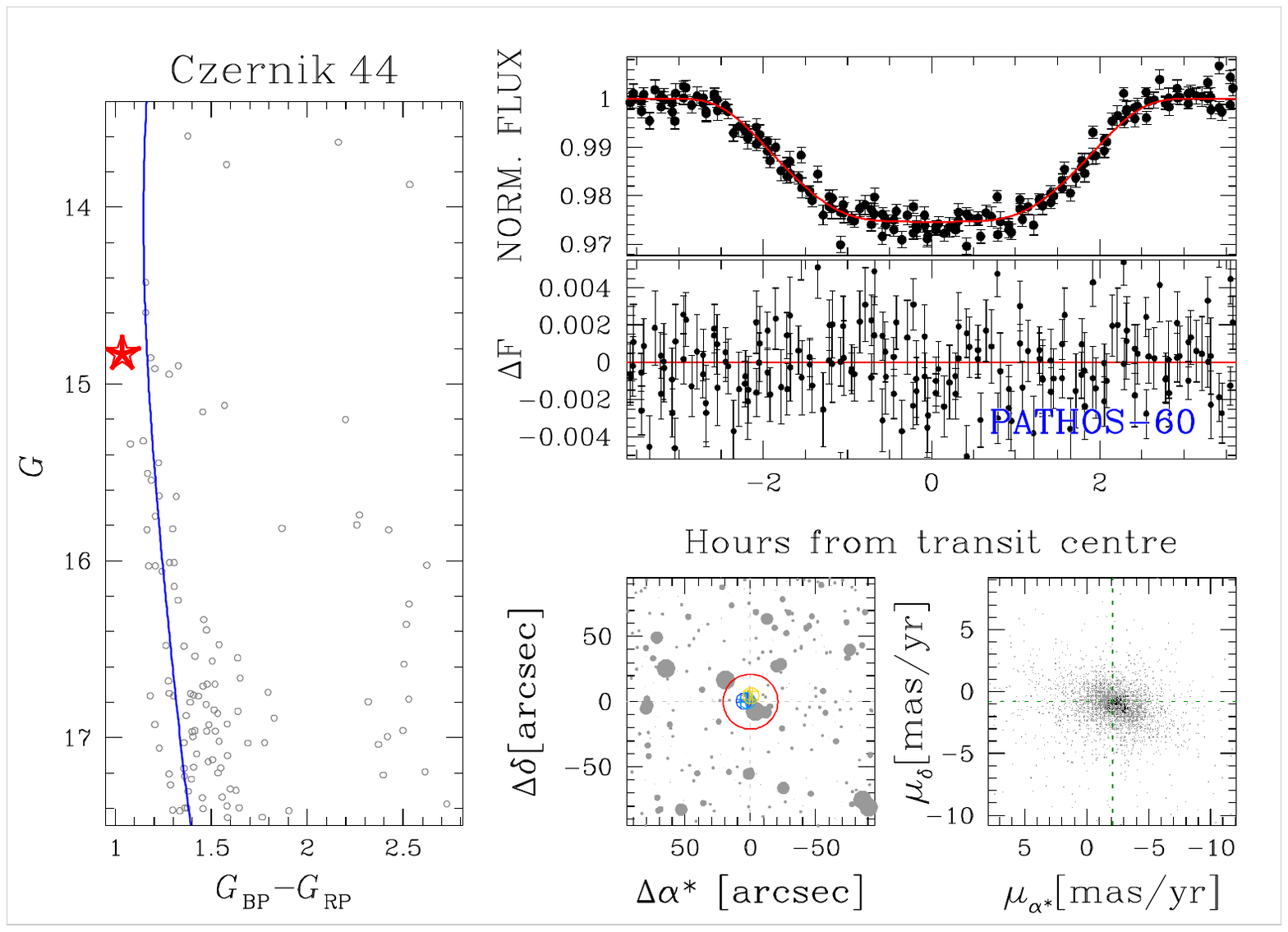}
\includegraphics[bb=77 360 535 691, width=0.33\textwidth]{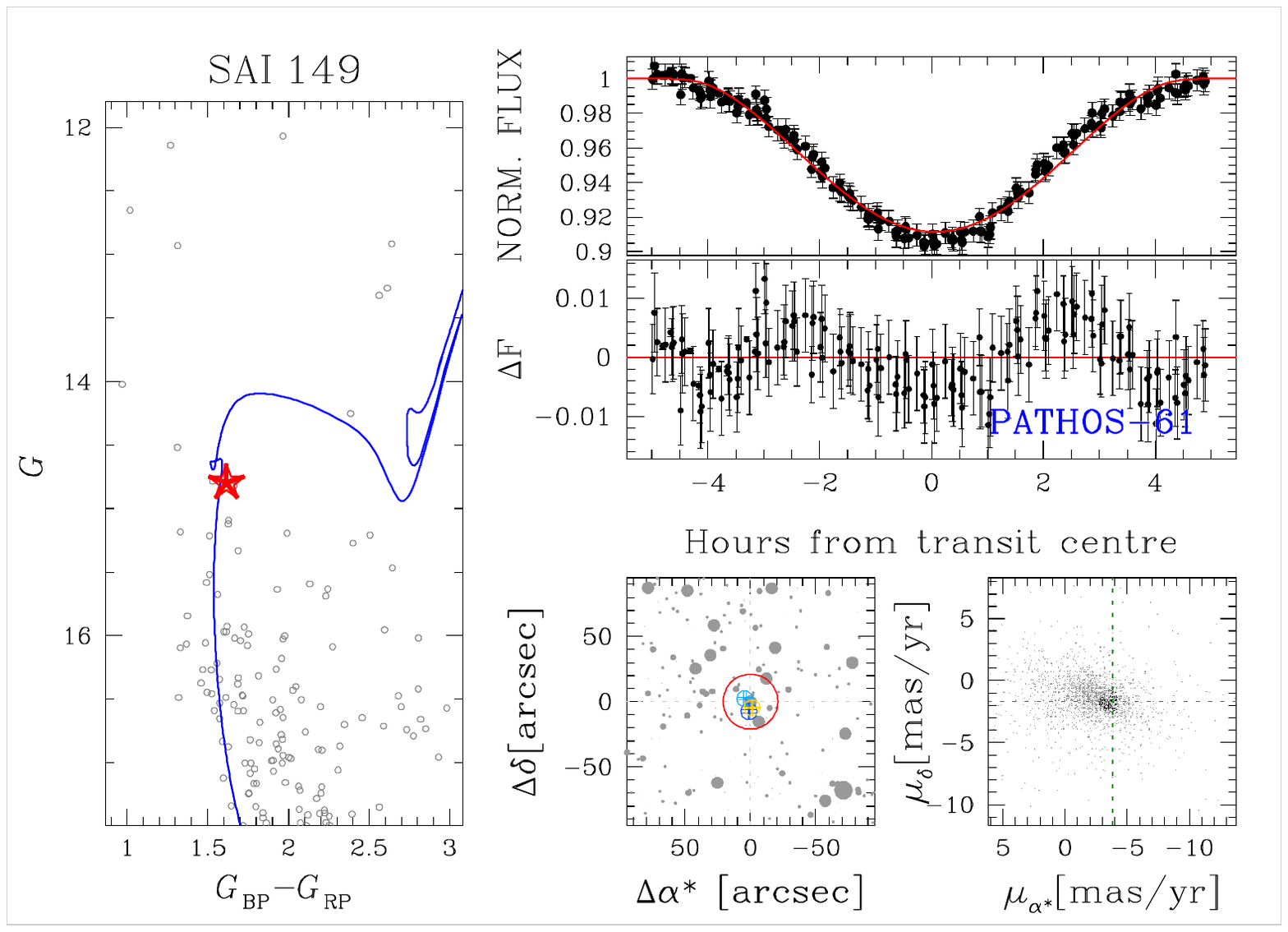} \\
\includegraphics[bb=77 360 535 691, width=0.33\textwidth]{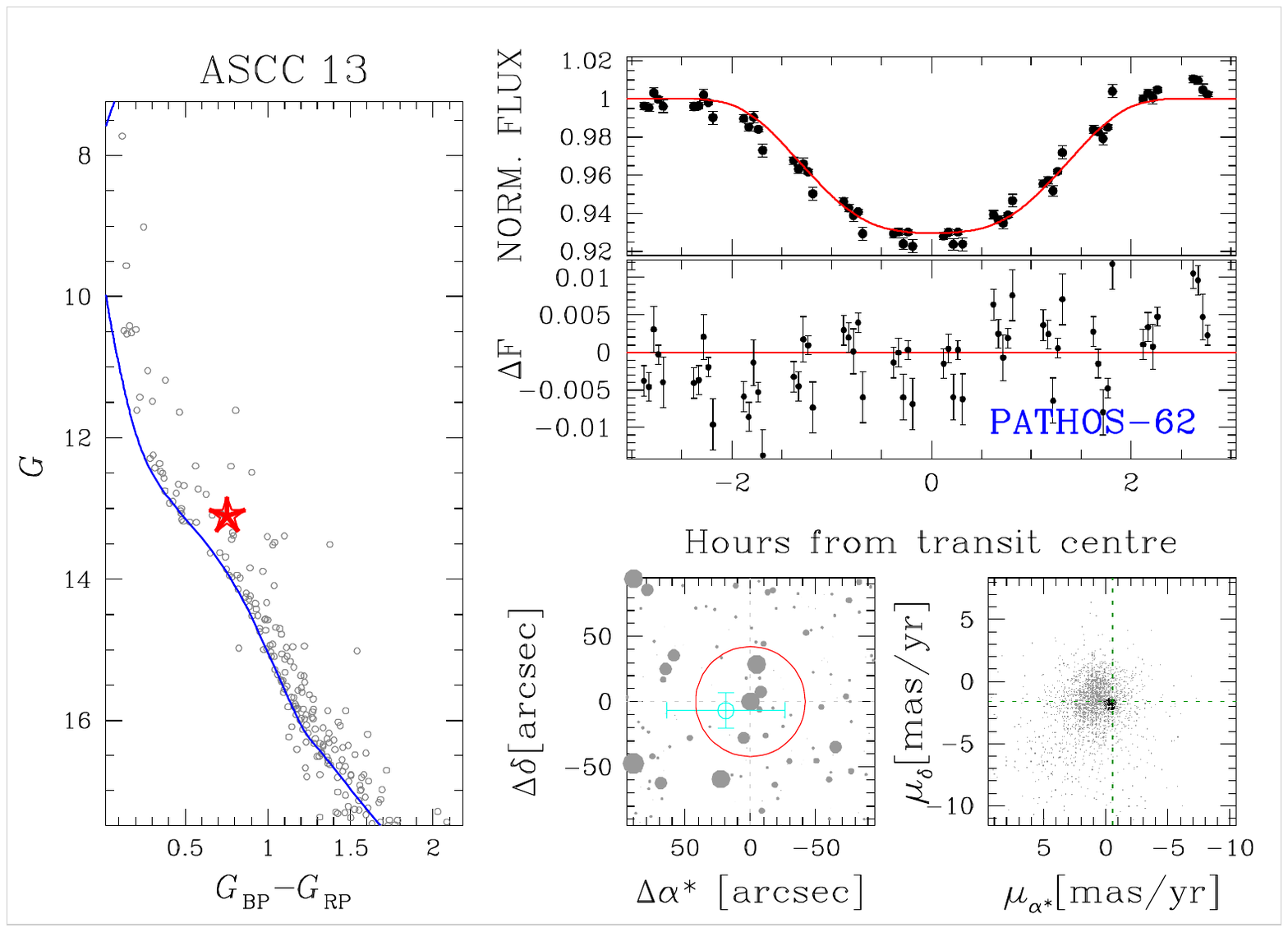}
\includegraphics[bb=77 360 535 691, width=0.33\textwidth]{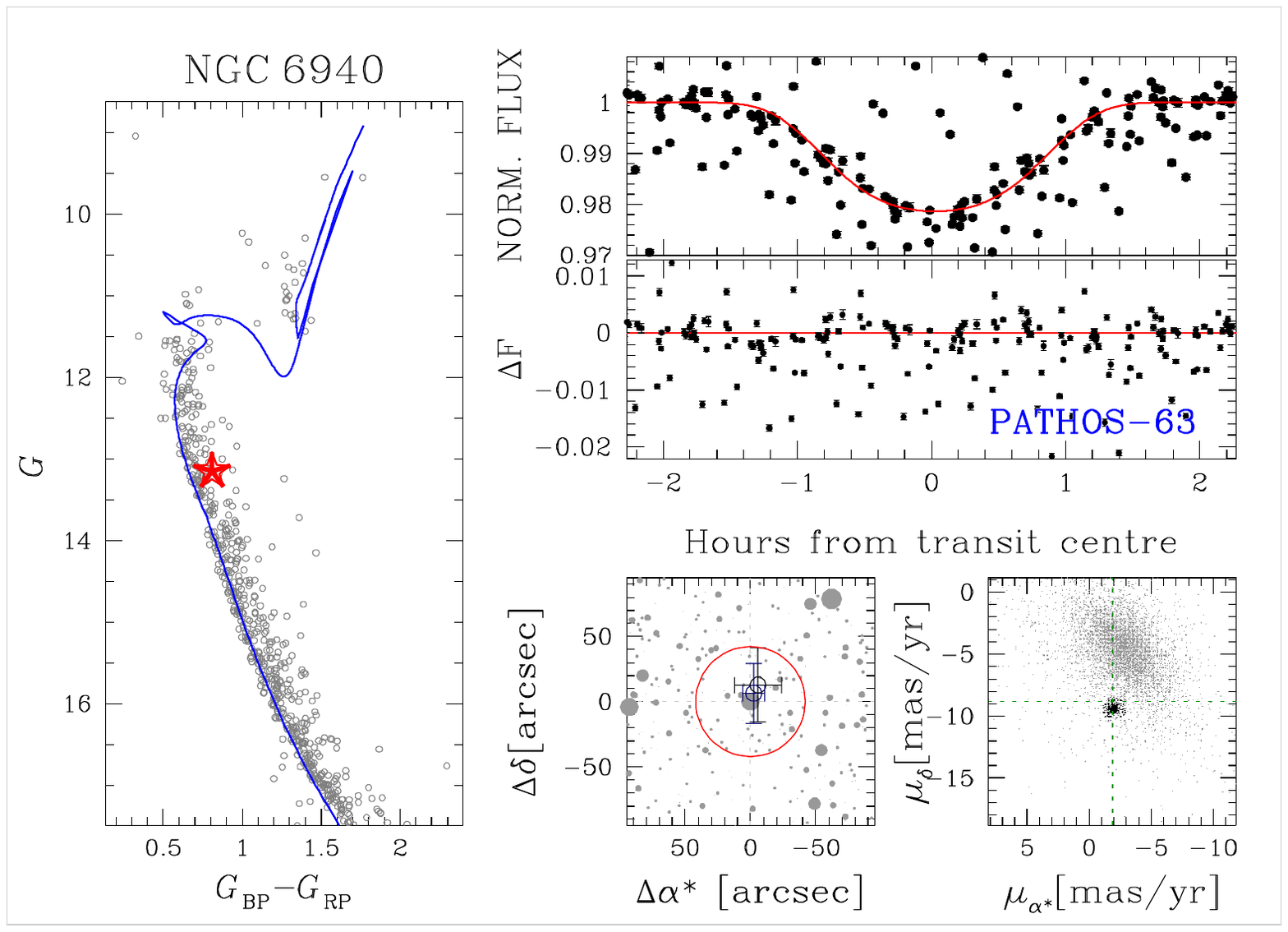}
\includegraphics[bb=77 360 535 691, width=0.33\textwidth]{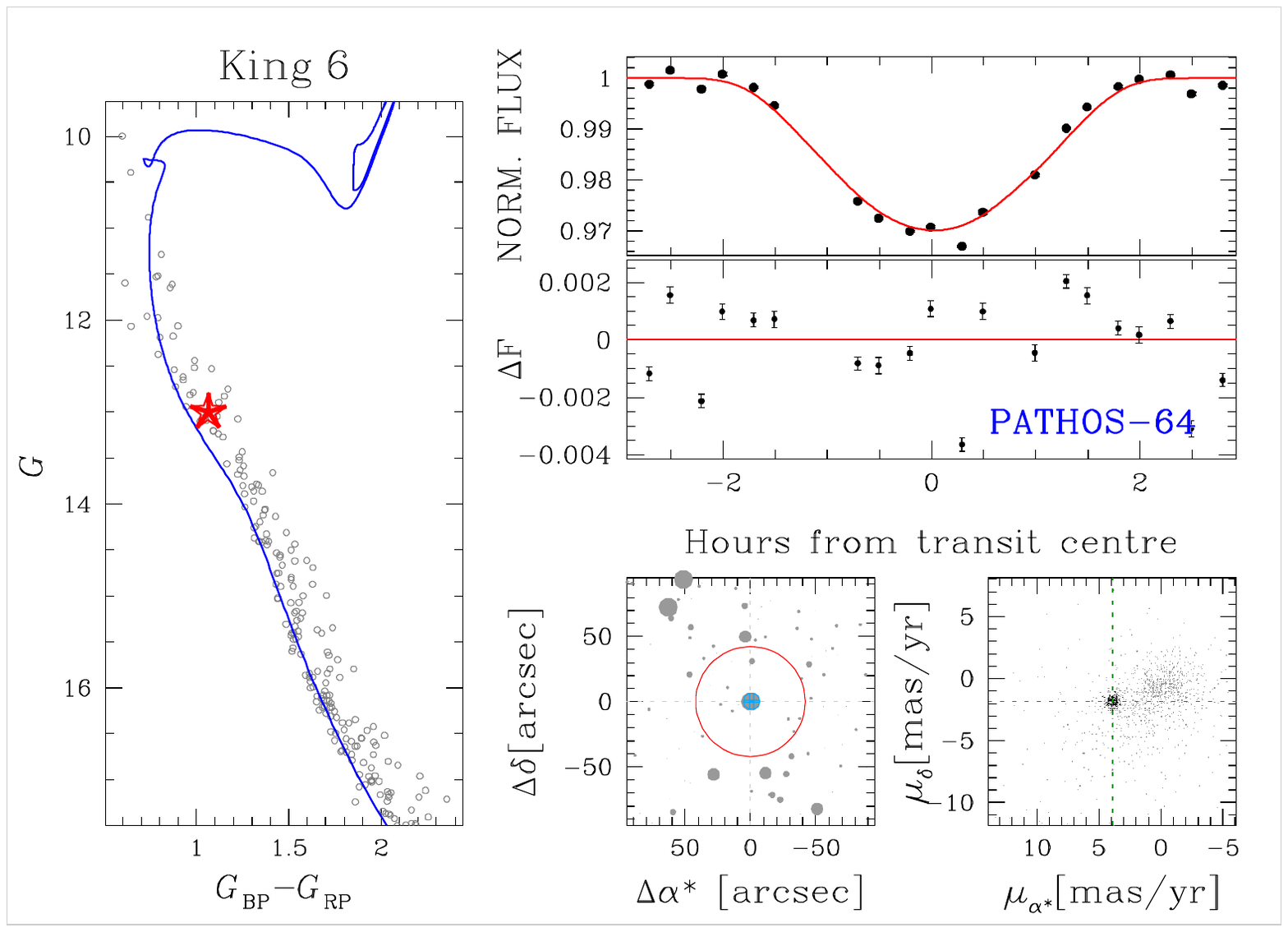} \\
\includegraphics[bb=77 360 535 691, width=0.33\textwidth]{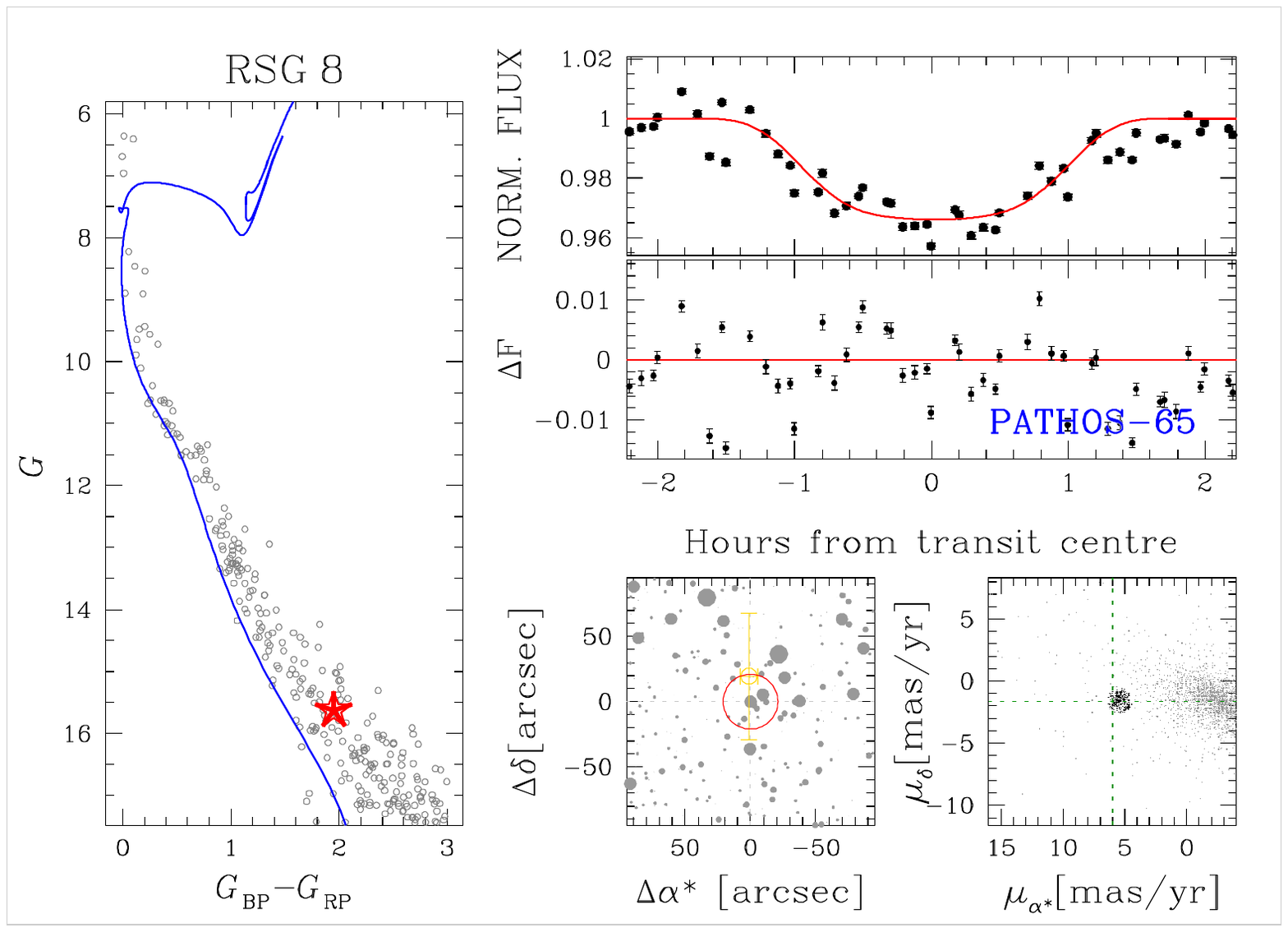}
\includegraphics[bb=77 360 535 691, width=0.33\textwidth]{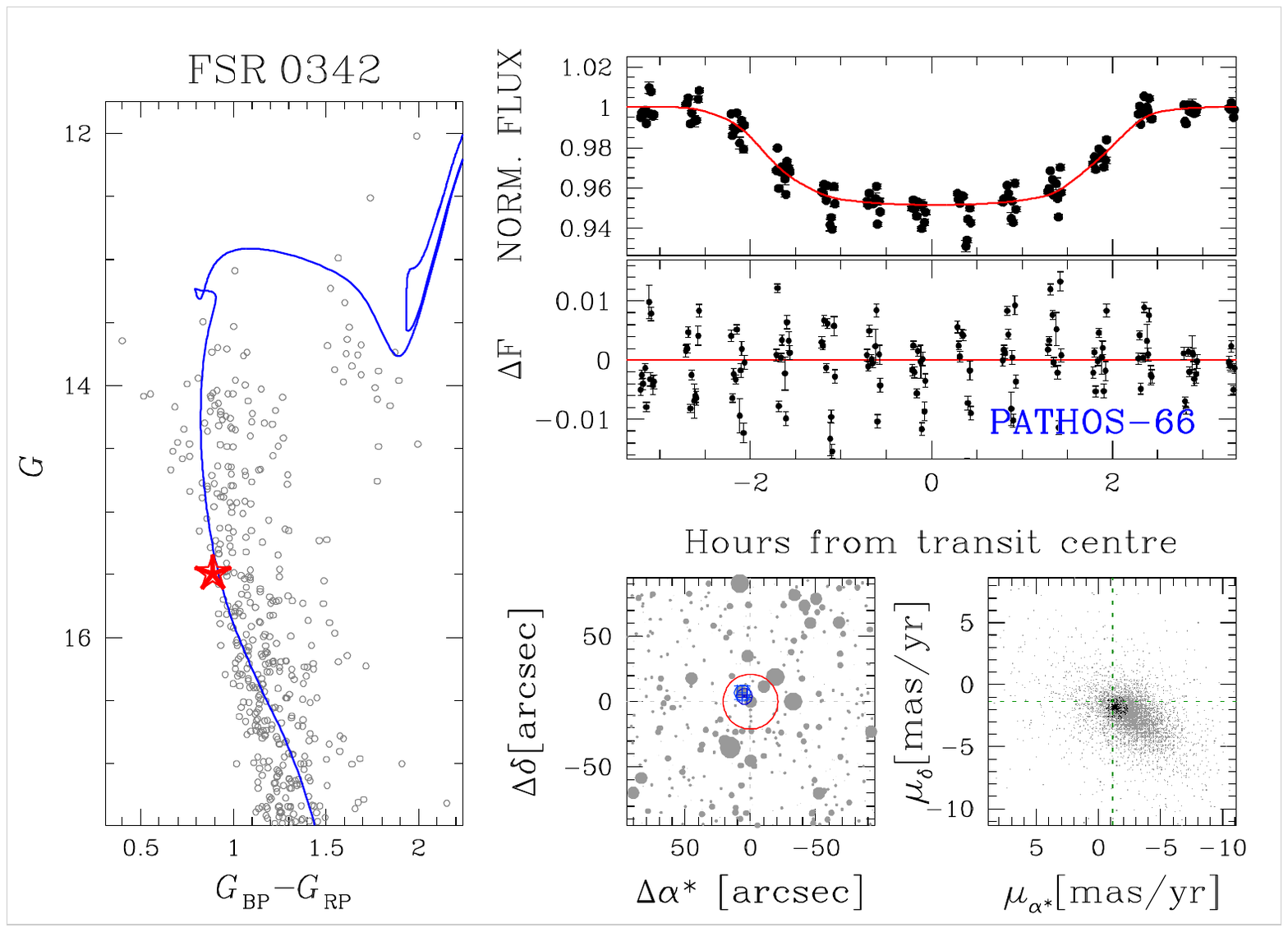}
\includegraphics[bb=77 360 535 691, width=0.33\textwidth]{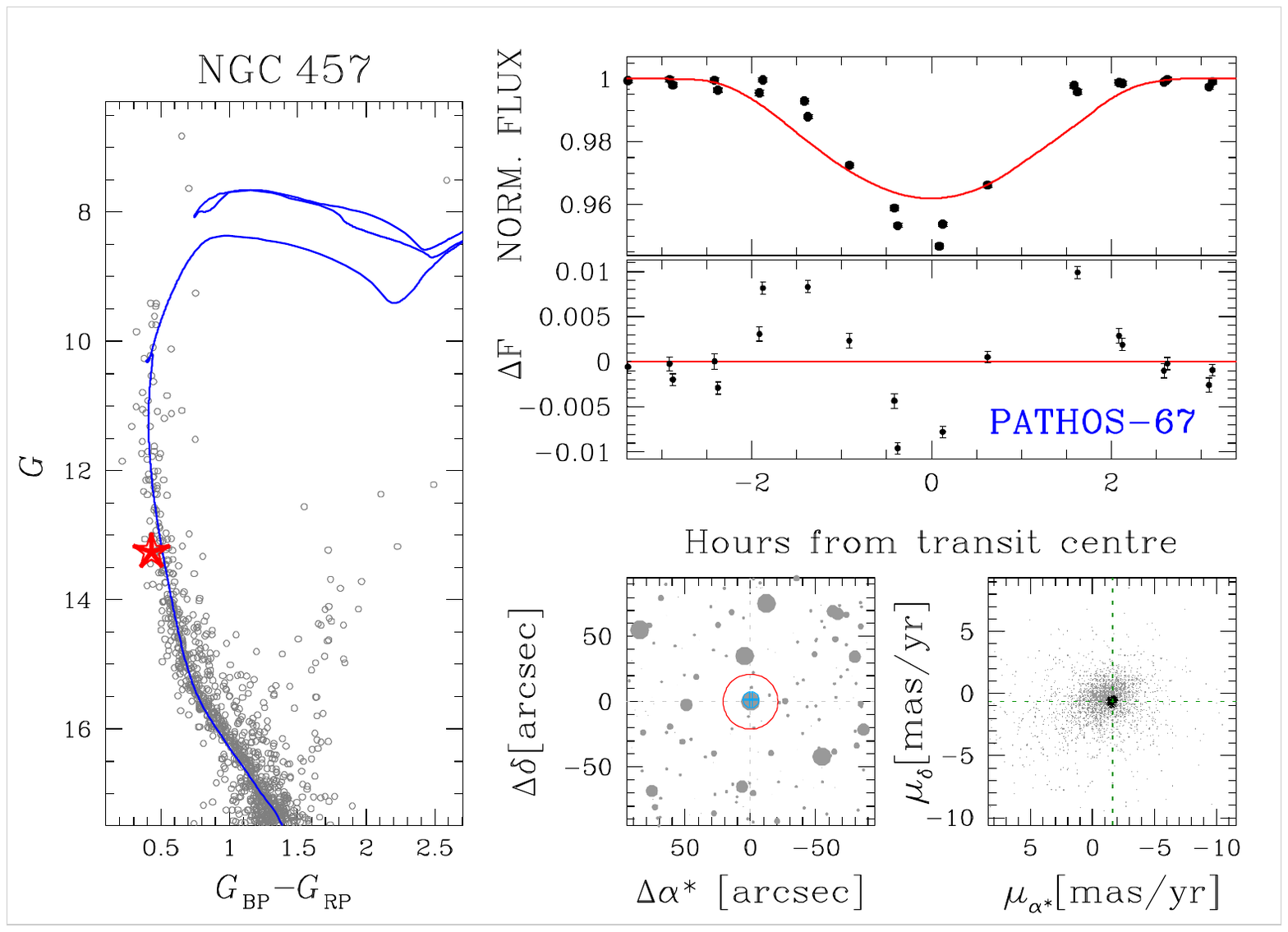} \\
\includegraphics[bb=77 360 535 691, width=0.33\textwidth]{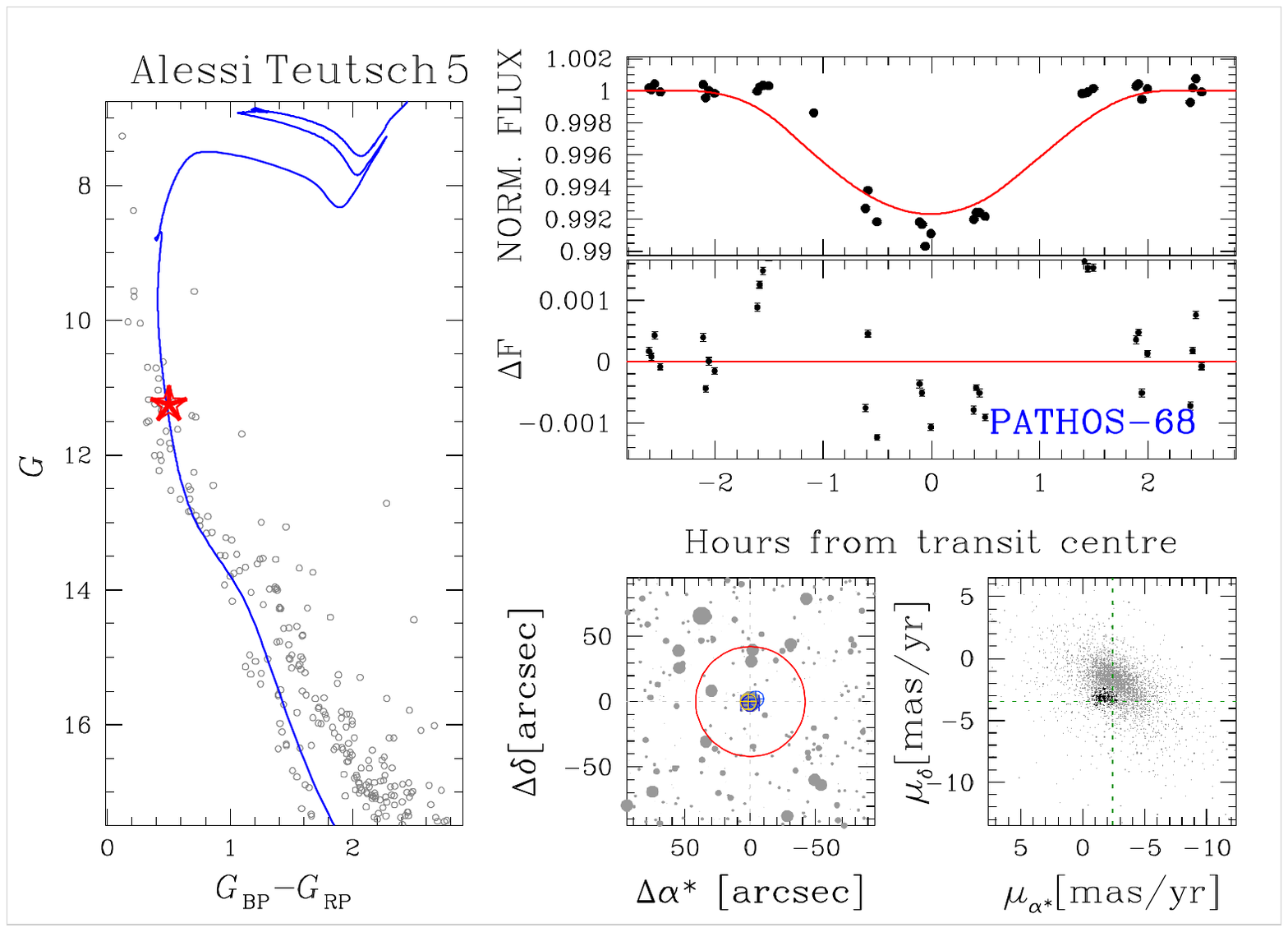}
\includegraphics[bb=77 360 535 691, width=0.33\textwidth]{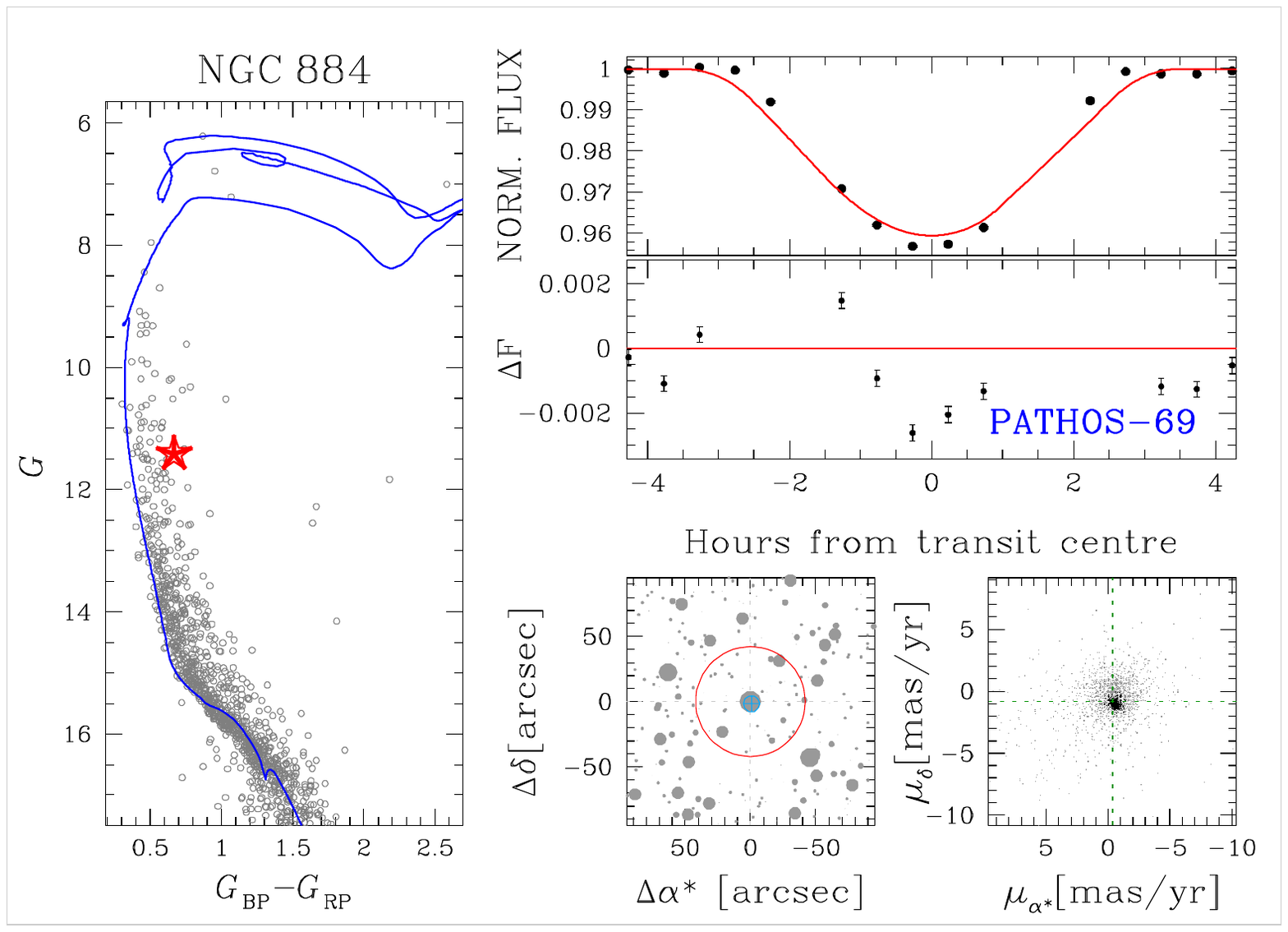}
\includegraphics[bb=77 360 535 691, width=0.33\textwidth]{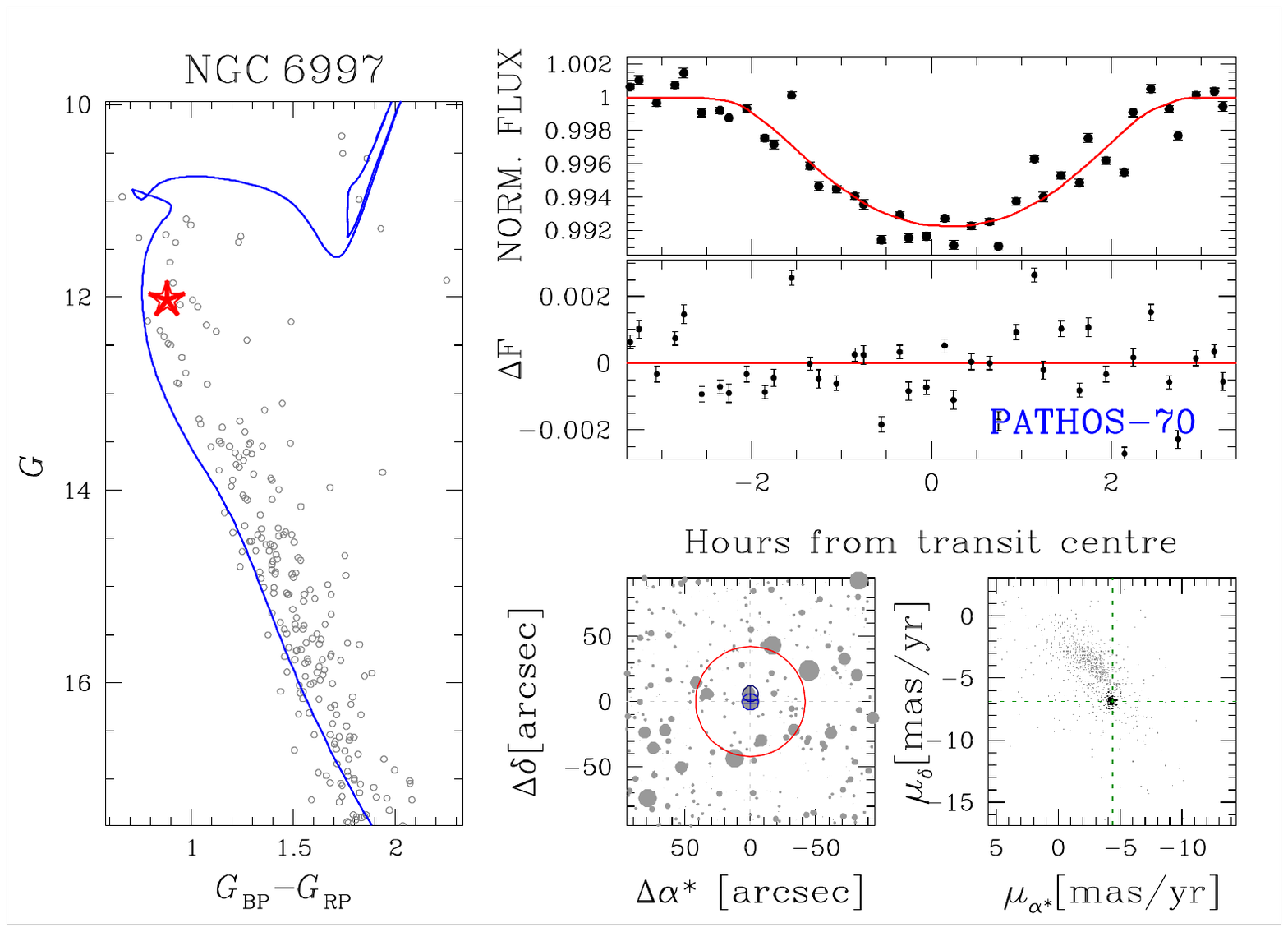} \\
\includegraphics[bb=77 360 535 691, width=0.33\textwidth]{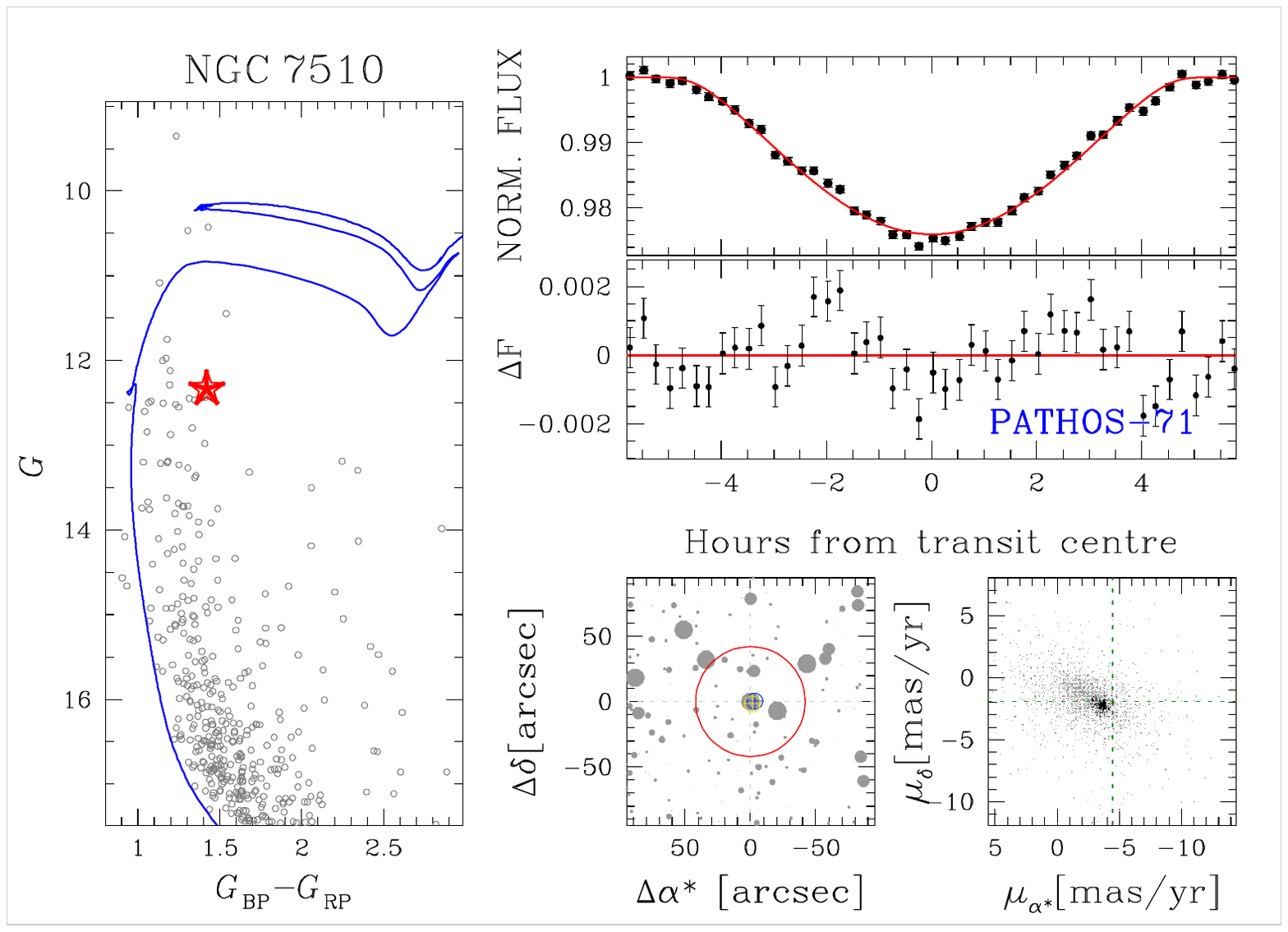}
\includegraphics[bb=77 360 535 691, width=0.33\textwidth]{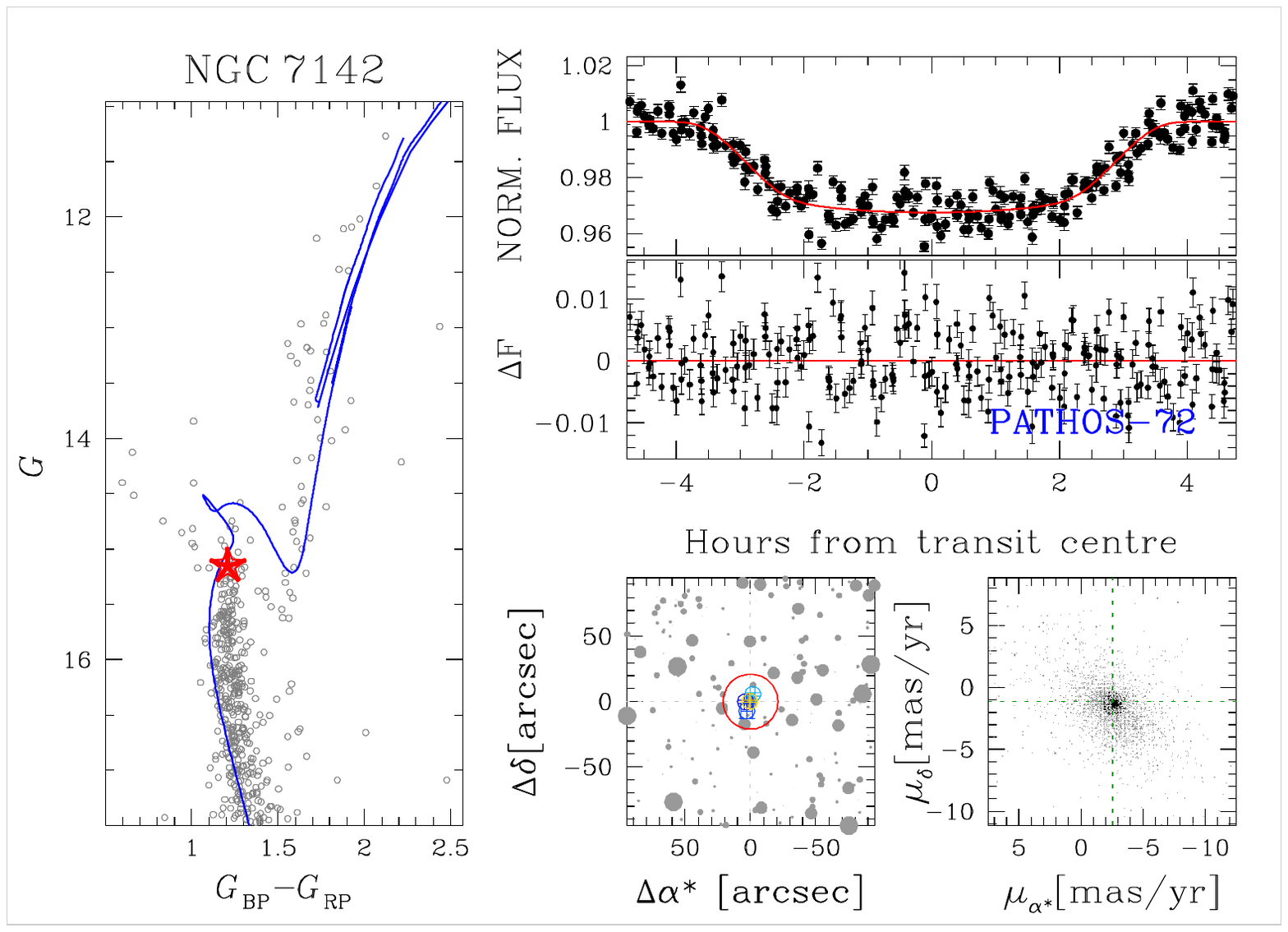}
\includegraphics[bb=77 360 535 691, width=0.33\textwidth]{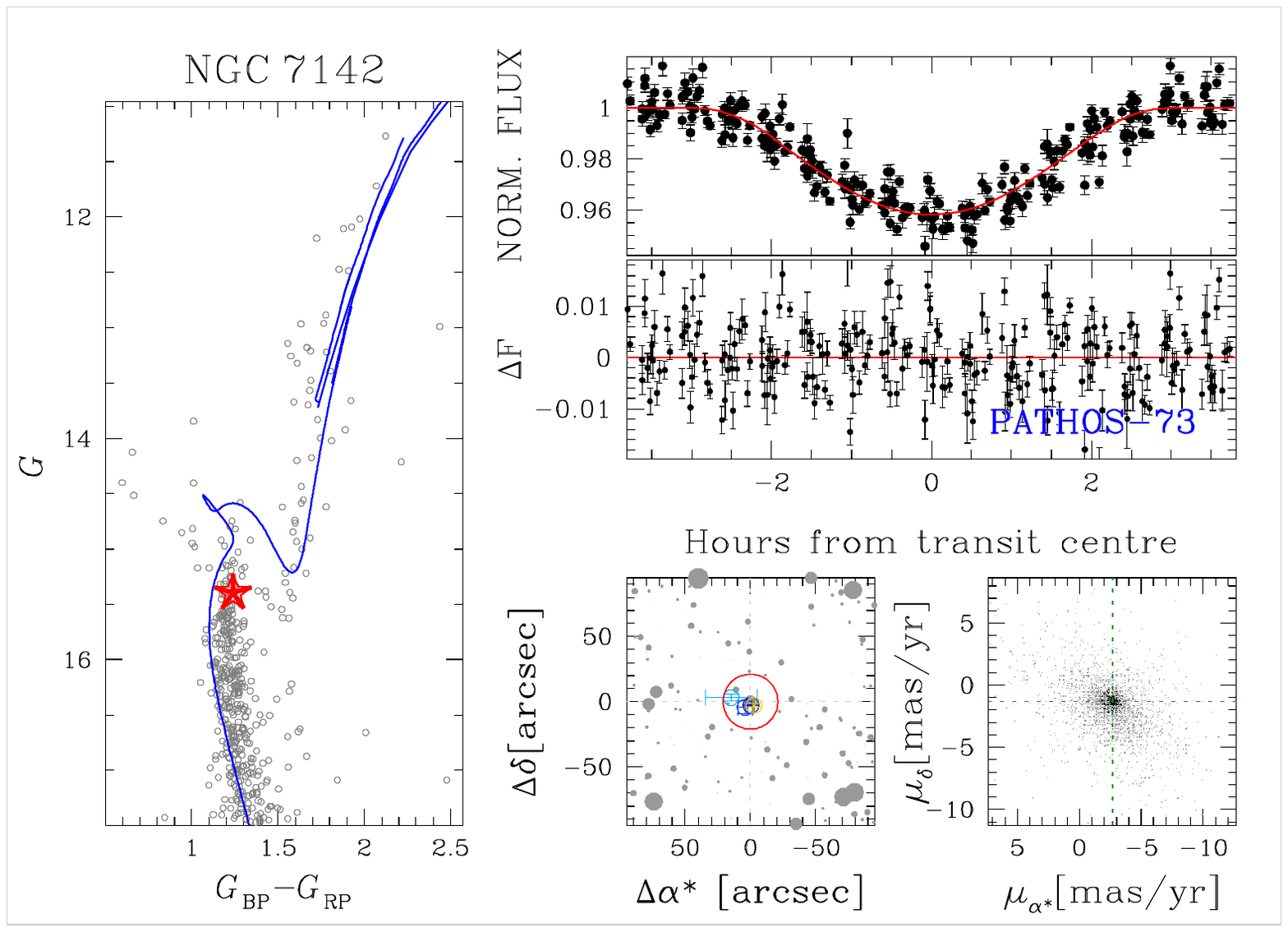} \\
\caption{As in Fig.~\ref{fig:A1a}, but for PATHOS-59 -- PATHOS-73.
  \label{fig:A1b}}
\end{figure*}

\begin{figure*}

\includegraphics[bb=77 360 535 691, width=0.33\textwidth]{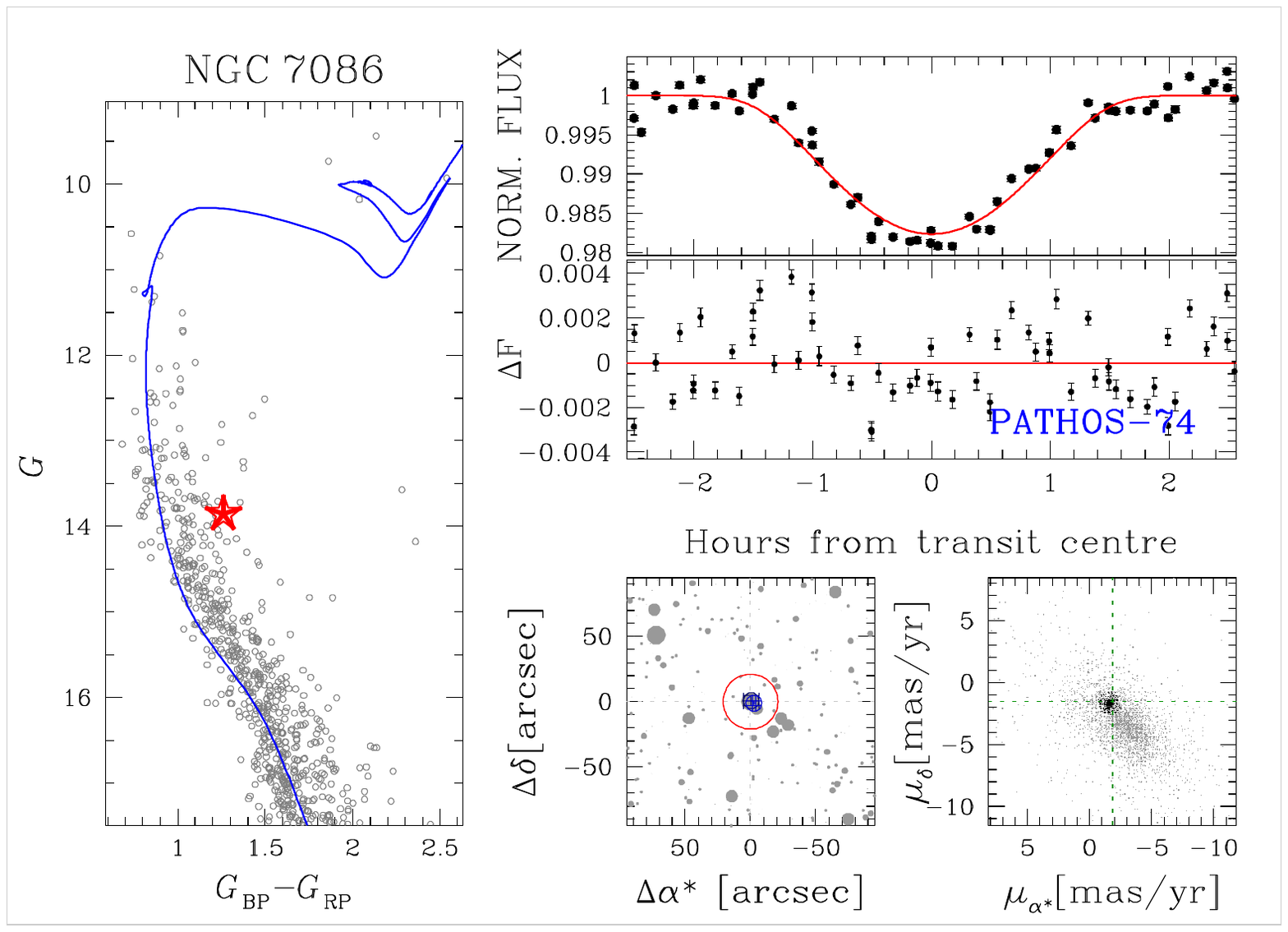}
\includegraphics[bb=77 360 535 691, width=0.33\textwidth]{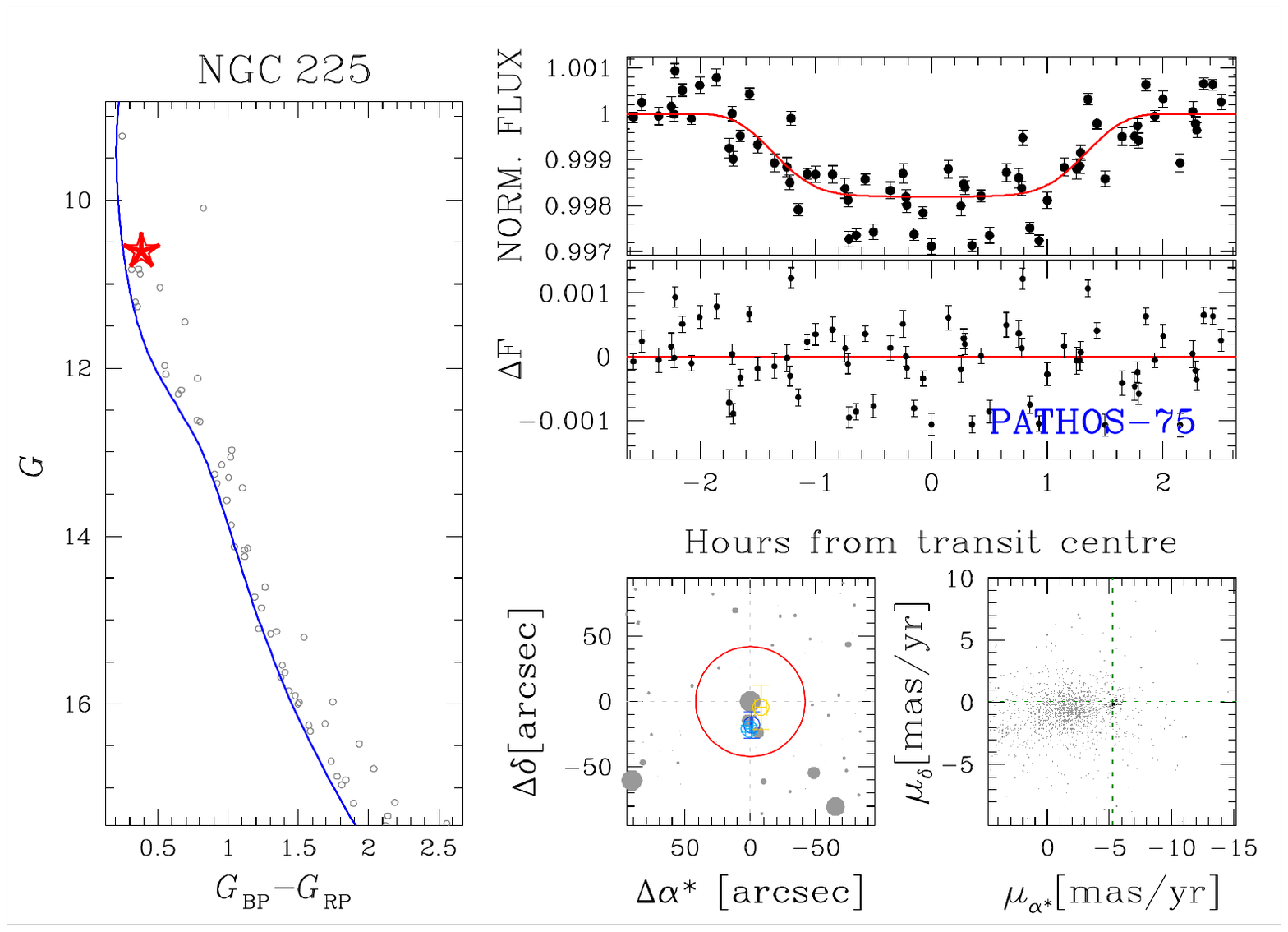}
\includegraphics[bb=77 360 535 691, width=0.33\textwidth]{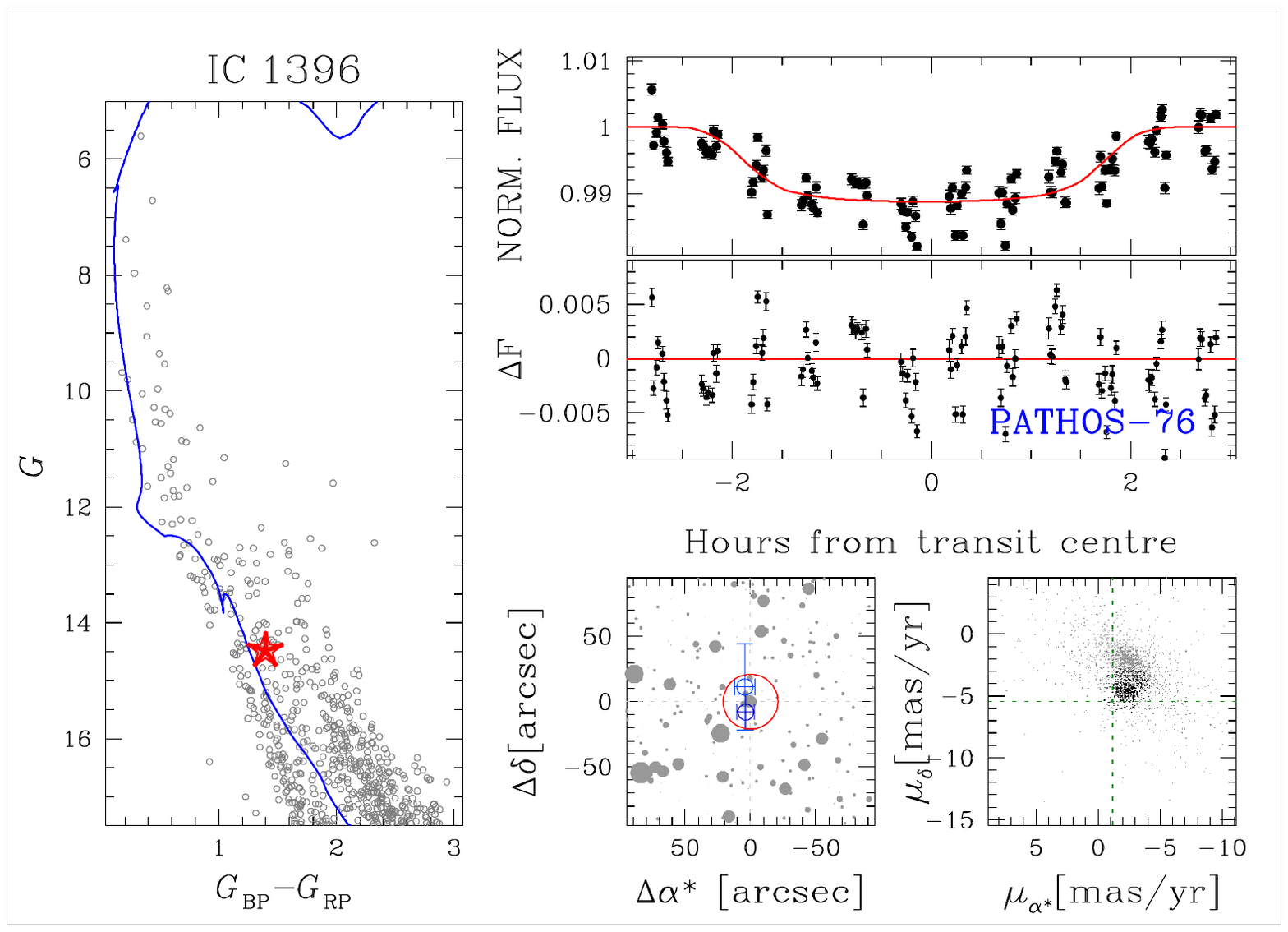} \\
\includegraphics[bb=77 360 535 691, width=0.33\textwidth]{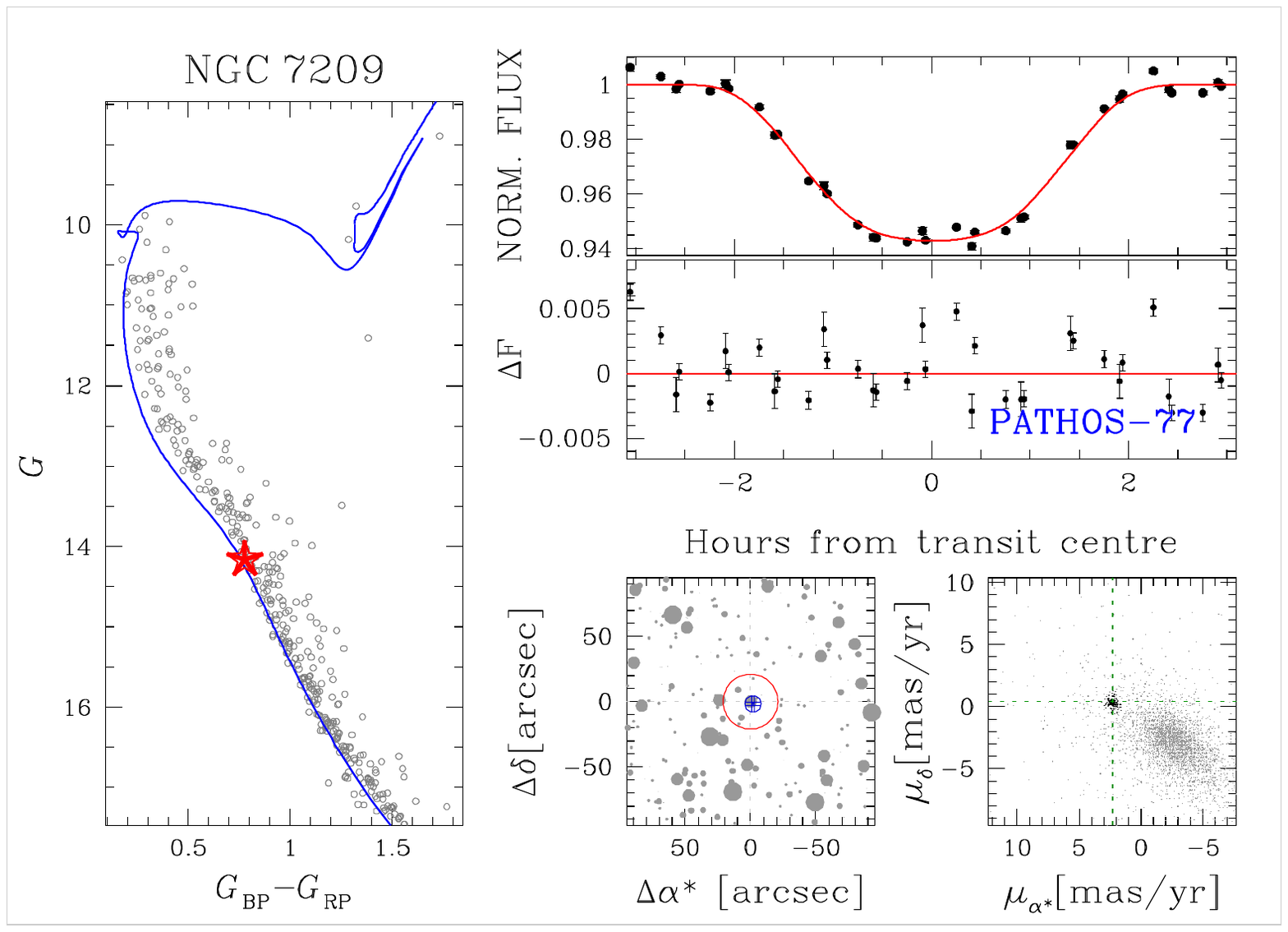}
\includegraphics[bb=77 360 535 691, width=0.33\textwidth]{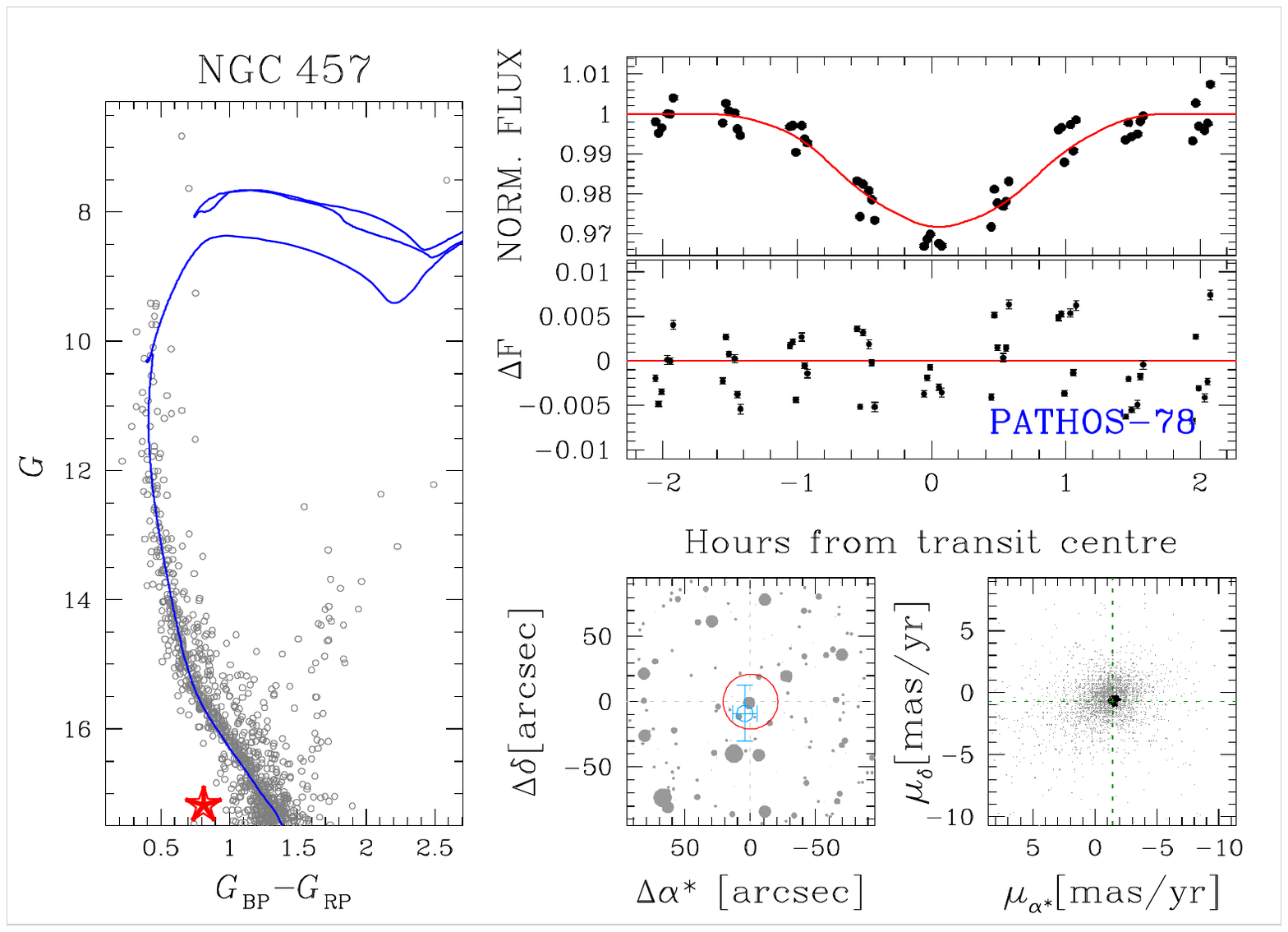}
\includegraphics[bb=77 360 535 691, width=0.33\textwidth]{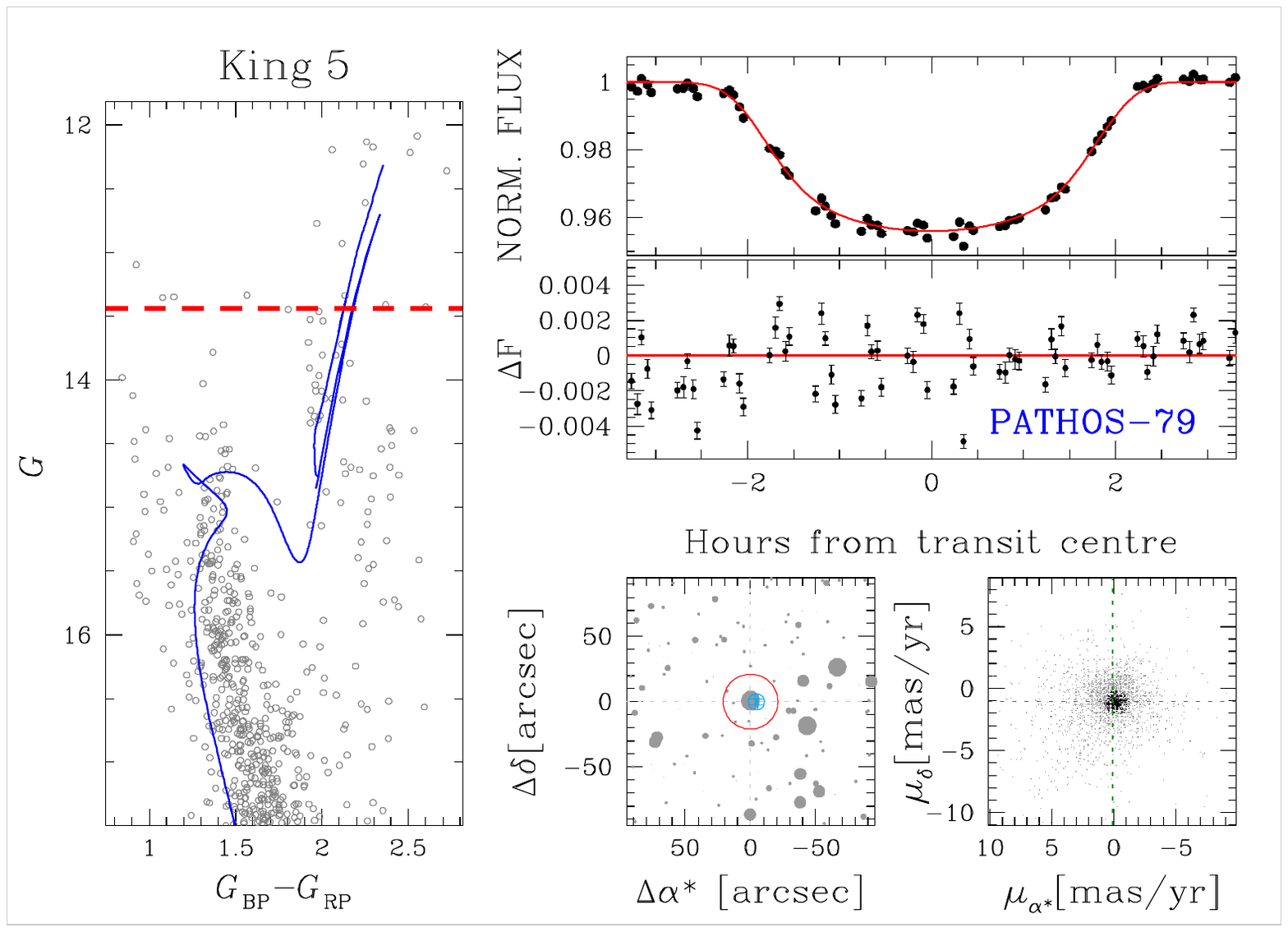} \\
\includegraphics[bb=77 360 535 691, width=0.33\textwidth]{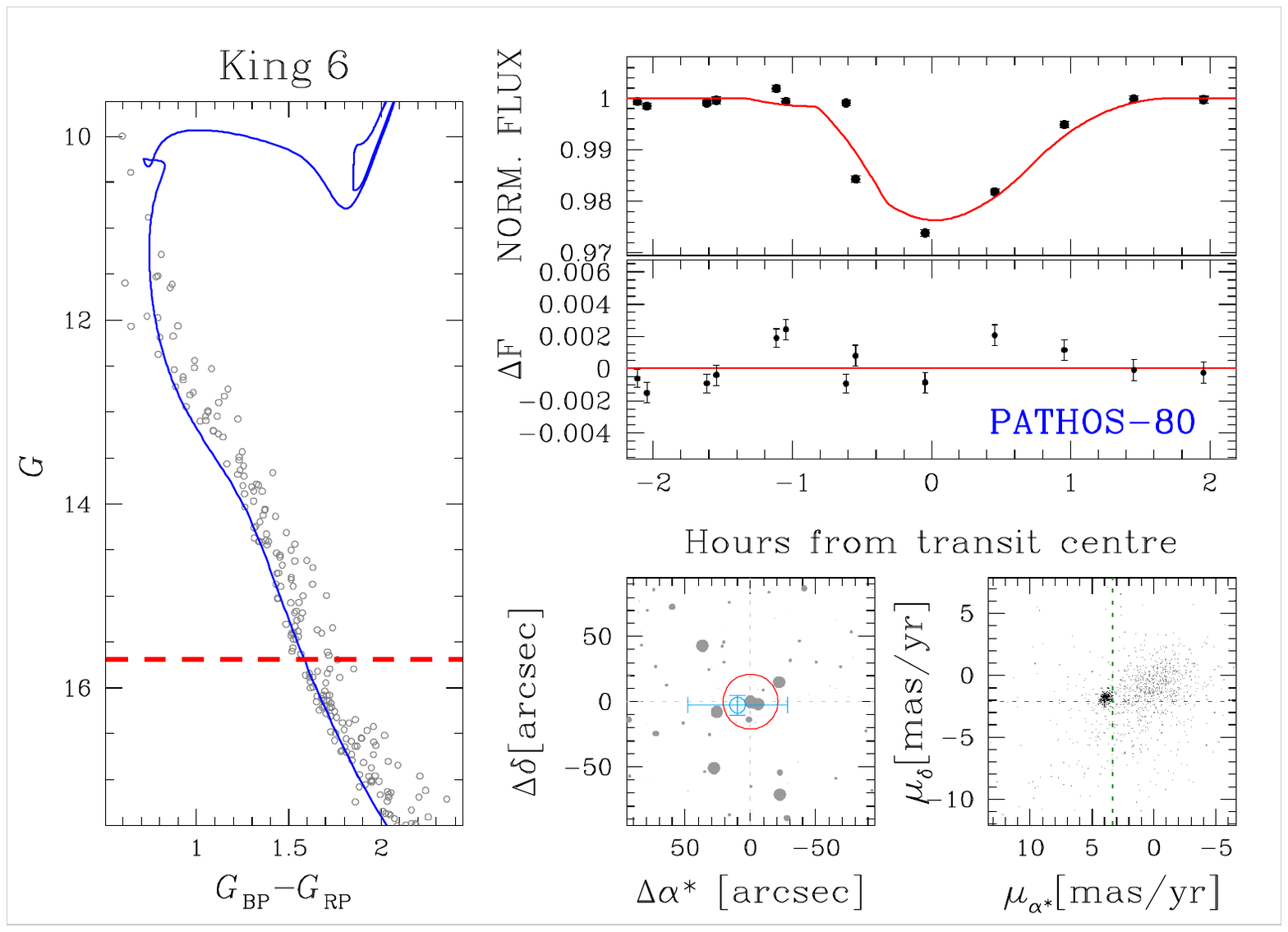}
\includegraphics[bb=77 360 535 691, width=0.33\textwidth]{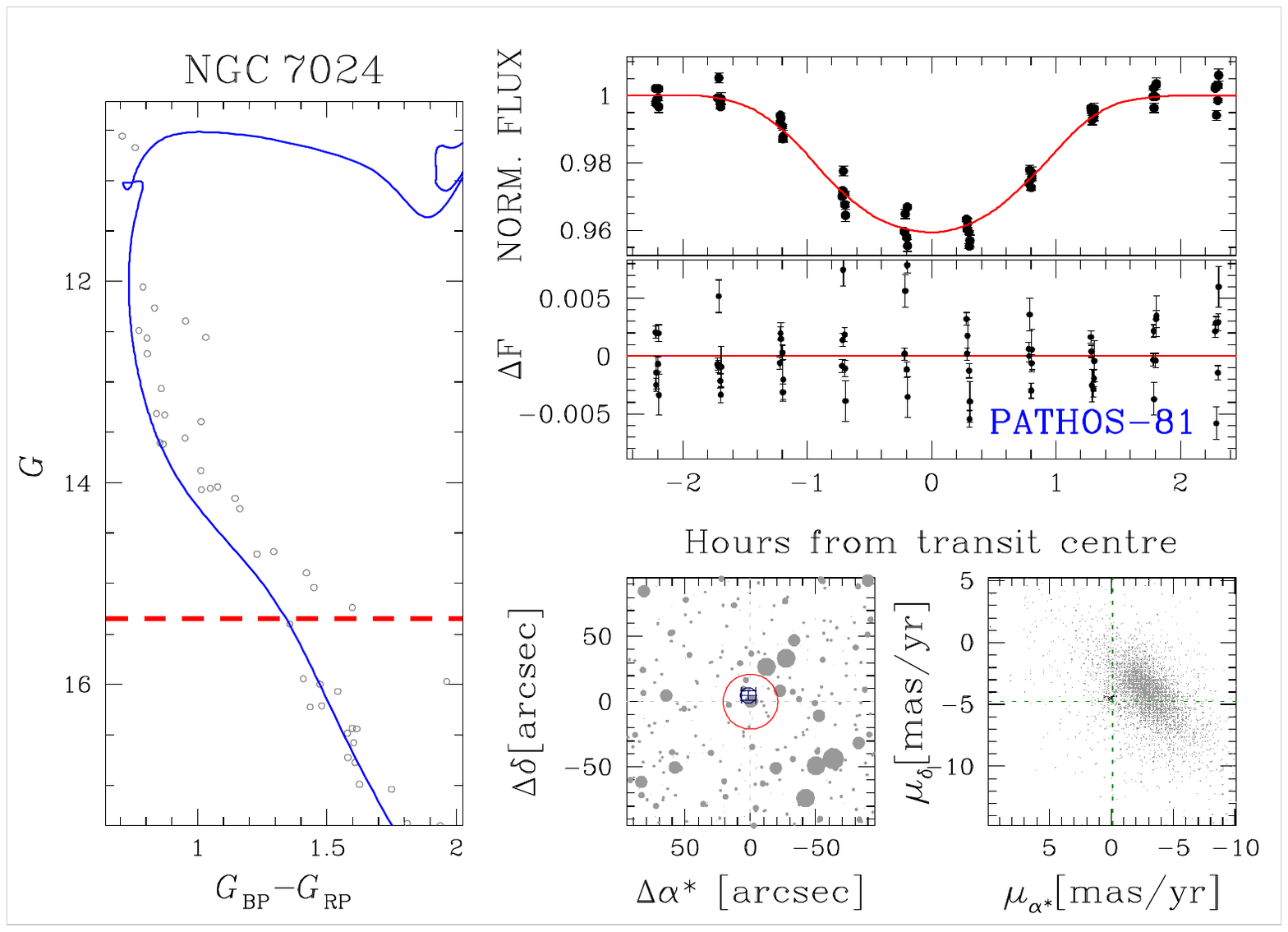}
\includegraphics[bb=77 360 535 691, width=0.33\textwidth]{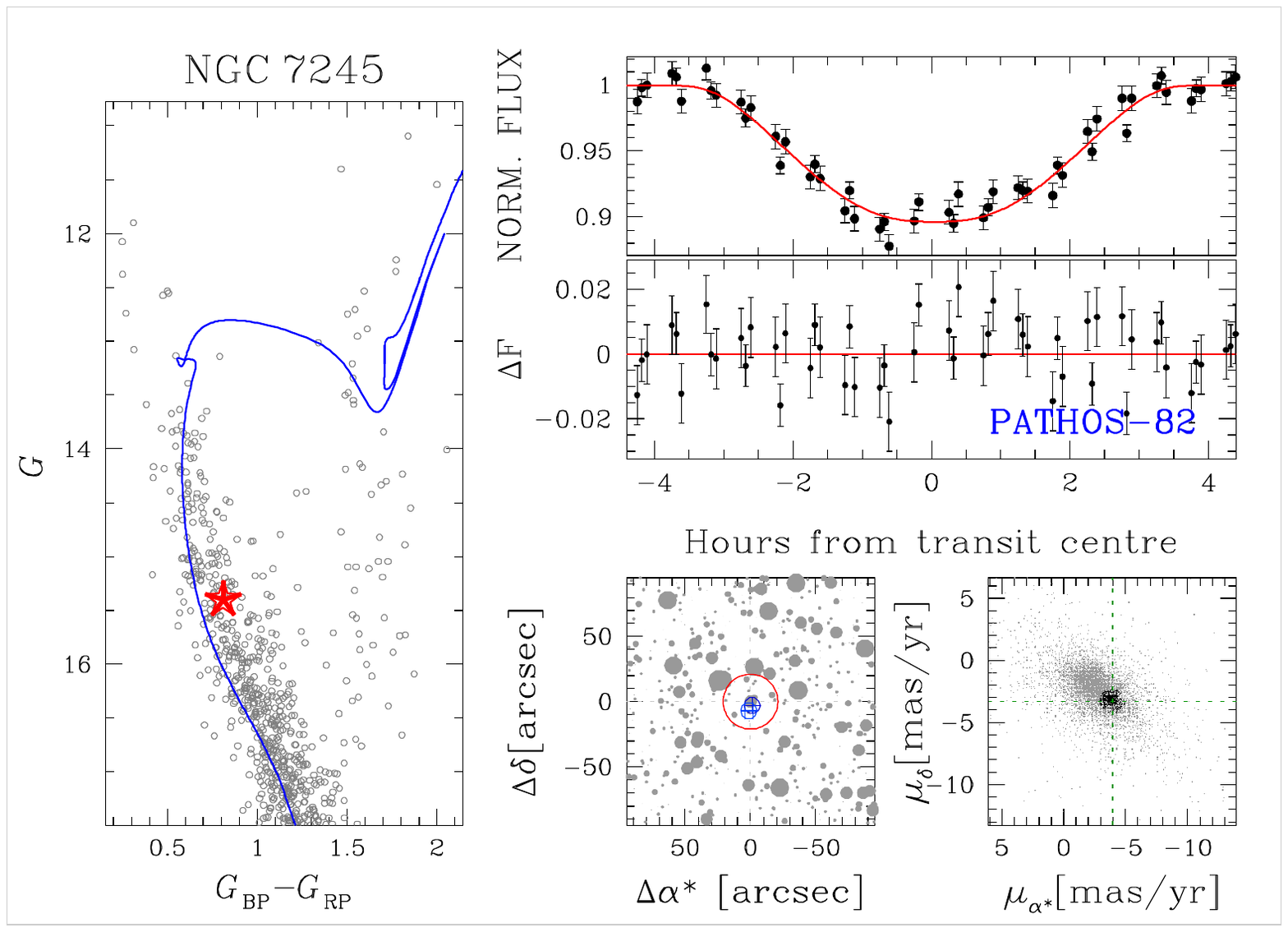} \\
\caption{As in Fig.~\ref{fig:A1a}, but for PATHOS-74 -- PATHOS-82.
  \label{fig:A1c}}
\end{figure*}

%%%%%%%%%%%%%%%%%%%%%%%%%%%%%%%%%%%%%%%%%%%%%%%%%%
\begin{table*}
  \caption{Star parameters and priors for the modelling.}
  \resizebox{0.95\textwidth}{!}{
  \input{figure/table2.tex}
  }
  \label{tab:2}
\end{table*}

%\begin{table*}
%  \renewcommand{\arraystretch}{1.5}
%  \caption{Results of transit modelling}
%  \label{tab:3}
%    \resizebox{.99\textwidth}{!}{
%      \input{figure/table3.tex}
%    }
%\end{table*}

\label{lastpage}
\end{document}

%% file: figure/table1.tex
\begin{tabular}{l  S[table-format=4.0(3)]   S[table-format=5.0(3)]  c  c }
  \hline
  Cluster Name & {Age}   & {Distance} & $E(B-V)$  & Reference \\
               & {(Myr)} &   {(pc)}   &             &    \\ 
  \hline
             ASCC\,13       &     44 \pm     4 &        1078 \pm   105 & $0.22 \pm 0.02$ &     (1) \\
           Alessi\,37       &    133 \pm    13 &         707 \pm    46 & $0.25 \pm 0.03$ &     (1) \\
   Alessi\,Teutsch\,5       &     74 \pm     7 &         876 \pm    70 & $0.50 \pm 0.05$ &     (2) \\
          Czernik\,44       &     32 \pm     3 &        4696 \pm  1500 & $1.13 \pm 0.11$ &     (2) \\
            FSR\,0342       &    376 \pm    38 &        2687 \pm   570 & $0.65 \pm 0.07$ &     (2) \\
         Gulliver\,49       &    200 \pm    20 &        1622 \pm   425 & $1.15 \pm 0.12$ &     (4) \\
             IC\,1396       &      1 \pm     1 &         913 \pm    76 & $0.42 \pm 0.04$ &     (2) \\
              King\,5       &   1230 \pm   123 &        2523 \pm   510 & $0.67 \pm 0.07$ &     (2) \\
              King\,6       &    382 \pm    38 &         727 \pm    50 & $0.59 \pm 0.06$ &     (1) \\
             King\,20       &    349 \pm    35 &        1093 \pm   305 & $0.67 \pm 0.07$ &     (1) \\
             NGC\,225       &    179 \pm    18 &         684 \pm    44 & $0.27 \pm 0.03$ &     (1) \\
             NGC\,457       &     24 \pm     2 &        2882 \pm   650 & $0.60 \pm 0.06$ &     (2) \\
             NGC\,752       &   1479 \pm   148 &         441 \pm    20 & $0.05 \pm 0.01$ &     (1) \\
             NGC\,884       &     16 \pm     2 &        2341 \pm   445 & $0.56 \pm 0.06$ &     (2) \\
            NGC\,1027       &    355 \pm    35 &        1097 \pm    98 & $0.45 \pm 0.05$ &     (2) \\
            NGC\,6811       &    863 \pm    86 &        1112 \pm   101 & $0.07 \pm 0.01$ &     (1) \\
            NGC\,6871       &     10 \pm     1 &        1841 \pm   285 & $0.60 \pm 0.06$ &     (2) \\
            NGC\,6910       &     34 \pm     3 &        1350 \pm   260 & $1.20 \pm 0.12$ &     (2) \\
            NGC\,6940       &   1023 \pm   102 &        1025 \pm    94 & $0.21 \pm 0.02$ &     (1) \\
            NGC\,6997       &    552 \pm    55 &         865 \pm    70 & $0.53 \pm 0.05$ &     (1) \\
            NGC\,7024       &    266 \pm    27 &        1182 \pm   130 & $0.63 \pm 0.06$ &     (2) \\
            NGC\,7086       &    116 \pm    12 &        1616 \pm   225 & $0.77 \pm 0.08$ &     (2) \\
            NGC\,7142       &   1778 \pm   178 &        2376 \pm   446 & $0.45 \pm 0.05$ &     (2) \\
            NGC\,7209       &    341 \pm    34 &        1178 \pm   125 & $0.18 \pm 0.02$ &     (1) \\
            NGC\,7245       &    355 \pm    35 &        3307 \pm   882 & $0.48 \pm 0.05$ &     (2) \\
            NGC\,7510       &     50 \pm     5 &        3177 \pm   765 & $0.95 \pm 0.09$ &     (2) \\
            NGC\,7654       &     79 \pm     8 &        1600 \pm   220 & $0.65 \pm 0.07$ &     (2) \\
            NGC\,7789       &   1841 \pm   184 &        2074 \pm   366 & $0.22 \pm 0.02$ &     (2) \\
               RSG\,5       &     50 \pm     5 &         336 \pm    11 & $0.04 \pm 0.00$ &     (3) \\
               RSG\,8       &    316 \pm    32 &         446 \pm    20 & $0.04 \pm 0.00$ &     (3) \\
              SAI\,25       &    243 \pm    24 &        2194 \pm   498 & $1.17 \pm 0.12$ &     (2) \\
             SAI\,149       &    251 \pm    25 &        3000 \pm   690 & $1.24 \pm 0.12$ &     (2) \\
  \hline
  \multicolumn{4}{l}{{(1)~\citet{2019A&A...623A.108B}; (2)~\citet{2016A&A...585A.101K}; (3)~\citet{2016A&A...595A..22R}}}; (4) This work \\
\end{tabular}

%% file: figure/table3.tex
\begin{tabular}{l c l r r r r r r r r r l}
  \hline
  TIC & PATHOS & Cluster & \multicolumn{1}{c}{$P$} & \multicolumn{1}{c}{$T_0$}  & \multicolumn{1}{c}{$R_{\rm p}/R_{\star}$} & \multicolumn{1}{c}{$b$} &  \multicolumn{1}{c}{$a$}  &  \multicolumn{1}{c}{$\rho_{\star}$}   &  \multicolumn{1}{c}{$i$}   &  \multicolumn{1}{c}{$R_{\rm p}$}  &  \multicolumn{1}{c}{$R_{\rm p}$} &    Note  \\
      &        &        & \multicolumn{1}{c}{(d)} & \multicolumn{1}{c}{(BTJD)} &                                         &                         &  \multicolumn{1}{c}{(au)} & \multicolumn{1}{c}{($\rho_{\sun}$)}  & \multicolumn{1}{c}{(deg)}  & \multicolumn{1}{c}{($R_{\rm J}$)} & \multicolumn{1}{c}{($R_{\earth}$)} &   \\
  \hline
  0013416465 &  44 &          NGC\,6910 &  $      5.5776_{ -0.0524}^{+  0.0587}$ & $   1688.665_{-0.117}^{+0.105}$ & $0.282_{-0.081}^{+0.144}$ & $ 1.00_{ -0.13}^{+0.18}$ & $0.0860_{-0.0005}^{+0.0006}$ & $ 0.41_{ -0.01}^{+  0.01}$ & $ 84.2_{-1.1}^{+0.8}$  & $  5.16_{-1.49}^{+2.64}$     & $   57.79_{ -16.7}^{+  29.6}$ &    \\
  0013866376 &  45 &          NGC\,6910 &  $      9.3972_{ -0.0031}^{+  0.0032}$ & $   1686.476_{-0.005}^{+0.005}$ & $0.075_{-0.006}^{+0.014}$ & $ 0.93_{ -0.02}^{+0.03}$ & $0.1575_{-0.0018}^{+0.0018}$ & $ 0.12_{ -0.01}^{+  0.01}$ & $ 84.2_{-0.2}^{+0.2}$  & $  2.68_{-0.22}^{+0.52}$     & $   30.06_{  -2.5}^{+   5.8}$ &    \\
  0013875852 &  46 &          NGC\,6910 &  $      8.3300_{ -0.0014}^{+  0.0014}$ & $   1684.099_{-0.005}^{+0.005}$ & $0.255_{-0.063}^{+0.166}$ & $ 0.98_{ -0.10}^{+0.21}$ & $0.1110_{-0.0003}^{+0.0003}$ & $ 0.43_{ -0.01}^{+  0.01}$ & $ 85.7_{-0.9}^{+0.5}$  & $  4.54_{-1.13}^{+2.95}$     & $   50.91_{ -12.7}^{+  33.1}$ &    \\
  0050361536 &  47 &          NGC\,1027 &  $     12.9725_{ -0.0086}^{+  0.0088}$ & $   1793.465_{-0.006}^{+0.006}$ & $0.327_{-0.100}^{+0.106}$ & $ 1.22_{ -0.11}^{+0.11}$ & $0.1482_{-0.0014}^{+0.0013}$ & $ 0.15_{ -0.01}^{+  0.01}$ & $ 84.3_{-0.5}^{+0.5}$  & $  8.28_{-2.54}^{+2.68}$     & $   92.75_{ -28.4}^{+  30.0}$ &    \\
  0051022999 &  48 &          NGC\,1027 &  $      7.7336_{ -0.0019}^{+  0.0020}$ & $   1794.506_{-0.003}^{+0.003}$ & $0.219_{-0.006}^{+0.006}$ & $ 0.08_{ -0.05}^{+0.08}$ & $0.0736_{-0.0003}^{+0.0003}$ & $ 1.77_{ -0.01}^{+  0.01}$ & $ 89.8_{-0.2}^{+0.2}$  & $  1.69_{-0.04}^{+0.04}$     & $   18.99_{  -0.5}^{+   0.5}$ &    \\
  0065557265 &  49 &          NGC\,7789 &  $      1.6828_{ -0.1092}^{+  0.0002}$ & $   1785.339_{-0.063}^{+0.011}$ & $0.114_{-0.094}^{+0.138}$ & $ 0.75_{ -0.11}^{+0.22}$ & $0.0318_{-0.0013}^{+0.0004}$ & $ 0.55_{ -0.34}^{+  0.32}$ & $ 79.8_{-1.8}^{+2.7}$  & $  1.42_{-1.18}^{+3.29}$     & $   15.93_{ -13.2}^{+  36.9}$ &    \\
  0067424670 &  50 &           NGC\,752 &  $      1.1328_{ -0.0001}^{+  0.0001}$ & $   1790.941_{-0.001}^{+0.001}$ & $0.188_{-0.006}^{+0.006}$ & $ 0.71_{ -0.03}^{+0.03}$ & $0.0200_{-0.0001}^{+0.0001}$ & $ 1.92_{ -0.01}^{+  0.01}$ & $ 82.8_{-0.3}^{+0.4}$  & $  1.39_{-0.04}^{+0.05}$     & $   15.53_{  -0.5}^{+   0.5}$ &    \\
  0106235729 &  51 &          NGC\,6871 &  $      4.0550_{ -0.0003}^{+  0.0003}$ & $   1684.418_{-0.002}^{+0.002}$ & $0.366_{-0.051}^{+0.060}$ & $ 1.11_{ -0.06}^{+0.07}$ & $0.0730_{-0.0003}^{+0.0003}$ & $ 0.40_{ -0.01}^{+  0.01}$ & $ 81.9_{-0.5}^{+0.5}$  & $  7.08_{-0.99}^{+1.16}$     & $   79.41_{ -11.1}^{+  13.0}$ &    \\
  0154304816 &  52 &         Alessi\,37 &  $      3.8552_{ -0.0009}^{+  0.0009}$ & $   1742.164_{-0.002}^{+0.002}$ & $0.259_{-0.119}^{+0.162}$ & $ 1.12_{ -0.15}^{+0.17}$ & $0.0537_{-0.0001}^{+0.0001}$ & $ 0.57_{ -0.01}^{+  0.01}$ & $ 82.5_{-1.2}^{+1.0}$  & $  3.39_{-1.57}^{+2.12}$     & $   38.02_{ -17.6}^{+  23.8}$ &    \\
  0185779182 &  53 &             RSG\,5 &  $     70.0427_{-19.5648}^{+ 19.4350}$ & $   1711.871_{-0.004}^{+0.004}$ & $0.228_{-0.011}^{+0.012}$ & $ 0.49_{ -0.25}^{+0.12}$ & $0.3046_{-0.0598}^{+0.0541}$ & $ 1.74_{ -0.17}^{+  0.17}$ & $ 89.7_{-0.1}^{+0.1}$  & $  1.69_{-0.10}^{+0.12}$     & $   18.99_{  -1.1}^{+   1.3}$ &    \\
  0251494772 &  54 &            SAI\,25 &  $      5.3034_{ -0.0023}^{+  0.0024}$ & $   1795.088_{-0.006}^{+0.005}$ & $0.121_{-0.004}^{+0.004}$ & $ 0.72_{ -0.04}^{+0.04}$ & $0.0894_{-0.0014}^{+0.0014}$ & $ 0.05_{ -0.01}^{+  0.01}$ & $ 81.4_{-0.9}^{+0.8}$  & $  4.70_{-0.29}^{+0.34}$     & $   52.64_{  -3.3}^{+   3.8}$ &    \\
  0251975224 &  55 &           King\,20 &  $      3.5618_{ -0.0008}^{+  0.0008}$ & $   1956.781_{-0.004}^{+0.003}$ & $0.187_{-0.004}^{+0.004}$ & $ 0.08_{ -0.06}^{+0.08}$ & $0.0531_{-0.0001}^{+0.0001}$ & $ 0.46_{ -0.01}^{+  0.01}$ & $ 89.4_{-0.6}^{+0.4}$  & $  2.73_{-0.06}^{+0.06}$     & $   30.65_{  -0.7}^{+   0.7}$ &    \\
  0260167199 &  56 &           IC\,1396 &  $     17.6348_{ -0.0077}^{+  0.0078}$ & $   1741.326_{-0.005}^{+0.005}$ & $0.312_{-0.106}^{+0.120}$ & $ 1.13_{ -0.13}^{+0.13}$ & $0.1494_{-0.0004}^{+0.0004}$ & $ 0.46_{ -0.01}^{+  0.01}$ & $ 87.1_{-0.3}^{+0.3}$  & $  4.44_{-1.51}^{+1.70}$     & $   49.72_{ -16.9}^{+  19.0}$ &    \\
  0269519402 &  57 &       Gulliver\,49 &  $      3.3736_{ -0.0000}^{+  0.0000}$ & $   1806.271_{-0.001}^{+0.001}$ & $0.120_{-0.002}^{+0.002}$ & $ 0.39_{ -0.07}^{+0.05}$ & $0.0630_{-0.0005}^{+0.0005}$ & $ 0.18_{ -0.01}^{+  0.01}$ & $ 85.9_{-0.7}^{+0.8}$  & $  2.95_{-0.09}^{+0.09}$     & $   33.07_{  -1.0}^{+   1.0}$ &    \\
  0270022396 &  58 &          NGC\,7654 &  $      3.7786_{ -0.0000}^{+  0.0000}$ & $   1807.429_{-0.001}^{+0.001}$ & $0.347_{-0.056}^{+0.055}$ & $ 1.12_{ -0.07}^{+0.07}$ & $0.0666_{-0.0002}^{+0.0002}$ & $ 0.36_{ -0.01}^{+  0.01}$ & $ 81.1_{-0.5}^{+0.6}$  & $  6.66_{-1.08}^{+1.05}$     & $   74.71_{ -12.1}^{+  11.8}$ &    \\
  0270618239 &  59 &          NGC\,6811 &  $      2.6377_{ -0.0002}^{+  0.0002}$ & $   1683.548_{-0.002}^{+0.002}$ & $0.217_{-0.003}^{+0.003}$ & $ 0.03_{ -0.02}^{+0.03}$ & $0.0381_{-0.0001}^{+0.0001}$ & $ 1.08_{ -0.01}^{+  0.01}$ & $ 89.8_{-0.2}^{+0.1}$  & $  2.10_{-0.03}^{+0.03}$     & $   23.56_{  -0.3}^{+   0.4}$ &    \\
  0270920839 &  60 &        Czernik\,44 &  $      5.2236_{ -0.0001}^{+  0.0001}$ & $   1768.090_{-0.002}^{+0.002}$ & $0.151_{-0.010}^{+0.008}$ & $ 0.81_{ -0.03}^{+0.02}$ & $0.1088_{-0.0012}^{+0.0011}$ & $ 0.12_{ -0.02}^{+  0.02}$ & $ 82.6_{-0.6}^{+0.6}$  & $  5.42_{-0.35}^{+0.39}$     & $   60.80_{  -3.9}^{+   4.3}$ &    \\
  0271443321 &  61 &           SAI\,149 &  $      5.9718_{ -0.0001}^{+  0.0001}$ & $   1765.415_{-0.002}^{+0.002}$ & $0.393_{-0.026}^{+0.030}$ & $ 0.89_{ -0.04}^{+0.05}$ & $0.0970_{-0.0019}^{+0.0019}$ & $ 0.07_{ -0.01}^{+  0.01}$ & $ 80.9_{-0.5}^{+0.4}$  & $ 14.17_{-1.00}^{+1.15}$     & $  158.79_{ -11.2}^{+  12.9}$ &    \\
  0285249796 &  62 &           ASCC\,13 &  $      4.7520_{ -0.0009}^{+  0.0009}$ & $   1816.836_{-0.002}^{+0.002}$ & $0.257_{-0.008}^{+0.008}$ & $ 0.64_{ -0.04}^{+0.03}$ & $0.0664_{-0.0001}^{+0.0001}$ & $ 0.52_{ -0.01}^{+  0.01}$ & $ 86.2_{-0.2}^{+0.2}$  & $  3.72_{-0.12}^{+0.12}$     & $   41.70_{  -1.3}^{+   1.4}$ &    \\
  0298292983 &  63 &          NGC\,6940 &  $      1.2816_{ -0.0131}^{+  0.0001}$ & $   1701.051_{-0.052}^{+0.002}$ & $0.146_{-0.016}^{+0.005}$ & $ 0.76_{ -0.08}^{+0.05}$ & $0.0267_{-0.0002}^{+0.0001}$ & $ 0.33_{ -0.02}^{+  0.02}$ & $ 77.2_{-1.1}^{+1.5}$  & $  2.37_{-0.28}^{+0.14}$     & $   26.60_{  -3.2}^{+   1.6}$ &    \\
  0316246231 &  64 &            King\,6 &  $      8.5751_{ -0.0020}^{+  0.0020}$ & $   1795.982_{-0.002}^{+0.002}$ & $0.360_{-0.065}^{+0.073}$ & $ 1.12_{ -0.08}^{+0.09}$ & $0.0992_{-0.0002}^{+0.0002}$ & $ 0.43_{ -0.01}^{+  0.01}$ & $ 85.2_{-0.4}^{+0.4}$  & $  5.63_{-1.01}^{+1.14}$     & $   63.14_{ -11.3}^{+  12.8}$ &    \\
  0323717669 &  65 &             RSG\,8 &  $      4.2378_{ -0.0015}^{+  0.0018}$ & $   1955.852_{-0.007}^{+0.006}$ & $0.169_{-0.008}^{+0.008}$ & $ 0.14_{ -0.10}^{+0.14}$ & $0.0453_{-0.0002}^{+0.0002}$ & $ 2.66_{ -0.01}^{+  0.01}$ & $ 89.5_{-0.5}^{+0.4}$  & $  1.05_{-0.05}^{+0.05}$     & $   11.77_{  -0.6}^{+   0.5}$ &    \\
  0326483210 &  66 &          FSR\,0342 &  $      3.8130_{ -0.0004}^{+  0.0004}$ & $   1741.329_{-0.003}^{+0.003}$ & $0.200_{-0.010}^{+0.007}$ & $ 0.35_{ -0.10}^{+0.07}$ & $0.0603_{-0.0005}^{+0.0005}$ & $ 0.33_{ -0.01}^{+  0.01}$ & $ 87.2_{-0.6}^{+0.8}$  & $  3.55_{-0.17}^{+0.14}$     & $   39.80_{  -2.0}^{+   1.6}$ &    \\
  0332258412 &  67 &           NGC\,457 &  $      7.1465_{ -0.0035}^{+  0.0035}$ & $   1794.905_{-0.004}^{+0.004}$ & $0.371_{-0.068}^{+0.074}$ & $ 1.10_{ -0.09}^{+0.09}$ & $0.1235_{-0.0005}^{+0.0005}$ & $ 0.22_{ -0.01}^{+  0.01}$ & $ 83.3_{-0.5}^{+0.6}$  & $ 10.19_{-1.88}^{+2.03}$     & $  114.24_{ -21.0}^{+  22.8}$ &    \\
  0334949878 &  68 & Alessi\,Teutsch\,5 &  $     19.4784_{ -0.0002}^{+  0.0002}$ & $   1741.392_{-0.001}^{+0.001}$ & $0.349_{-0.048}^{+0.055}$ & $ 1.26_{ -0.05}^{+0.06}$ & $0.2126_{-0.0011}^{+0.0010}$ & $ 0.28_{ -0.01}^{+  0.01}$ & $ 86.4_{-0.2}^{+0.2}$  & $  7.76_{-1.06}^{+1.22}$     & $   87.01_{ -11.9}^{+  13.7}$ &    \\
  0348608380 &  69 &           NGC\,884 &  $     16.3026_{ -0.2298}^{+  0.5559}$ & $   1806.922_{-0.003}^{+0.003}$ & $0.380_{-0.058}^{+0.064}$ & $ 1.11_{ -0.07}^{+0.08}$ & $0.2476_{-0.0024}^{+0.0056}$ & $ 0.19_{ -0.02}^{+  0.02}$ & $ 85.9_{-0.3}^{+0.3}$  & $ 12.60_{-1.93}^{+2.22}$     & $  141.26_{ -21.6}^{+  24.9}$ &    \\
  0356973763 &  70 &          NGC\,6997 &  $     10.0250_{ -0.4778}^{+  0.0035}$ & $   1720.499_{-0.328}^{+0.155}$ & $0.104_{-0.017}^{+0.113}$ & $ 0.92_{ -0.18}^{+0.09}$ & $0.1182_{-0.0029}^{+0.0014}$ & $ 0.13_{ -0.02}^{+  0.02}$ & $ 84.5_{-0.9}^{+1.1}$  & $  2.71_{-0.58}^{+2.94}$     & $   30.34_{  -6.5}^{+  33.0}$ &    \\
  0377619148 &  71 &          NGC\,7510 &  $     19.8963_{ -0.1048}^{+  0.0004}$ & $   1784.408_{-0.002}^{+0.002}$ & $0.373_{-0.077}^{+0.077}$ & $ 1.17_{ -0.09}^{+0.09}$ & $0.2713_{-0.0065}^{+0.0062}$ & $ 0.05_{ -0.01}^{+  0.01}$ & $ 84.1_{-0.4}^{+0.5}$  & $ 18.65_{-3.85}^{+3.81}$     & $  209.06_{ -43.2}^{+  42.7}$ &    \\
  0408094816 &  72 &          NGC\,7142 &  $      5.4037_{ -0.0001}^{+  0.0001}$ & $   1743.980_{-0.002}^{+0.003}$ & $0.172_{-0.005}^{+0.005}$ & $ 0.52_{ -0.13}^{+0.08}$ & $0.0711_{-0.0003}^{+0.0003}$ & $ 0.09_{ -0.01}^{+  0.02}$ & $ 84.8_{-1.2}^{+1.5}$  & $  4.41_{-0.37}^{+0.38}$     & $   49.40_{  -4.2}^{+   4.2}$ &    \\
  0408358709 &  73 &          NGC\,7142 &  $      6.3320_{ -0.0001}^{+  0.0001}$ & $   1741.434_{-0.002}^{+0.002}$ & $0.337_{-0.090}^{+0.108}$ & $ 1.03_{ -0.13}^{+0.13}$ & $0.0778_{-0.0003}^{+0.0003}$ & $ 0.14_{ -0.01}^{+  0.02}$ & $ 82.1_{-1.0}^{+1.2}$  & $  7.40_{-2.12}^{+2.29}$     & $   82.92_{ -23.8}^{+  25.7}$ &    \\
  0417058223 &  74 &          NGC\,7086 &  $      5.4767_{ -0.0005}^{+  0.0004}$ & $   1712.263_{-0.002}^{+0.002}$ & $0.347_{-0.078}^{+0.086}$ & $ 1.18_{ -0.09}^{+0.10}$ & $0.0853_{-0.0003}^{+0.0003}$ & $ 0.31_{ -0.01}^{+  0.01}$ & $ 82.3_{-0.6}^{+0.6}$  & $  7.04_{-1.57}^{+1.74}$     & $   78.89_{ -17.6}^{+  19.5}$ &    \\
  0420288086 &  75 &           NGC\,225 &  $      6.5343_{ -0.0002}^{+  0.0002}$ & $   1769.585_{-0.005}^{+0.006}$ & $0.043_{-0.002}^{+0.002}$ & $ 0.85_{ -0.02}^{+0.02}$ & $0.0950_{-0.0006}^{+0.0006}$ & $ 0.25_{ -0.01}^{+  0.01}$ & $ 84.7_{-0.1}^{+0.2}$  & $  0.93_{-0.04}^{+0.05}$     & $   10.46_{  -0.5}^{+   0.5}$ & (1) \\
  0421630760 &  76 &           IC\,1396 &  $      4.1869_{ -0.0008}^{+  0.0008}$ & $   1742.859_{-0.004}^{+0.005}$ & $0.095_{-0.003}^{+0.003}$ & $ 0.10_{ -0.07}^{+0.10}$ & $0.0556_{-0.0001}^{+0.0001}$ & $ 0.47_{ -0.01}^{+  0.01}$ & $ 89.4_{-0.7}^{+0.5}$  & $  1.30_{-0.05}^{+0.05}$     & $   14.61_{  -0.5}^{+   0.5}$ &    \\
  0427943483 &  77 &          NGC\,7209 &  $      6.2981_{ -0.0007}^{+  0.0007}$ & $   1738.788_{-0.001}^{+0.001}$ & $0.238_{-0.004}^{+0.004}$ & $ 0.70_{ -0.02}^{+0.02}$ & $0.0741_{-0.0005}^{+0.0005}$ & $ 0.55_{ -0.01}^{+  0.01}$ & $ 86.6_{-0.1}^{+0.1}$  & $  3.14_{-0.06}^{+0.07}$     & $   35.21_{  -0.7}^{+   0.8}$ &    \\
  0602870459 &  78 &           NGC\,457 &  $      3.1259_{ -0.0010}^{+  0.0010}$ & $   1792.949_{-0.004}^{+0.004}$ & $0.333_{-0.097}^{+0.109}$ & $ 1.09_{ -0.13}^{+0.13}$ & $0.0469_{-0.0002}^{+0.0002}$ & $ 0.55_{ -0.02}^{+  0.02}$ & $ 81.5_{-1.0}^{+1.0}$  & $  4.43_{-1.29}^{+1.45}$     & $   49.62_{ -14.4}^{+  16.3}$ &    \\
  0645455722 &  79 &            King\,5 &  $      2.7663_{ -0.0004}^{+  0.0004}$ & $   1794.690_{-0.001}^{+0.001}$ & $0.190_{-0.007}^{+0.009}$ & $ 0.35_{ -0.24}^{+0.20}$ & $0.0484_{-0.0012}^{+0.0011}$ & $ 0.25_{ -0.06}^{+  0.04}$ & $ 86.2_{-2.7}^{+2.6}$  & $  3.65_{-0.27}^{+0.52}$     & $   40.88_{  -3.1}^{+   5.8}$ &    \\
  0645713782 &  80 &            King\,6 &  $      6.8572_{ -0.3488}^{+  0.0287}$ & $   1790.975_{-0.009}^{+0.006}$ & $0.317_{-0.126}^{+0.124}$ & $ 1.09_{ -0.17}^{+0.15}$ & $0.0715_{-0.0023}^{+0.0004}$ & $ 1.23_{ -0.01}^{+  0.01}$ & $ 86.1_{-0.6}^{+0.6}$  & $  2.92_{-1.16}^{+1.14}$     & $   32.76_{ -13.0}^{+  12.8}$ &    \\
  1961935435 &  81 &          NGC\,7024 &  $      2.7707_{ -0.0005}^{+  0.0005}$ & $   1713.398_{-0.002}^{+0.002}$ & $0.202_{-0.008}^{+0.012}$ & $ 0.77_{ -0.03}^{+0.04}$ & $0.0430_{-0.0001}^{+0.0001}$ & $ 0.55_{ -0.01}^{+  0.01}$ & $ 83.5_{-0.3}^{+0.3}$  & $  2.68_{-0.11}^{+0.16}$     & $   30.00_{  -1.2}^{+   1.8}$ &    \\
  2015243161 &  82 &          NGC\,7245 &  $     11.5026_{ -0.0042}^{+  0.0041}$ & $   1746.036_{-0.010}^{+0.010}$ & $0.325_{-0.024}^{+0.076}$ & $ 0.68_{ -0.07}^{+0.19}$ & $0.1267_{-0.0006}^{+0.0006}$ & $ 0.32_{ -0.01}^{+  0.01}$ & $ 87.3_{-0.7}^{+0.3}$  & $  5.87_{-0.45}^{+1.40}$     & $   65.83_{  -5.0}^{+  15.6}$ &    \\
\hline
\multicolumn{13}{l}{$^{(1)}$~Also in the TOI catalogue} \\
%multicolumn{13}{l}{$^{(2)}$~Radius too large, suspected eclipsing binary} \\
%multicolumn{13}{l}{$^{(3)}$~Another single deeper (suspected) transit ($\delta_{\rm T}\sim 5$~\%) is present in the light curve due to a likely stellar companion or second planet. } \\
%multicolumn{13}{l}{$^{(4)}$~Likely field star. } \\
\end{tabular}

%% file: figure/table7.tex
\begin{tabular} {c | c c c c  }
\hline
\multicolumn{1}{c}{}      & \multicolumn{4}{c}{Detection efficiency (\%)} \\
\hline
\hline
 \multicolumn{1}{c|}{} & \multicolumn{4}{c}{$0.85~R_{\rm Earth} < R_{\rm P} \leq 3.9~R_{\rm Earth}$ } \\
\hline
\multicolumn{1}{c|}{$T$}   & 0.5--   & 2.0-- & 10.0--  & 85.0--  \\
\multicolumn{1}{c|}{(mag)} & 2.0~d   &10.0~d & 85.0~d  & 365.0~d \\
\hline
 6--7  & $40.00\pm  23.66$ & $10.00\pm 10.00$ &$  <0.01            $ & $<0.01$        \\  
 7--8  & $40.38\pm   7.38$ & $14.42\pm  3.98$ &$  5.21\pm2.39      $ & $3.85\pm 1.96$ \\  
 8--9  & $24.79\pm   3.64$ & $10.26\pm  2.20$ &$  2.75\pm1.14      $ & $2.14\pm 0.97$ \\  
 9--10 & $16.86\pm   2.37$ & $ 8.00\pm  1.57$ &$  4.95\pm1.27      $ & $2.86\pm 0.92$ \\  
10--11 & $14.29\pm   2.16$ & $ 7.43\pm  1.51$ &$  4.62\pm1.22      $ & $4.86\pm 1.21$ \\  
11--12 & $10.29\pm   1.80$ & $ 4.29\pm  1.13$ &$  5.79\pm1.37      $ & $3.43\pm 1.01$ \\  
12--13 & $ 6.00\pm   1.35$ & $ 9.71\pm  1.75$ &$  4.10\pm1.16      $ & $4.57\pm 1.17$ \\  
13--14 & $ 6.00\pm   1.35$ & $ 6.57\pm  1.41$ &$  4.13\pm1.17      $ & $2.00\pm 0.76$ \\  
14--15 & $ 5.43\pm   1.28$ & $ 4.29\pm  1.13$ &$  2.41\pm0.92      $ & $1.14\pm 0.57$ \\  
15--16 & $ 2.86\pm   0.92$ & $ 3.14\pm  0.96$ &$  0.35\pm0.35      $ & $1.43\pm 0.64$ \\  
16--17 & $ 2.29\pm   0.82$ & $ 2.00\pm  0.76$ &$  0.69\pm0.49      $ & $0.86\pm 0.50$ \\  
17--18 & $ 1.91\pm   0.97$ & $ 0.96\pm  0.68$ &$  <0.01            $ & $<0.01$        \\  
\hline
\hline
 \multicolumn{1}{c|}{} & \multicolumn{4}{c}{$3.9~R_{\rm Earth} < R_{\rm P} \leq 11.2~R_{\rm Earth}$}  \\
\hline
%\multicolumn{1}{c|}{$T$}   & 0.5--   & 2.0-- & 10.0--  & 85.0--  \\
%\multicolumn{1}{c|}{(mag)} & 2.0~d   &10.0~d & 85.0~d  & 365.0~d \\
%\hline
 6--7  & $80.00\pm  37.95  $ & $ 60.00\pm30.98 $ &$ 10.00\pm    10.00 $ & $ <0.01 $ \\  
 7--8  & $82.69\pm  12.05  $ & $ 56.73\pm 9.25 $ &$  6.73\pm     2.63 $ & $ 3.85 \pm 1.96 $ \\  
 8--9  & $75.64\pm   7.54  $ & $ 52.56\pm 5.85 $ &$  5.13\pm     1.52 $ & $ 2.56 \pm 1.06 $ \\  
 9--10 & $73.14\pm   6.02  $ & $ 40.57\pm 4.04 $ &$  5.43\pm     1.28 $ & $ 3.14 \pm 0.96 $ \\  
10--11 & $64.00\pm   5.48  $ & $ 40.86\pm 4.05 $ &$  2.86\pm     0.92 $ & $ 5.71 \pm 1.31 $ \\  
11--12 & $52.29\pm   4.77  $ & $ 30.37\pm 3.37 $ &$  8.00\pm     1.57 $ & $ 3.43 \pm 1.01 $ \\  
12--13 & $47.14\pm   4.45  $ & $ 22.57\pm 2.81 $ &$  3.71\pm     1.05 $ & $ 2.86 \pm 0.92 $ \\  
13--14 & $44.86\pm   4.31  $ & $ 28.57\pm 3.24 $ &$  4.29\pm     1.13 $ & $ 3.43 \pm 1.01 $ \\  
14--15 & $45.43\pm   4.34  $ & $ 21.14\pm 2.71 $ &$  6.00\pm     1.35 $ & $ 1.14 \pm 0.57 $ \\  
15--16 & $38.00\pm   3.87  $ & $ 18.29\pm 2.49 $ &$  2.57\pm     0.87 $ & $ 2.00 \pm 0.76 $ \\  
16--17 & $24.29\pm   2.94  $ & $ 11.43\pm 1.91 $ &$  0.86\pm     0.50 $ & $ 1.43 \pm 0.64 $ \\  
17--18 & $ 8.13\pm   2.05  $ & $  1.91\pm 0.97 $ &$  <0.01            $ & $ <0.01 $ \\  
\hline
\hline
 \multicolumn{1}{c|}{} & \multicolumn{4}{c}{$1.0~R_{\rm J} < R_{\rm P} \leq 2.5~R_{\rm J}$} \\ 
\hline
%\multicolumn{1}{c|}{$T$}   & 0.5--   & 2.0-- & 10.0--  & 85.0--  \\
%\multicolumn{1}{c|}{(mag)} & 2.0~d   &10.0~d & 85.0~d  & 365.0~d \\
%\hline
 6--7  & $80.00\pm  37.95$ & $ 70.00\pm 34.50 $ &$  <0.01$          & $<0.01   $ \\  
 7--8  & $97.12\pm  13.57$ & $ 87.50\pm 12.56 $ &$ 11.54 \pm 3.52 $ & $2.88\pm 1.69   $ \\  
 8--9  & $90.17\pm   8.56$ & $ 78.21\pm  7.72 $ &$ 11.54 \pm 2.35 $ & $2.14\pm 0.97   $ \\  
 9--10 & $90.00\pm   6.99$ & $ 78.00\pm  6.30 $ &$ 12.29 \pm 1.99 $ & $4.00\pm 1.09   $ \\  
10--11 & $90.00\pm   6.99$ & $ 69.71\pm  5.81 $ &$  7.71 \pm 1.54 $ & $3.71\pm 1.05   $ \\  
11--12 & $84.57\pm   6.68$ & $ 70.57\pm  5.86 $ &$  8.29 \pm 1.60 $ & $6.29\pm 1.38   $ \\  
12--13 & $85.14\pm   6.71$ & $ 68.86\pm  5.76 $ &$  6.29 \pm 1.38 $ & $4.86\pm 1.21   $ \\  
13--14 & $78.86\pm   6.35$ & $ 69.14\pm  5.78 $ &$  8.57 \pm 1.63 $ & $3.14\pm 0.96   $ \\  
14--15 & $69.43\pm   5.80$ & $ 61.43\pm  5.32 $ &$  6.57 \pm 1.41 $ & $2.57\pm 0.87   $ \\  
15--16 & $51.14\pm   4.70$ & $ 42.86\pm  4.18 $ &$  3.71 \pm 1.05 $ & $1.43\pm 0.64   $ \\  
16--17 & $53.71\pm   4.86$ & $ 29.71\pm  3.32 $ &$  3.71 \pm 1.05 $ & $0.57\pm 0.41   $ \\  
17--18 & $46.41\pm   5.70$ & $ 28.23\pm  4.16 $ &$  <0.01 $         & $<0.01   $ \\  
\hline
\end{tabular}

%% file: figure/table5.tex
\begin{tabular}{l c c l c c}
  \hline
  TIC & PATHOS & FPP & Scenario~(prob.) & FPP$_{\rm c}$ & RUWE \\
\hline
\multicolumn{6}{c}{Southern Ecliptic hemisphere} \\
\hline
0039291805 &  3    &  0.92 &   BEB~(0.51) &  0.87 & 1.023 \\
0088977253 &  6    &  0.76 &   EB~(0.47)  &  0.72 & 1.079 \\
0125414447 &  9    &  0.98 &   BEB~(0.79) &  0.93 & 0.929 \\
0126600730 & 10    &  0.97 &   BEB~(0.97) &  0.92 & 0.971 \\
0159059181 & 16    &  0.99 &   BEB~(0.52) &  0.94 & 1.230 \\
0306385801 & 20    &  0.51 &   EB~(0.51)  &  0.48 & 0.851 \\
0308538095 & 21    &  0.00 &   pl~(1.00)  &  0.00 & 0.895 \\
0372913337 & 23    &  0.74 &   BEB~(0.73) &  0.70 & 1.133 \\
0410450228 & 25    &  0.46 &   pl~(0.54)  &  0.44 & 0.901 \\
0432564189 & 28    &  0.96 &   BEB~(0.92) &  0.91 & 1.078 \\
0460205581 & 30    &  0.21 &   pl~(0.79)  &  0.20 & 0.985 \\
0460950389 & 31    &  0.22 &   pl~(0.78)  &  0.21 & 0.891 \\
0748919024 & 33    &  0.00 &   pl~(1.00)  &  0.00 & 2.638 \\
1036769612 & 34    &  0.63 &   BEB~(0.63) &  0.60 & 1.062 \\
\hline
\multicolumn{6}{c}{Northern Ecliptic hemisphere} \\
\hline
0051022999 & 48    & 0.95 & BEB~(0.88)  & 0.90 &  0.947 \\
0065557265 & 49    & 1.00 & BEB~(0.41)  & 0.95 &  1.066 \\
0067424670 & 50    & 0.53 & pl~(0.47)   & 0.50 &  0.989 \\
0185779182 & 53    & 0.36 & pl~(0.64)   & 0.34 &  1.086 \\
0270618239 & 59    & 0.28 & pl~(0.72)   & 0.27 &  1.226 \\
0298292983 & 63    & 1.00 & EB~(0.67)   & 0.95 &  1.112 \\
0323717669 & 65    & 0.98 & BEB~(0.96)  & 0.93 &  1.893 \\
0420288086 & 75    & 0.00 & pl~(1.00)   & 0.00 &  0.971 \\
0421630760 & 76    & 1.00 & BEB~(0.93)  & 0.95 &  1.029 \\
\hline
\end{tabular}

%% file: figure/table6a.tex
\begin{tabular}{r | c c c | c c c | c c c }
  \hline
 & \multicolumn{3}{c|}{$0.85~R_{\rm Earth} < R_{\rm P} \leq 3.9~R_{\rm Earth}$ }& \multicolumn{3}{c|}{$3.9~R_{\rm Earth} < R_{\rm P} \leq 11.2~R_{\rm Earth}$}  & \multicolumn{3}{c}{$1.0~R_{\rm J} < R_{\rm P} \leq 2.5~R_{\rm J}$} \\ 
\hline
\multicolumn{1}{c|}{Period} & $f^{\rm min}_{\star}$ & $f^{\rm mid}_{\star}$ & $f^{\rm max}_{\star}$ & $f^{\rm min}_{\star}$ & $f^{\rm mid}_{\star}$ & $f^{\rm max}_{\star}$ & $f^{\rm min}_{\star}$ & $f^{\rm mid}_{\star}$ & $f^{\rm max}_{\star}$\\
\multicolumn{1}{c|}{(d)}  &    (\%) & (\%) & (\%) &    (\%) & (\%) & (\%) &    (\%) & (\%) & (\%) \\
\hline
0.5--2.0    &  $<0.034$ & $<0.063$  & $<0.086$      &  $<0.004$        & $<0.007$         & $<0.009$         &  $<0.002$         & $<0.004$         & $<0.006$          \\  
2.0--10.0   &  $<0.092$ & $<0.191$  & $<0.269$      &  $0.034\pm0.025$ & $0.070\pm0.052$  & $0.098\pm 0.073$ &  $0.023\pm 0.013$ & $0.047\pm0.027$  & $0.066 \pm 0.038$ \\  
10.0--85.0  &  $<0.533$ & $<1.505$  & $<2.218$      &  $<0.327$        & $<0.925$         & $<1.364$         &  $0.555\pm0.336 $ & $1.568\pm0.955 $ & $2.312\pm1.413  $ \\  
85.0--365.0 &  $<2.716$ & $<5.200$  & $<7.174$      &  $<2.197$        & $<4.205$         & $<5.806$         &  $<2.102$         & $<4.022 $         & $<5.553 $         \\  
\hline
\end{tabular}

%% file: figure/table4a.tex
\begin{tabular}{l c c c c l | l c c c c l}
\hline
Object & Cluster/Association   & Age & Period & $R_{\rm P}$  & Reference  & Object & Cluster/Association   & Age & Period & $R_{\rm P}$  & Reference \\
       &                       & (Myr) & (d) & ($R_{\rm Earth}$) &         &   &                             & (Myr) & (d) & ($R_{\rm Earth}$) &        \\
\hline 
K-66b            & NGC\,6811     & $863 ^{+30   }_{-30 }$     & $17.815815 ^{+0.000075}_{-0.000075}$     & $2.80^{+0.16}_{-0.16}$                & \citet{2013Natur.499...55M}     &  K2-33b           &  Upp-Sco      & $9.3^{+1.1}_{-1.1}$       & $5.424865 ^{+0.000035 }_{-0.000031}$      & $5.04    ^{+0.34}_{-0.37}$            & \citet{2016AJ....152...61M}  \\
K-67b            & NGC\,6811     & $863 ^{+30   }_{-30 }$     & $15.72590  ^{+0.00011 }_{-0.00011 }$     & $2.94^{+0.16}_{-0.16}$                & \citet{2013Natur.499...55M}     &  TOI-200b         &  Tuc-Hor      & $40 ^{+5  }_{-5 }$        & $8.1387   ^{+0.0005   }_{-0.0005  }$      & $5.63    ^{+0.22}_{-0.21}$            & \citet{2019AA...630A..81B}   \\
HD\,283869b      & Hyades        & $728 ^{+71   }_{-71 }$     & $106       ^{+74.0   }_{-25.0   } $      & $1.96^{+0.13}_{-0.16}$                 & \citet{2018AJ....156...46V}    &  TOI-451b         &  Psc-Eri      & $134^{+6.5}_{-6.5}$        & $1.858703^{+0.000025}_{-0.000035}$       & $1.91^{+0.12}_{-0.12}$                & \citet{2021AJ....161...65N}  \\
K2-25b           & Hyades        & $728 ^{+71   }_{-71 }$     & $3.484552  ^{+0.000031}_{-0.000037}$     & $3.43^{+0.95}_{-0.31}$                 & \citet{2016ApJ...818...46M}    &  TOI-451c         &  Psc-Eri      & $134^{+6.5}_{-6.5}$        & $9.192522^{+0.00006}_{-0.00010}$         & $3.10^{+0.13}_{-0.13}$                & \citet{2021AJ....161...65N}  \\
K2-136Ab         & Hyades        & $728 ^{+71   }_{-71 }$     & $7.975292  ^{+0.000833}_{-0.000770}$     & $0.99^{+0.06}_{-0.04}$                 & \citet{2018AJ....155....4M}    &  TOI-451d         &  Psc-Eri      & $134^{+6.5}_{-6.5}$        & $16.364988^{+0.00044}_{-0.000044}$       & $4.07^{+0.15}_{-0.15}$                & \citet{2021AJ....161...65N}  \\
K2-136Ac         & Hyades        & $728 ^{+71   }_{-71 }$     & $17.307137 ^{+0.000252}_{-0.000284}$     & $2.91^{+0.11}_{-0.10}$                 & \citet{2018AJ....155....4M}    &  TOI-1098b        &  Melange-1    & $250^{+50}_{-70}$         & $10.18271^{+0.0004}_{-0.00004}$          & $3.2^{+0.1}_{-0.1}$                   & \citet{2021arXiv210206066T}   \\
K2-136Ad         & Hyades        & $728 ^{+71   }_{-71 }$     & $25.575065 ^{+0.002418}_{-0.002357}$     & $1.45^{+0.11}_{-0.08}$                 & \citet{2018AJ....155....4M}    &  HD\,63433b       &  UMa          & $414^{+23 }_{-23}$        & $7.10801  ^{+0.00046  }_{-0.00034 }$      & $2.15    ^{+0.10}_{-0.10}$            & \citet{2020AJ....160..179M}  \\
K2-95b           & Praesepe      & $670 ^{+67   }_{-67 }$     & $10.135091 ^{+0.000495}_{-0.000488}$     & $3.7 ^{+0.2 }_{-0.2 }$                & \citet{2017AJ....153...64M}     &  HD\,63433c       &  UMa          & $414^{+23 }_{-23}$        & $20.5455  ^{+0.0011   }_{-0.0011  }$      & $2.64    ^{+0.12}_{-0.12}$            & \citet{2020AJ....160..179M}  \\
K2-100b          & Praesepe      & $670 ^{+67   }_{-67 }$     & $1.673915  ^{+0.000011}_{-0.000011}$     & $3.5 ^{+0.2 }_{-0.2 }$                & \citet{2017AJ....153...64M}     &  HIP\,67522b      &  Sco-Cen      & $17 ^{+2  }_{-2 }$        & $6.959503 ^{+0.000016 }_{-0.000015}$      & $10.07   ^{+0.47}_{-0.47}$            & \citet{2020AJ....160...33R}  \\
K2-101b          & Praesepe      & $670 ^{+67   }_{-67 }$     & $14.677303 ^{+0.000824}_{-0.000809}$     & $3.0 ^{+0.1 }_{-0.1 }$                & \citet{2017AJ....153...64M}     &  HIP\,67522c      &  Sco-Cen      & $17 ^{+2  }_{-2 }$        & $54.0     ^{+70.0     }_{-24.0    }$      & $8.01    ^{+0.75}_{-0.71}$            & \citet{2020AJ....160...33R}  \\
K2-102b          & Praesepe      & $670 ^{+67   }_{-67 }$     & $9.915651  ^{+0.001194}_{-0.001175}$     & $1.3 ^{+0.1 }_{-0.1 }$                & \citet{2017AJ....153...64M}     &  AU~Mic~b         &  AU~Mic       & $22 ^{+3  }_{-3 }$        & $8.46321  ^{+0.00004  }_{-0.00004 }$      & $4.20    ^{+0.20}_{-0.20}$            & \citet{2020Natur.582..497P}  \\
K2-103b          & Praesepe      & $670 ^{+67   }_{-67 }$     & $21.169687 ^{+0.001636}_{-0.001655}$     & $2.2 ^{+0.2 }_{-0.1 }$                & \citet{2017AJ....153...64M}     &  AU~Mic~c         &  AU~Mic       & $22 ^{+3  }_{-3 }$        & $30.0     ^{+6.0      }_{-6.0     }$      & $2.35    ^{+0.67}_{-0.67}$            & \citet{2020Natur.582..497P}  \\
K2-104b          & Praesepe      & $670 ^{+67   }_{-67 }$     & $1.974189  ^{+0.000110}_{-0.000109}$     & $1.9 ^{+0.2 }_{-0.1 }$                & \citet{2017AJ....153...64M}     &  V~1298c          &  Tau          & $23 ^{+4  }_{-4 }$        & $8.24958  ^{+0.00072  }_{-0.00072 }$      & $5.59    ^{+0.36}_{-0.32}$            & \citet{2019ApJ...885L..12D}  \\
K2-264b          & Praesepe      & $670 ^{+67   }_{-67 }$     & $5.839770  ^{+0.000063}_{-0.000063}$     & $2.27^{+0.20}_{-0.16}$                & \citet{2018AJ....156..195R}     &  V~1298d          &  Tau          & $23 ^{+4  }_{-4 }$        & $12.4032  ^{+0.0015   }_{-0.0015  }$      & $6.41    ^{+0.45}_{-0.40}$            & \citet{2019ApJ...885L..12D}  \\
K2-264c          & Praesepe      & $670 ^{+67   }_{-67 }$     & $19.663650 ^{+0.000303}_{-0.000306}$     & $2.77^{+0.20}_{-0.18}$                & \citet{2018AJ....156..195R}     &  V~1298b          &  Tau          & $23 ^{+4  }_{-4 }$        & $24.1396  ^{+0.0018   }_{-0.0018  }$      & $10.27   ^{+0.58}_{-0.53}$            & \citet{2019ApJ...885L..12D}  \\
K2-231b          &  Ruprecht\,147 & $3000^{+250  }_{-250}$    & $13.841901 ^{+0.001352}_{-0.001352}$     & $2.5 ^{+0.2 }_{-0.2 }$                & \citet{2019AJ....158...77C}     &  V~1298e          &  Tau          & $23 ^{+4  }_{-4 }$        & $60       ^{+60       }_{-18      }$      & $8.74    ^{+0.84}_{-0.72}$            & \citet{2019ApJ...885L..12D}  \\
K2-284b          &  Cas-Tau      & $120^{+640}_{-20}$        & $4.79507  ^{+0.00012  }_{-0.00012 }$      & $2.77    ^{+0.12}_{-0.12}$            & \citet{2018AJ....156..302D}     &    \\
\hline
\end{tabular}

%% file: figure/table2.tex
\begin{tabular}{l c l S[table-format=4.0(4)] S[table-format=4.0(4)] S[table-format=3.0(1)] c c l l c c c}
  \hline
  TIC & PATHOS & Cluster & {$\alpha$} & {$\delta$} & {$T$} & $R_\star$ & $M_\star$ & \multicolumn{1}{c}{Period} &  \multicolumn{1}{c}{$T_0$} & LD$_{c1}$ & LD$_{c2}$ & $df$  \\
  &        &            & {(deg.)}   &  {(deg.)}  &  {(mag.) }   & ($R_{\sun}$) & ($M_{\sun}$) &  \multicolumn{1}{c}{(d)}    &  \multicolumn{1}{c}{(BTJD)} &         &       &    \\
  \hline
0013416465 & 44 &          NGC\,6910 &  305.4104 &   40.6532   & 14.0 &  $  1.88\pm 0.05 $ & $ 2.74\pm 0.05$ & $\mathcal{U}(5.5, 6.0)   $   &  $\mathcal{U}(1688, 1689)       $ &  $0.15\pm0.10$ &  $0.18\pm0.10$         & $0.02\pm0.05$     \\
0013866376 & 45 &          NGC\,6910 &  305.6858 &   40.5015   & 11.6 &  $  3.68\pm 0.08 $ & $ 5.90\pm 0.19$ & $\mathcal{U}(9, 9.5)     $   &  $\mathcal{U}(1686.0, 1687.0)   $ &  $0.06\pm0.10$ &  $0.10\pm0.10$         & $0.04\pm0.01$     \\
0013875852 & 46 &          NGC\,6910 &  305.7604 &   40.9818   & 13.9 &  $  1.84\pm 0.05 $ & $ 2.64\pm 0.05$ & $\mathcal{U}(8.0, 8.5)   $   &  $\mathcal{U}(1683.5, 1684.5)   $ &  $0.15\pm0.10$ &  $0.18\pm0.10$         & $0.14\pm0.05$     \\
0050361536 & 47 &          NGC\,1027 &   40.4690 &   61.7671   & 11.6 &  $  2.64\pm 0.12 $ & $ 2.58\pm 0.07$ & $\mathcal{U}(12.6, 13.4) $   &  $\mathcal{U}(1793, 1794)       $ &  $0.15\pm0.10$ &  $0.20\pm0.10$         & $0.01\pm0.05$     \\
0051022999 & 48 &          NGC\,1027 &   41.9287 &   61.6554   & 16.2 &  $  0.80\pm 0.03 $ & $ 0.89\pm 0.00$ & $\mathcal{U}(7.5, 8.0)   $   &  $\mathcal{U}(1794.0, 1795.0)   $ &  $0.37\pm0.10$ &  $0.33\pm0.10$         & $0.92\pm0.05$     \\
0065557265 & 49 &          NGC\,7789 &  359.1359 &   57.0475   & 14.0 &  $  2.06\pm 0.67 $ & $ 1.54\pm 0.03$ & $\mathcal{U}(1.5, 2.0)   $   &  $\mathcal{U}(1785.2, 1785.5)   $ &  $0.28\pm0.10$ &  $0.28\pm0.10$         & $0.01\pm0.05$     \\
0067424670 & 50 &           NGC\,752 &   29.0476 &   37.9182   & 13.6 &  $  0.76\pm 0.03 $ & $ 0.83\pm 0.05$ & $\mathcal{U}(0.9, 1.3)   $   &  $\mathcal{U}(1790.5, 1791.5)   $ &  $0.39\pm0.10$ &  $0.35\pm0.10$         & $0.08\pm0.05$     \\
0106235729 & 51 &          NGC\,6871 &  301.0821 &   35.7889   & 13.0 &  $  2.01\pm 0.04 $ & $ 3.16\pm 0.06$ & $\mathcal{U}(3.7,4.3)    $   &  $\mathcal{U}(1684, 1685)       $ &  $0.13\pm0.10$ &  $0.16\pm0.10$         & $0.03\pm0.05$     \\
0154304816 & 52 &         Alessi\,37 &  342.4106 &   46.3500   & 12.7 &  $  1.35\pm 0.03 $ & $ 1.39\pm 0.05$ & $\mathcal{U}(3.5, 4.0)   $   &  $\mathcal{U}(1742.0, 1743.0)   $ &  $0.27\pm0.10$ &  $0.28\pm0.10$         & $0.02\pm0.05$     \\
0185779182 & 53 &             RSG\,5 &  302.5300 &   45.0274   & 13.3 &  $  0.76\pm 0.05 $ & $ 0.77\pm 0.07$ & $\mathcal{U}(0.1,99)     $   &  $\mathcal{U}(1711, 1712)       $ &  $0.42\pm0.10$ &  $0.36\pm0.10$         & $0.02\pm0.03$     \\
0251494772 & 54 &            SAI\,25 &   45.3145 &   57.1103   & 13.3 &  $  4.42\pm 0.35 $ & $ 3.39\pm 0.16$ & $\mathcal{U}(5.0, 5.5)   $   &  $\mathcal{U}(1795, 1796)       $ &  $0.15\pm0.10$ &  $0.19\pm0.10$         & $0.01\pm0.05$     \\
0251975224 & 55 &           King\,20 &  353.1928 &   58.3744   & 14.0 &  $  1.49\pm 0.04 $ & $ 1.57\pm 0.06$ & $\mathcal{U}(3.3, 3.9)   $   &  $\mathcal{U}(1956, 1957)       $ &  $0.23\pm0.10$ &  $0.26\pm0.10$         & $0.04\pm0.02$     \\
0260167199 & 56 &           IC\,1396 &  325.3457 &   57.1089   & 13.5 &  $  1.43\pm 0.03 $ & $ 1.34\pm 0.05$ & $\mathcal{U}(17.3, 18.0) $   &  $\mathcal{U}(1741.0, 1742.0)   $ &  $0.33\pm0.10$ &  $0.31\pm0.10$         & $0.10\pm0.05$     \\
0269519402 & 57 &       Gulliver\,49 &  350.6974 &   61.9366   & 13.7 &  $  2.54\pm 0.07 $ & $ 2.93\pm 0.07$ & $\mathcal{U}(3.0, 3.5)   $   &  $\mathcal{U}(1806.0, 1806.5)   $ &  $0.15\pm0.10$ &  $0.18\pm0.10$         & $0.12\pm0.05$     \\
0270022396 & 58 &          NGC\,7654 &  351.2741 &   61.6596   & 13.1 &  $  2.00\pm 0.05 $ & $ 2.76\pm 0.06$ & $\mathcal{U}(3.6, 3.9)   $   &  $\mathcal{U}(1807.3, 1807.5)   $ &  $0.15\pm0.10$ &  $0.18\pm0.10$         & $0.18\pm0.05$     \\
0270618239 & 59 &          NGC\,6811 &  293.6701 &   46.2956   & 14.5 &  $  0.99\pm 0.03 $ & $ 1.07\pm 0.05$ & $\mathcal{U}(2.2, 3.0)   $   &  $\mathcal{U}(1683.0, 1684.0)   $ &  $0.32\pm0.30$ &  $0.30\pm0.30$         & $0.05\pm0.05$     \\
0270920839 & 60 &        Czernik\,44 &  353.5841 &   62.0029   & 14.3 &  $  3.94\pm 0.09 $ & $ 6.30\pm 0.22$ & $\mathcal{U}(5.0, 5.5)   $   &  $\mathcal{U}(1767.5, 1768.5)   $ &  $0.05\pm0.10$ &  $0.08\pm0.10$         & $0.70\pm0.05$     \\
0271443321 & 61 &           SAI\,149 &  354.4998 &   60.5393   & 13.9 &  $  5.00\pm 0.32 $ & $ 3.41\pm 0.17$ & $\mathcal{U}(5.7, 6.2)   $   &  $\mathcal{U}(1765.0, 1766.0)   $ &  $0.15\pm0.10$ &  $0.21\pm0.10$         & $0.04\pm0.05$     \\
0285249796 & 62 &           ASCC\,13 &   78.2136 &   44.5317   & 12.7 &  $  1.49\pm 0.03 $ & $ 1.73\pm 0.05$ & $\mathcal{U}(4.5, 5.0)   $   &  $\mathcal{U}(1816, 1817)       $ &  $0.20\pm0.10$ &  $0.24\pm0.10$         & $0.31\pm0.05$     \\
0298292983 & 63 &          NGC\,6940 &  308.6667 &   27.8038   & 12.7 &  $  1.67\pm 0.05 $ & $ 1.56\pm 0.05$ & $\mathcal{U}(1.2, 1.3)   $   &  $\mathcal{U}(1700.9, 1701.2)   $ &  $0.24\pm0.10$ &  $0.26\pm0.10$         & $0.13\pm0.05$     \\
0316246231 & 64 &            King\,6 &   51.9527 &   56.5553   & 12.4 &  $  1.62\pm 0.03 $ & $ 1.77\pm 0.05$ & $\mathcal{U}(8.3, 8.7)   $   &  $\mathcal{U}(1795.5, 1796.5)   $ &  $0.20\pm0.10$ &  $0.24\pm0.10$         & $0.01\pm0.05$     \\
0323717669 & 65 &             RSG\,8 &  348.0553 &   59.4878   & 14.6 &  $  0.64\pm 0.03 $ & $ 0.69\pm 0.05$ & $\mathcal{U}(4.0, 4.5)   $   &  $\mathcal{U}(1955.0, 1956.0)   $ &  $0.45\pm0.10$ &  $0.36\pm0.10$         & $0.70\pm0.05$     \\
0326483210 & 66 &          FSR\,0342 &  331.8775 &   53.3684   & 15.0 &  $  1.83\pm 0.05 $ & $ 2.01\pm 0.07$ & $\mathcal{U}(3.5, 4.0)   $   &  $\mathcal{U}(1741.0, 1742.0)   $ &  $0.17\pm0.10$ &  $0.22\pm0.10$         & $0.57\pm0.05$     \\
0332258412 & 67 &           NGC\,457 &   19.5199 &   58.7259   & 13.0 &  $  2.86\pm 0.06 $ & $ 4.93\pm 0.06$ & $\mathcal{U}(7.0, 7.5)   $   &  $\mathcal{U}(1794.5, 1795.5)   $ &  $0.07\pm0.10$ &  $0.12\pm0.10$         & $0.02\pm0.05$     \\
0334949878 & 68 & Alessi\,Teutsch\,5 &  331.7424 &   60.5581   & 11.0 &  $  2.38\pm 0.06 $ & $ 3.38\pm 0.05$ & $\mathcal{U}(19.0, 20.0) $   &  $\mathcal{U}(1741.0, 1742.0)   $ &  $0.13\pm0.10$ &  $0.16\pm0.10$         & $0.01\pm0.05$     \\
0348608380 & 69 &           NGC\,884 &   35.5582 &   57.3225   & 11.0 &  $  4.13\pm 0.08 $ & $ 7.63\pm 0.06$ & $\mathcal{U}(16, 99)     $   &  $\mathcal{U}(1806.0, 1807.5)   $ &  $0.02\pm0.10$ &  $0.01\pm0.10$         & $0.01\pm0.05$     \\
0356973763 & 70 &          NGC\,6997 &  314.4065 &   44.5798   & 11.6 &  $  2.51\pm 0.15 $ & $ 2.24\pm 0.06$ & $\mathcal{U}(9.5, 10.3)  $   &  $\mathcal{U}(1720, 1721)       $ &  $0.17\pm0.10$ &  $0.22\pm0.10$         & $0.03\pm0.05$     \\
0377619148 & 71 &          NGC\,7510 &  348.0732 &   60.4962   & 11.6 &  $  7.15\pm 0.46 $ & $ 6.77\pm 0.47$ & $\mathcal{U}(19.5, 20.0) $   &  $\mathcal{U}(1784, 1785)       $ &  $0.06\pm0.10$ &  $0.11\pm0.10$         & $0.01\pm0.05$     \\
0408094816 & 72 &          NGC\,7142 &  326.3290 &   65.7443   & 14.5 &  $  2.57\pm 0.26 $ & $ 1.64\pm 0.07$ & $\mathcal{U}(5.3,5.9)    $   &  $\mathcal{U}(1743.5, 1744.5)   $ &  $0.28\pm0.10$ &  $0.28\pm0.10$         & $0.04\pm0.05$     \\
0408358709 & 73 &          NGC\,7142 &  326.6784 &   65.7431   & 14.7 &  $  2.17\pm 0.32 $ & $ 1.57\pm 0.06$ & $\mathcal{U}(6.0,6.5)    $   &  $\mathcal{U}(1741.0, 1742.0)   $ &  $0.27\pm0.10$ &  $0.28\pm0.10$         & $0.02\pm0.05$     \\
0417058223 & 74 &          NGC\,7086 &  322.8993 &   51.7531   & 13.2 &  $  2.09\pm 0.05 $ & $ 2.76\pm 0.07$ & $\mathcal{U}(5.2, 5.8)   $   &  $\mathcal{U}(1712.0, 1713.0)   $ &  $0.15\pm0.10$ &  $0.18\pm0.10$         & $0.44\pm0.05$     \\
0420288086 & 75 &           NGC\,225 &   10.9627 &   61.8356   & 10.4 &  $  2.20\pm 0.06 $ & $ 2.69\pm 0.06$ & $\mathcal{U}(6.3, 6.9)   $   &  $\mathcal{U}(1769.0, 1770.0)   $ &  $0.15\pm0.10$ &  $0.19\pm0.10$         & $0.01\pm0.05$     \\
0421630760 & 76 &           IC\,1396 &  324.6780 &   57.8705   & 13.7 &  $  1.41\pm 0.03 $ & $ 1.32\pm 0.05$ & $\mathcal{U}(4.0, 4.5)   $   &  $\mathcal{U}(1742.0, 1743.0)   $ &  $0.33\pm0.10$ &  $0.32\pm0.10$         & $0.10\pm0.05$     \\
0427943483 & 77 &          NGC\,7209 &  331.5198 &   46.8716   & 13.7 &  $  1.35\pm 0.03 $ & $ 1.37\pm 0.05$ & $\mathcal{U}(6.0, 6.5)   $   &  $\mathcal{U}(1738, 1739)       $ &  $0.27\pm0.10$ &  $0.28\pm0.10$         & $0.01\pm0.05$     \\
0602870459 & 78 &           NGC\,457 &   19.2565 &   58.4526   & 16.7 &  $  1.37\pm 0.05 $ & $ 1.41\pm 0.07$ & $\mathcal{U}(3.0, 3.5)   $   &  $\mathcal{U}(1792.5, 1793.5)   $ &  $0.27\pm0.10$ &  $0.28\pm0.10$         & $0.95\pm0.05$     \\
0645455722 & 79 &            King\,5 &   48.2841 &   52.2956   & 13.0 &  $ 18.44\pm 0.58 $ & $ 1.98\pm 0.14$ & $\mathcal{U}(2.5, 3.0)   $   &  $\mathcal{U}(1794.5, 1795.0)   $ &  $0.43\pm0.30$ &  $0.38\pm0.30$         & $0.65\pm0.05$     \\
0645713782 & 80 &            King\,6 &   52.1609 &   56.5984   & 15.2 &  $  0.94\pm 0.03 $ & $ 1.04\pm 0.05$ & $\mathcal{U}(6.3, 6.9)   $   &  $\mathcal{U}(1790.5, 1791.5)   $ &  $0.32\pm0.10$ &  $0.30\pm0.10$         & $0.75\pm0.05$     \\
1961935435 & 81 &          NGC\,7024 &  316.4611 &   41.5474   & 14.9 &  $  1.35\pm 0.03 $ & $ 1.38\pm 0.05$ & $\mathcal{U}(2.5, 3.0)   $   &  $\mathcal{U}(1713.0, 1714.0)   $ &  $0.27\pm0.10$ &  $0.28\pm0.10$         & $0.78\pm0.05$     \\
2015243161 & 82 &          NGC\,7245 &  333.8243 &   54.2961   & 14.9 &  $  1.85\pm 0.05 $ & $ 2.06\pm 0.07$ & $\mathcal{U}(11.2, 11.8) $   &  $\mathcal{U}(1745.5, 1746.5)   $ &  $0.17\pm0.10$ &  $0.22\pm0.10$         & $0.46\pm0.05$     \\
\hline
\end{tabular}